\documentclass[prd,aps,twocolumn,preprintnumbers, showpacs,
nofootinbib,superscriptaddress,notitlepage]{revtex4-1}
\usepackage{amssymb, amsthm}
\usepackage{amsmath}    
\usepackage{graphicx}   
\usepackage{footnote}   

\usepackage{slashed}    
\usepackage{hyperref}   
\usepackage{physics}
\usepackage[utf8]{inputenc}
\usepackage{color}
\usepackage[normalem]{ulem}

\newcommand{\fm}{\text{ fm}}
\newcommand{\GeV}{\text{ GeV}}

\begin{document}

\preprint{MSUHEP-20-010}

\title{Pion and Kaon Distribution Amplitudes in the Continuum Limit}

\author{Rui Zhang}
\email{zhang60@msu.edu}
\affiliation{Department of Physics and Astronomy, Michigan State University, East Lansing, MI 48824}
\affiliation{Department of Computational Mathematics, Science \& Engineering, Michigan State University, East Lansing, MI 48824}

\author{Carson Honkala}
\affiliation{Honors College, Michigan State University, East Lansing, MI 48824}

\author{Huey-Wen Lin}
\email{hwlin@pa.msu.edu}
\affiliation{Department of Physics and Astronomy, Michigan State University, East Lansing, MI 48824}
\affiliation{Department of Computational Mathematics, Science \& Engineering, Michigan State University, East Lansing, MI 48824}

\author{Jiunn-Wei Chen}
\email{jwc@phys.ntu.edu.tw}
\affiliation{Department of Physics, Center for Theoretical Physics, and Leung Center for Cosmology and Particle Astrophysics, National Taiwan University, Taipei, Taiwan 106}

\begin{abstract}
We present a lattice-QCD calculation of the pion, kaon and $\eta_s$ distribution amplitudes using large-momentum effective theory (LaMET).
Our calculation is carried out using three ensembles with 2+1+1 flavors of highly improved staggered quarks (HISQ), generated by MILC collaboration, at 310-MeV pion mass with 0.06, 0.09 and 0.12~fm lattice spacings.
We use clover fermion action for the valence quarks and tune the quark mass to match the lightest light and strange masses in the sea.
The resulting lattice matrix elements are nonperturbatively renormalized in regularization-independent momentum-subtraction (RI/MOM) scheme and extrapolated to the continuum.
We use two approaches to extract the $x$-dependence of the meson distribution amplitudes:
1) we fit the renormalized matrix elements in coordinate space to an assumed distribution form through a one-loop matching kernel; 
2) we use a machine-learning algorithm trained on pseudo lattice-QCD data to make predictions on the lattice data. 
We found the results are consistent between these methods with the latter method giving a less smooth shape. 
Both approaches suggest that as the quark mass increases, the distribution amplitude becomes narrower.
Our pion distribution amplitude has broader distribution than predicted by light-front constituent-quark model, and the moments of our pion distributions agree with previous lattice-QCD results using the operator production expansion.
\end{abstract}

\maketitle

\section{Introduction}

Meson distribution amplitudes (DAs) $\phi_M$ hold the key to understanding how light-quark hadron masses emerge from QCD, 
an important topic of study at a future electron-ion collider~\cite{Aguilar:2019teb}.
Meson DA are also important inputs in many hard exclusive processes at large momentum transfers $Q^2 \gg\Lambda_{\text{QCD}}^2$~\cite{Beneke:1999br,Beneke:2001ev}.  
In such processes, the cross section can be factorized into a short-distance hard-scattering part and long-distance universal quantities such as the lightcone DAs.
Unlike the hard-scattering subprocess, which can be calculated perturbatively, the lightcone DAs need to be determined from fits to experimental data or to be calculated nonperturbatively from lattice QCD.  

Such direct computations have become possible recently, thanks to large-momentum effective theory (LaMET)~\cite{Ji:2013dva,Ji:2014gla,Ji:2017rah}. 
The LaMET method calculates equal-time spatial correlators (whose Fourier transforms are called quasi-distributions) on the lattice and takes the infinite-momentum limit to extract the true lightcone distribution.
For large momenta feasible in lattice simulations, LaMET can be used to relate Euclidean quasi-distributions to physical ones through a factorization theorem, which involves a matching and power corrections that are suppressed by the hadron momentum~\cite{Ji:2014gla}.
The proof of factorization was developed in  Refs.~\cite{Ma:2017pxb,Izubuchi:2018srq,Liu:2019urm}.

Since LaMET was proposed, a lot of progress has been achieved with respect to both the theoretical understanding of the formalism~\cite{Xiong:2013bka,Ji:2015jwa,Ji:2015qla,Xiong:2015nua,Ji:2014hxa,Monahan:2017hpu,Ji:2018hvs,Stewart:2017tvs,Constantinou:2017sej,Green:2017xeu,Izubuchi:2018srq,Xiong:2017jtn,Wang:2017qyg,Wang:2017eel,Xu:2018mpf,Chen:2016utp,Zhang:2017bzy,Ishikawa:2016znu,Chen:2016fxx,Ji:2017oey,Ishikawa:2017faj,Chen:2017mzz,Alexandrou:2017huk,Constantinou:2017sej,Green:2017xeu,Chen:2017mzz,Chen:2017mie,Lin:2017ani,Chen:2017lnm,Li:2016amo,Monahan:2016bvm,Radyushkin:2016hsy,Rossi:2017muf,Carlson:2017gpk,Ji:2017rah,Briceno:2018lfj,Hobbs:2017xtq,Jia:2017uul,Xu:2018eii,Jia:2018qee,Spanoudes:2018zya,Rossi:2018zkn,Liu:2018uuj,Ji:2018waw,Bhattacharya:2018zxi,Radyushkin:2018nbf,Zhang:2018diq,Li:2018tpe,Braun:2018brg,Detmold:2019ghl,Ebert:2019tvc,Ji:2019ewn,Sufian:2020vzb,Shugert:2020tgq,Green:2020xco,Chai:2020nxw,Shanahan:2020zxr,Lin:2020ssv,Braun:2020ymy,Bhattacharya:2020cen,Ji:2020ect,Ebert:2020gxr,Lin:2020ijm,Zhang:2020dkn,Bhat:2020ktg,Fan:2020nzz}
and its application to lattice calculations of nucleon and meson parton distribution functions (PDFs)~\cite{Lin:2014zya,Chen:2016utp,Lin:2017ani,Alexandrou:2015rja,Alexandrou:2016jqi,Alexandrou:2017huk,Chen:2017mzz,Alexandrou:2018pbm,Chen:2018xof,Chen:2018fwa,Alexandrou:2018eet,Lin:2018qky,Fan:2018dxu,Liu:2018hxv,Wang:2019tgg,Lin:2019ocg,Chen:2019lcm}, as well as meson distribution amplitudes~\cite{Zhang:2017bzy,Chen:2017gck,Bali:2018spj}. 
Despite limited volumes and relatively coarse lattice spacings, the state-of-the-art nucleon isovector quark PDFs, determined from lattice data at the physical point, have shown reasonable agreement~\cite{Chen:2018xof,Lin:2018qky,Alexandrou:2018pbm} with phenomenological results extracted from the experimental data~\cite{Dulat:2015mca,Ball:2017nwa,Harland-Lang:2014zoa,Nocera:2014gqa,Ethier:2017zbq}. 
Of course, a careful study of theoretical uncertainties and lattice artifacts is still needed to fully establish the reliability of the results.
Ongoing efforts include an analysis of finite-volume systematics~\cite{Lin:2019ocg} and exploration of machine-learning application~\cite{Zhang:2019qiq} that have been carried out recently. 

For meson DAs, the first lattice calculation of the leading-twist pion DA using LaMET was performed in Ref.~\cite{Zhang:2017bzy}.
The result favored a single-hump form for the pion DA.
The first calculation of the kaon DA was performed in Ref.~\cite{Chen:2017gck}.
The expected skewness was seen in the asymmetry of the kaon DA around the quark momentum fraction $x = 1/2$.
These results were improved by a Wilson-line renormalization that removes power divergences.
Also, the momentum-smearing technique proposed in Ref.~\cite{Bali:2016lva} was implemented to increase the overlap with the ground state of a moving hadron.
Despite these improvements, the DAs did not vanish in the unphysical region outside $x\in [0,1]$. 

In this paper, we further improve the meson-DA calculation by implementing nonperturbative renormalization (NPR) in regularization-independent momentum-subtraction (RI/MOM) scheme.
Also, computations are performed with three different lattice spacings and two different pion masses, allowing the continuum extrapolation and chiral extrapolation. 
Despite these improvements, the contribution in the unphysical region remains.
This is largely due to the omission of the long-range tail of the spatial correlator, which is cut off by the finite size of the lattice. To fix this problem, we would need larger hadron momentum instead of a larger lattice volume, because the long-range correlations of the matrix elements increase the undesired mixing with higher-twist operators. Alternatively, we  
explore the possibility of constraining the DA without the long-range correlation by fitting to a commonly used DA parametrization.
The model dependence of the parametrization can be later reduced by using a general set of basis functions, using machine learning to determine the functional form or by combining with other lattice inputs.

The continuum extrapolation performed in this work is relevant to several important questions regarding the LaMET and related approaches. 
First, how does the quasi-distribution approach avoid the power-divergent mixing of a twist-2 operator with a twist-2 operator of lower dimension as seen in moment calculations?
The answer is that this power-divergent mixing is due to the breaking of rotational symmetry on a lattice.
When the continuum limit is taken after the correlator is renormalized,
rotational symmetry can be recovered as an accidental symmetry. (This is possible because the nonlocal operators used for quasi-distributions are the lowest-dimension ones with the same symmetry properties~\cite{Chen:2017mie}.)
Hence, power-divergent mixing among twist-2 operators should no longer exist.

Second, the operator product expansion of the equal-time correlators gives rise to twist-2, twist-4 and higher-twist contributions.
Ref.~\cite{Rossi:2018zkn} argued that the matrix element of the twist-4 operator is set by the scale $a$; hence, its suppression factor compared with twist-2 is $\mathcal{O}(1/(P_z a)^2)$ instead of $\mathcal{O}(\Lambda^2_\text{QCD}/P_z^2)$ with the hadron momentum $P_z$. However, the twist-4 contribution that needs to be subtracted from the quasi-distribution operator can be written as equal-time correlators with two more mass dimensions than the original quasi-distribution operator~\cite{Chen:2016utp}. Hence, they do not cause power-divergent mixings that need to be subtracted before applying RI/MOM renormalization.

Proving the above statements requires a careful analysis of the mixing matrix, which is beyond the scope of this paper. In this work, we check whether the continuum extrapolation of the RI/MOM-renormalized matrix elements is consistent with the absence of power-divergent terms, which, by itself, is a necessary (but not sufficient) condition for the above statements to be true. If there were mixing with lower-dimension operators, the matrix element could still be renormalized, but one might get the undesired lower-dimension operator in the continuum limit instead of the one of interest. However, as discussed above, power-divergent mixing in quasi-distributions was not found in the studies of Refs.~\cite{Chen:2017mie,Chen:2016utp}.

The article is organized in the following way. In Sec.~\ref{sec:set_up} we present the lattice setup of this calculation, and the strategies used to extract the bare matrix elements from lattice DA correlators. Section~\ref{sec:results} shows the NPR procedure, and the continuum and chiral extrapolation of the renormalized matrix elements. The $x$ dependence of DAs are obtained from two approaches: the fit to a functional form and the prediction with a machine learning algorithm. Finally, we summarize the results and future prospects in Sec.~\ref{sec:summary}. 

\section{Lattice Setup}\label{sec:set_up}

In this work, we extend our previous work on the kaon distribution amplitude from a single a12m310 lattice~\cite{Chen:2017gck} to 3 lattice ensembles with different lattice spacings and extrapolate the results to continuum. The three ensembles have lattice spacings $a=0.582(4)$~fm, $a=0.888(8)$~fm and $a=0.1207(11)$~fm with $N_f=2+1+1$ flavors of highly improved staggered quarks (HISQ)~\cite{Follana:2006rc} generated by MILC collaboration~\cite{Bazavov:2012xda}.
One-step hypercubic (HYP) smearing of the gauge links is applied to improve the discretization effects.
We use clover action for the valence quarks with the clover parameters tuned to recover the lowest pion mass of the staggered quarks in the sea~\cite{Rajan:2017lxk,Bhattacharya:2015wna,Bhattacharya:2013ehc}: $M_\pi = 319.3(5)$~MeV, 312.7(6)~MeV and 305.3(4)~MeV on the three ensembles, respectively.
On each lattice configuration, we use multiple sources uniformly distributed in the time direction and randomly distributed in the spatial directions to reach high statistics.
We have 24 sources in total for the a06m310 and a09m310 ensembles and 32 sources for a12m310, corresponding to 2280, 5544 and 2912 measurements in total, respectively.

The hadron spectrum (HS) and distribution amplitude (DA) two-point correlators are calculated for different mesons:
\begin{align}
C_M^{\text{HS}}(P,t)& = \left\langle 0\left| \int d^3y\, e^{i\vec{P}\cdot\vec{y}} \bar{\psi}_1(\vec{y},t) \gamma_5 \psi_2(\vec{y},t)\right.\right.\nonumber\\
 &\left.\left.{\vphantom{\int d^3y}}\bar{\psi}_2(0,0) \gamma_5 \psi_1(0,0) \right|0 \right\rangle, \\
C_M^{\text{DA}}(z,P,t) &=
\left\langle 0\left| \int d^3y\, e^{i\vec{P}\cdot\vec{y}} \bar{\psi}_1(\vec{y},t) \gamma_z \gamma_5  U(\vec{y},\vec{y}+z\hat{z})\right.\right.\nonumber\\
 &\left.\left.{\vphantom{\int d^3y}}\psi_2(\vec{y}+z\hat{z},t) \bar{\psi}_2(0,0) \gamma_5 \psi_1(0,0) \right|0 \right\rangle,
\label{eq:meson_2pt}
\end{align}
where $M$ represents different mesons ($\pi$, $K$, $\eta_s$), $\{\psi_1,\psi_2\}$ are $\{u,u\}$ for $\pi$, $\{u,s\}$ for $K$ and $\{s,s\}$ for $\eta_s$ (only connected diagrams are computed in this work), $U(\vec{y},\vec{y}+z\hat{z})=\prod_{x=0}^{z-1} U_z(y+x\hat{z},t)$ is the Wilson line connecting lattice site $\vec{y}$ to $\vec{y}+z\hat{z}$, as defined in Ref.~\cite{Zhang:2017bzy,Chen:2017gck}.
The light-quark $u$ and strange-quark $s$ mass parameters used here are from Ref.~\cite{Gupta:2018qil}.

The DA matrix element (ME) and ground-state energies of the mesons can be extracted from the HS and DA two-point correlators by a two-state fit to the form:
\begin{align}
C_M^{\text{HS}}(P,t) &= A_{M,0}^{\text{HS}}(P) e^{-E_{M,0}(P)t} \nonumber\\
                     &+ A_{M,1}^{\text{HS}}(P) e^{-E_{M,1}(P)t} + ..., \\
C_M^{\text{DA}}(z,P,t) &= A_{M,0}^{\text{DA}}(P,z) e^{-E_{M,0}(P)t} \nonumber\\
                     &+ A_{M,1}^{\text{DA}}(P,z) e^{-E_{M,1}(P)t} + ..., 
\label{eq:fit_formula}
\end{align}
where $A_{M,0}(P)$ and $E_{M,0}(P)$ are the amplitude and energy, respectively, of ground state of a boosted meson with momentum $P_z=P$ while $A_{M,1}(P)$ and $E_{M,1}(P)$ are for the first excited state. $E_{M,0}(P=0)$ is the mass of the meson.

\begin{figure*}
\centering
\includegraphics[width=0.32\textwidth]{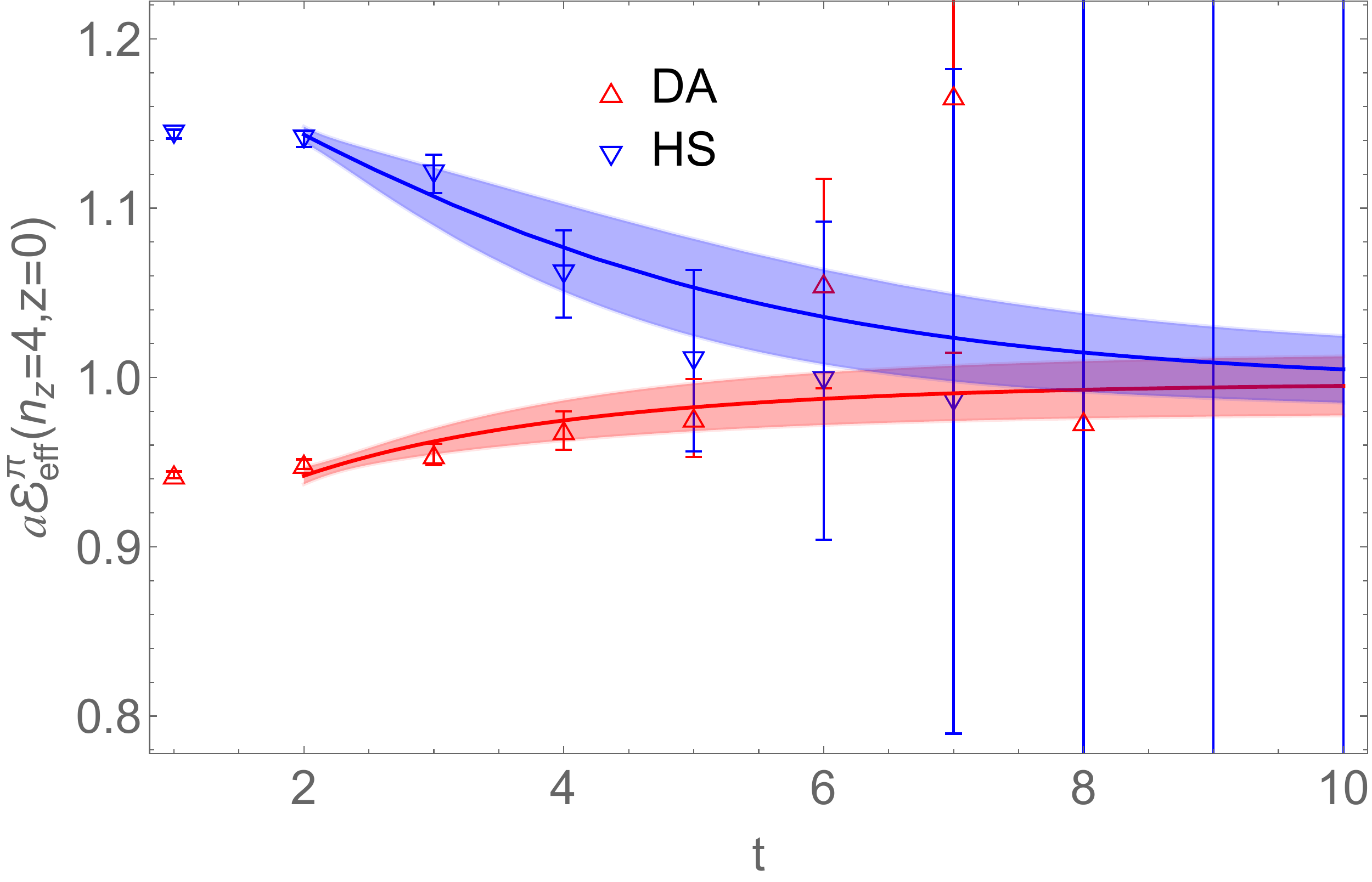}
\includegraphics[width=0.32\textwidth]{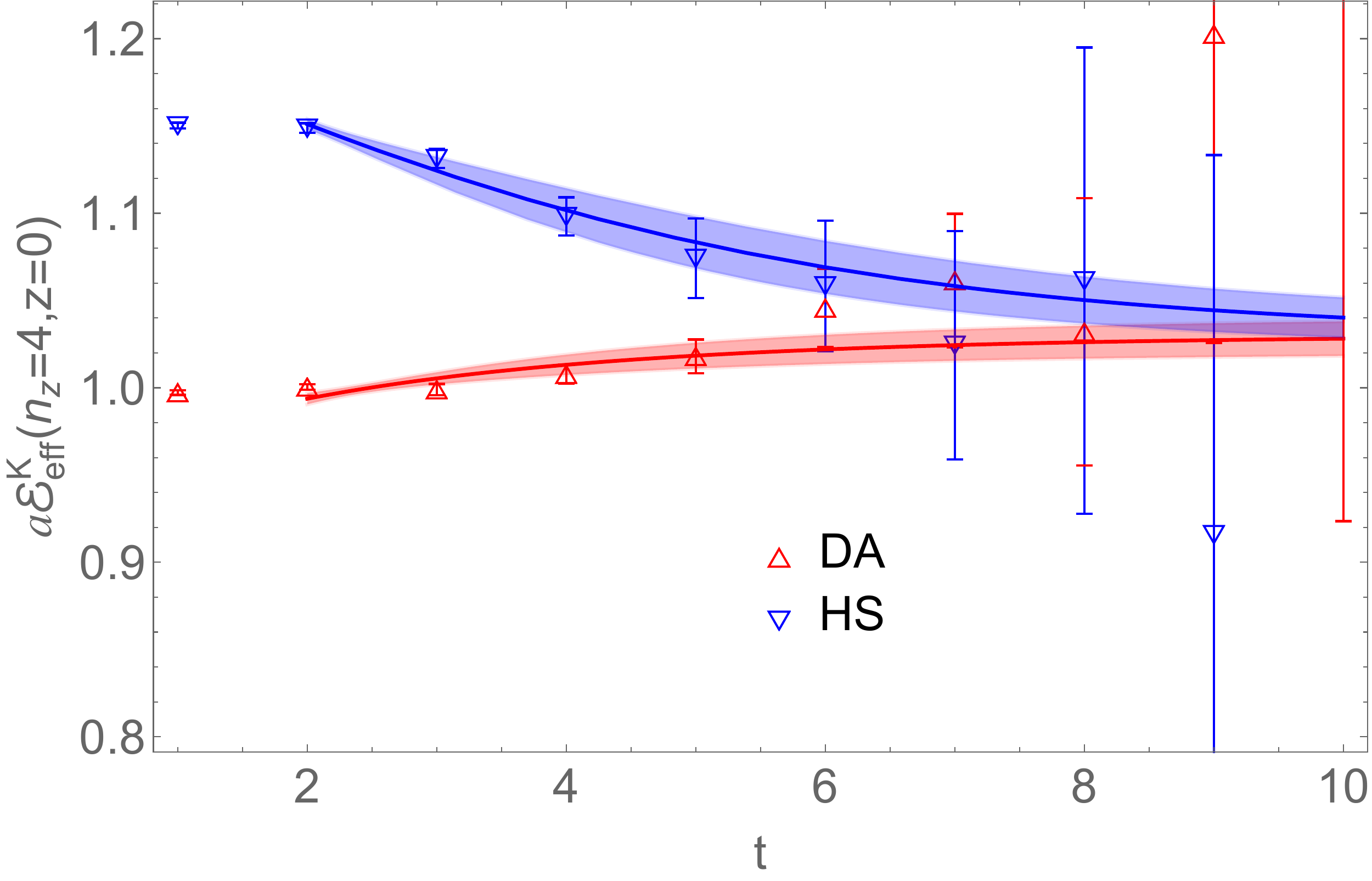}
\includegraphics[width=0.32\textwidth]{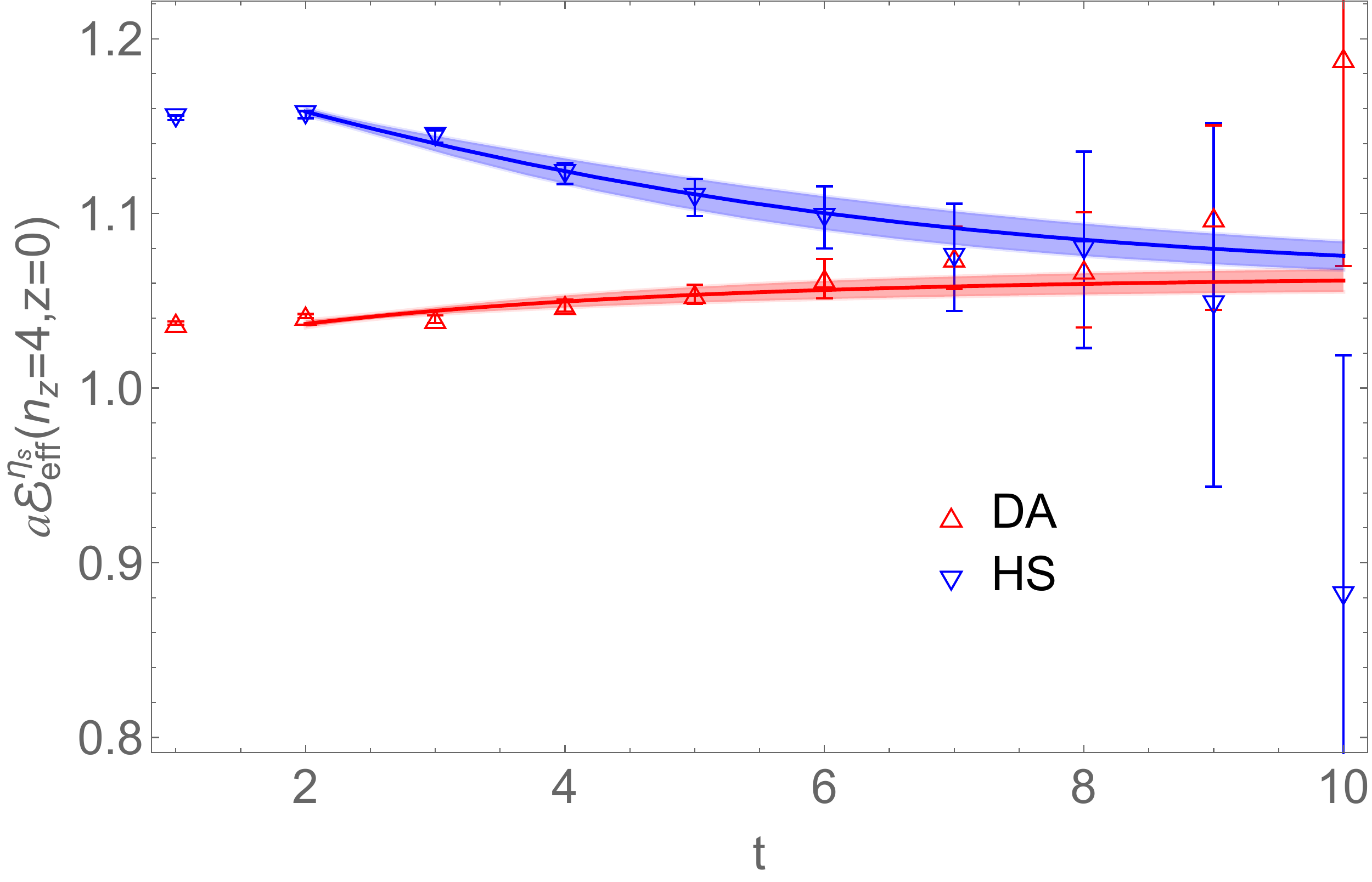}
\includegraphics[width=0.32\textwidth]{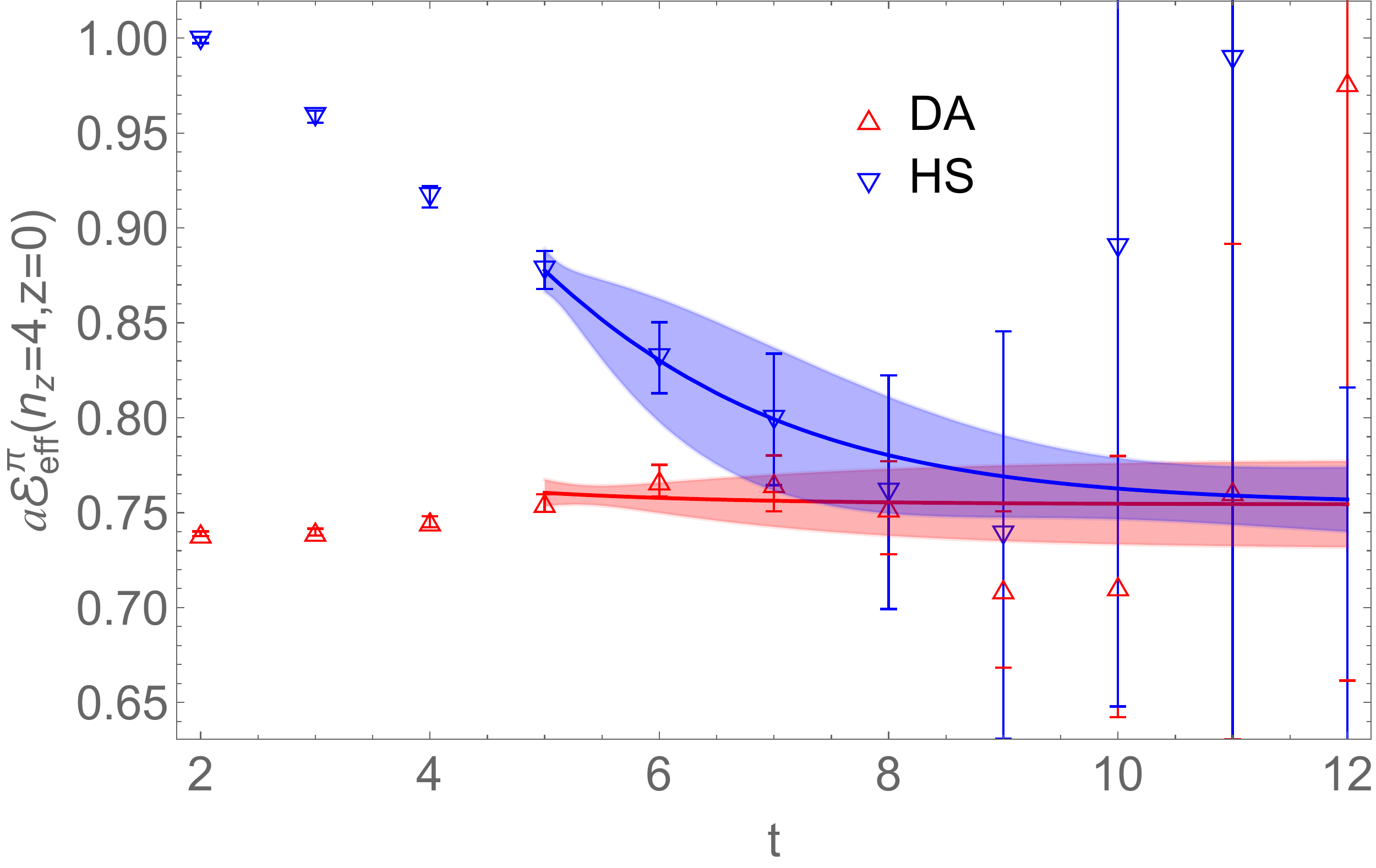}
\includegraphics[width=0.32\textwidth]{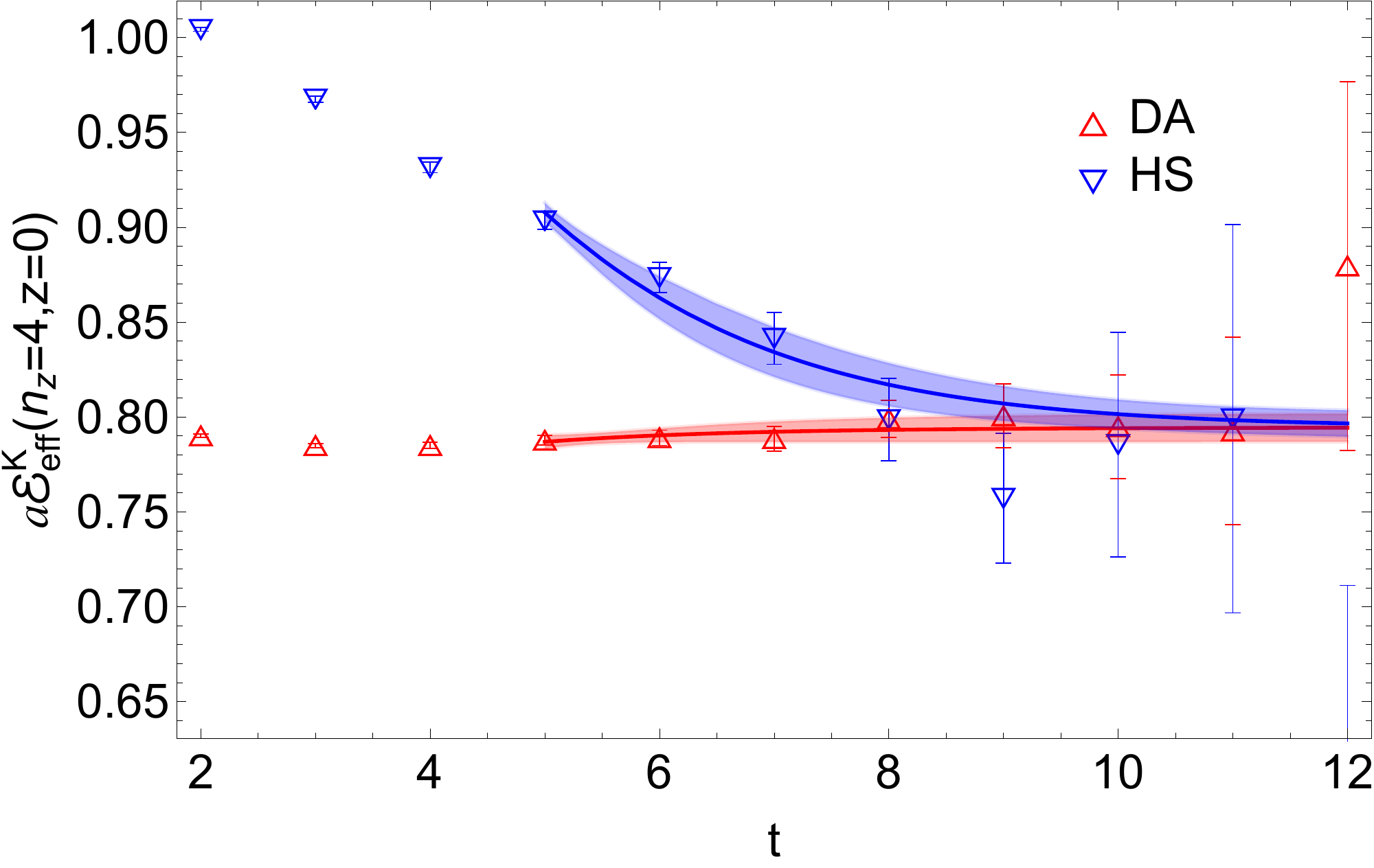}
\includegraphics[width=0.32\textwidth]{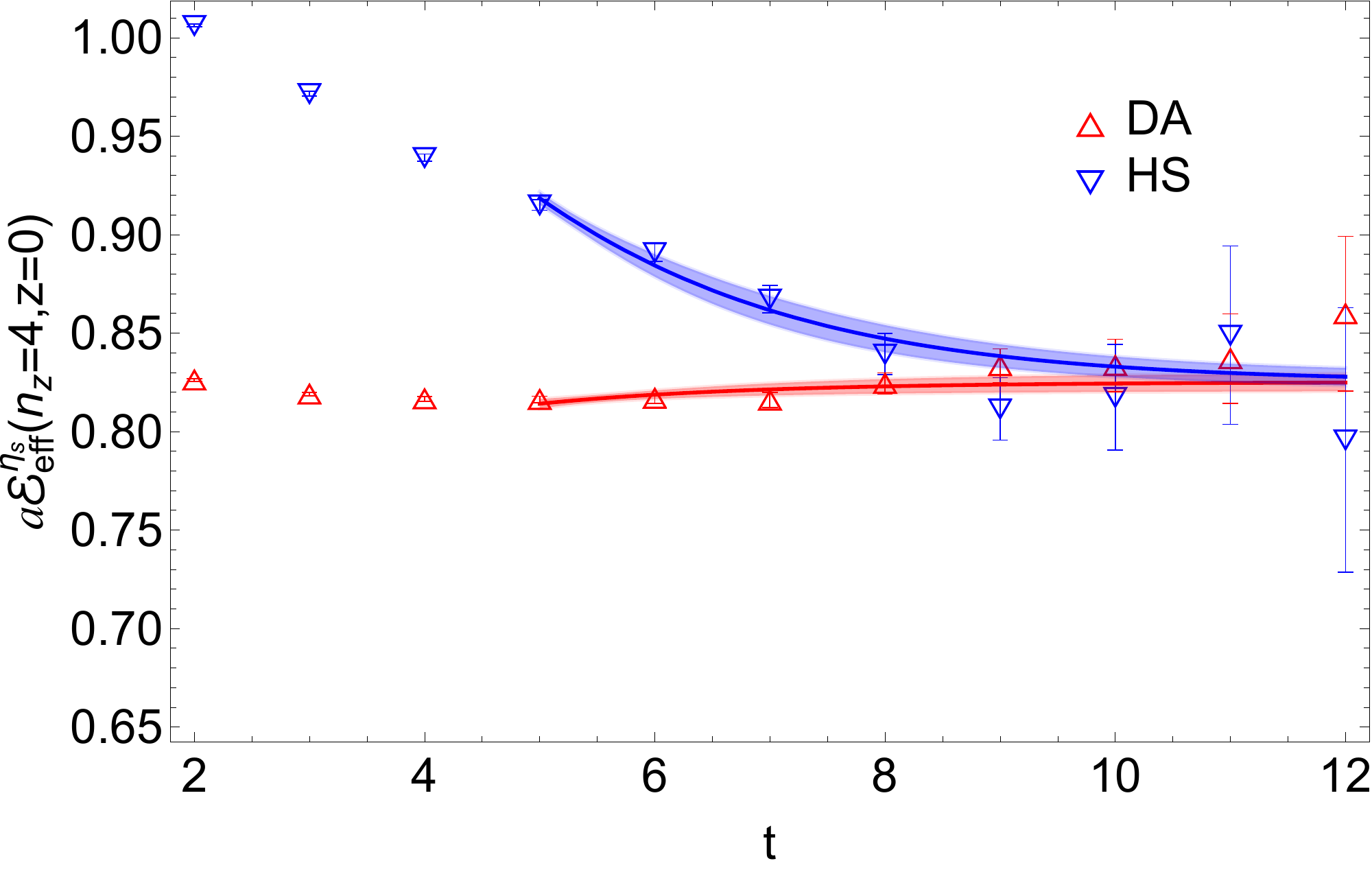}
\includegraphics[width=0.32\textwidth]{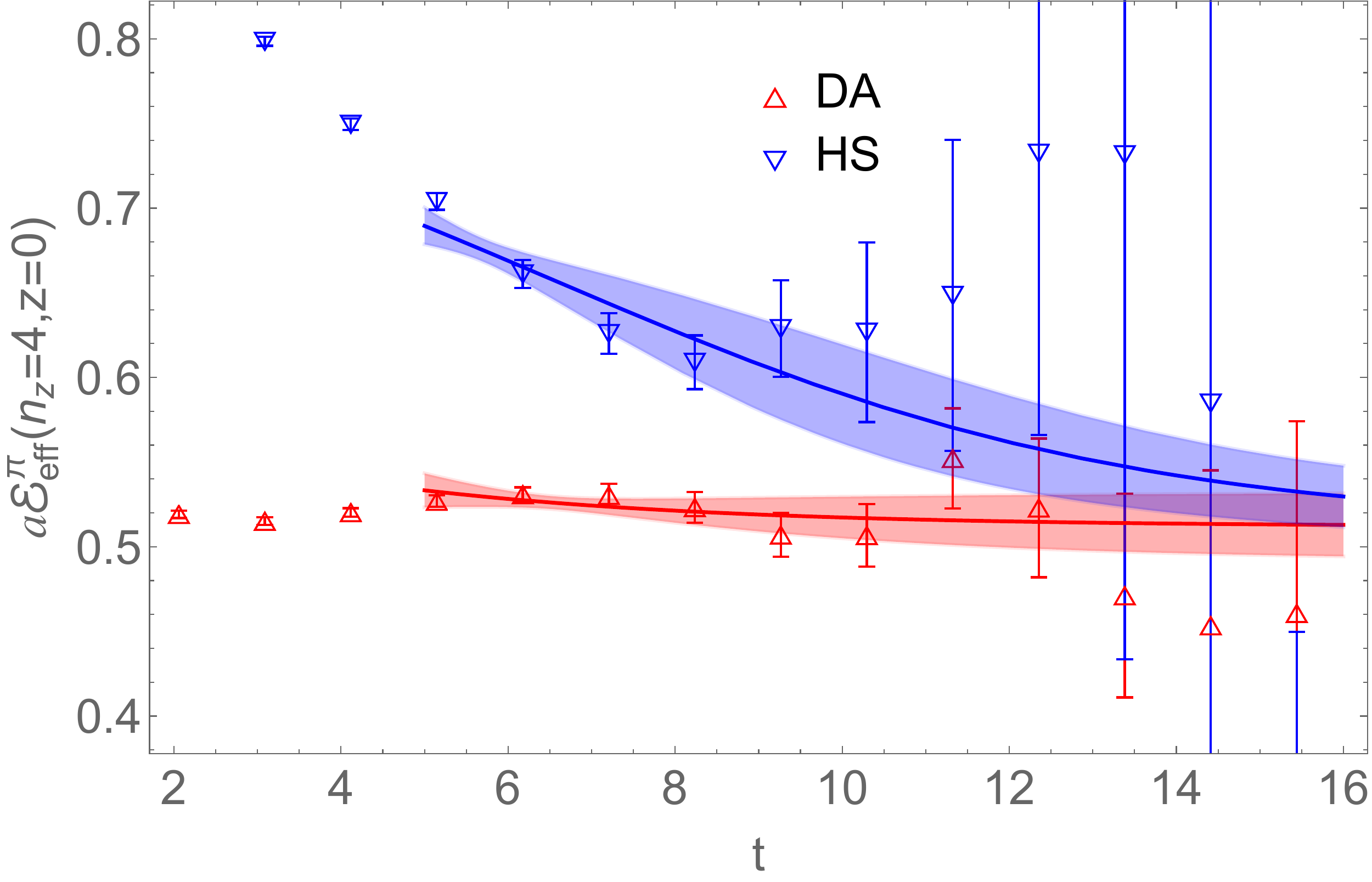}
\includegraphics[width=0.32\textwidth]{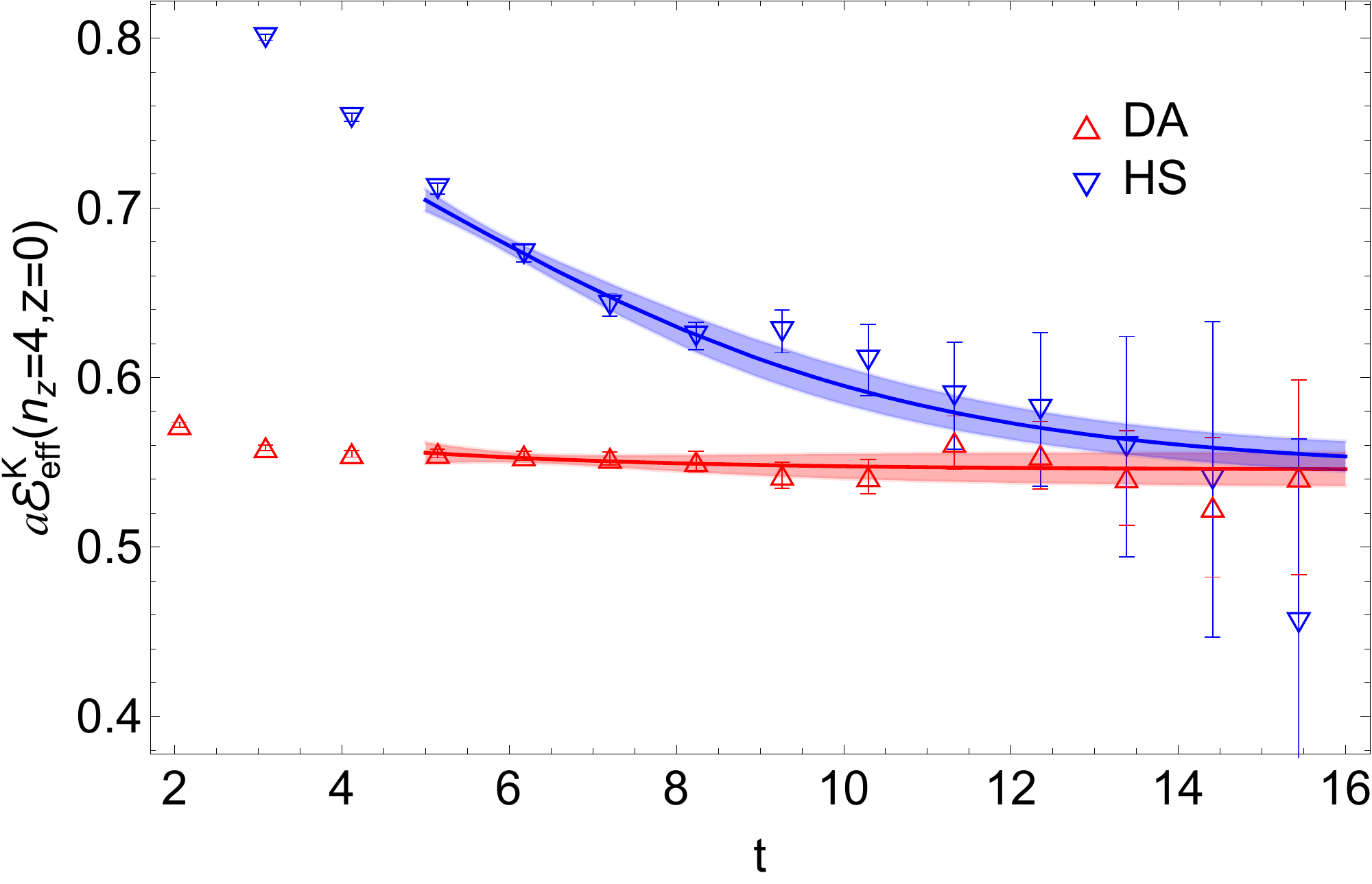}
\includegraphics[width=0.32\textwidth]{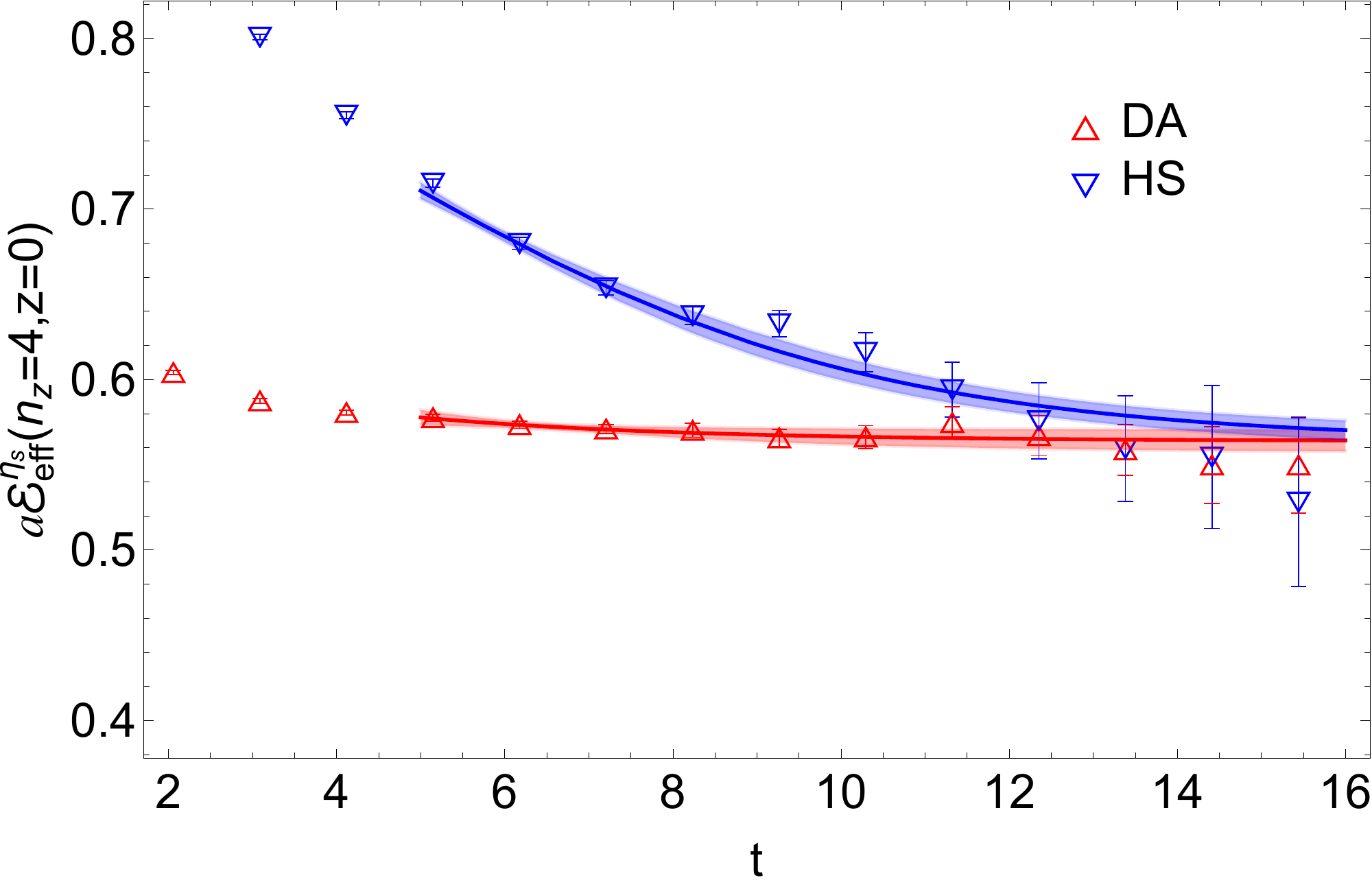}
\caption{The $\pi$ (left column), $K$ (middle column) and $\eta_s$ (right column) effective-mass plots at $z=0$, $P_z=4\frac{2\pi}{L}$ on ensembles a12m310, a09m310 and a06m310, respectively, from top to bottom.
The bands are reconstructed from the fitted parameters of real part of HS correlators and the imaginary part of DA correlators, which are represented by blue triangles and red squares, respectively.
The momentum $P_z=4\frac{2\pi}{L}$ is the largest momentum we used, and it is the noisiest data set. 
}
\label{fig:Meff}
\end{figure*}

We consider the energies to be the same for HS and DA. Therefore, we fit both the HS and DA correlators simultaneously to get the ground-state energy $E_{M,0}(P)$ and first excited-state energy $E_{M,1}(P)$ of the various momenta $P_z$.
The fit range $[t_{\text{min}},t_{\text{max}}]$ is determined by scanning different $t$ to get the smallest $\chi^2/\text{dof}$ for all the Wilson-line lengths $z$ and at different $P_z$.
When $\chi^2/\text{dof}$ for different fit ranges are close to each other, we prefer the smaller-$t$ region where the data is less noisy.
Selected effective masses at the largest meson momentum $P_z \equiv n_z\frac{2\pi}{L}$ with $n_z=4$ are shown in Fig.~\ref{fig:Meff} for HS and DA correlators.
The bands reconstructed from the fitted parameters agree with the data well. 
We check the dispersion relation, $E_{M,0}(P)^2 = E_{M,0}(P=0)^2 + c^2 P^2$,  where $c$ is the dispersion coefficient (often called ``the speed of light'').  
The dispersion relations for all three mesons on the three ensembles are shown in Fig.~\ref{fig:DR} of Appendix~\ref{Appendix B}.
On the two coarser lattices, $c$ is closer to 1 for lighter mesons, and it becomes closer to $1$ for finer lattices.
On the a06m310 lattice, the $c$ values for all three mesons are consistent with 1.

\begin{figure*}
\centering
\includegraphics[width=0.32\textwidth]{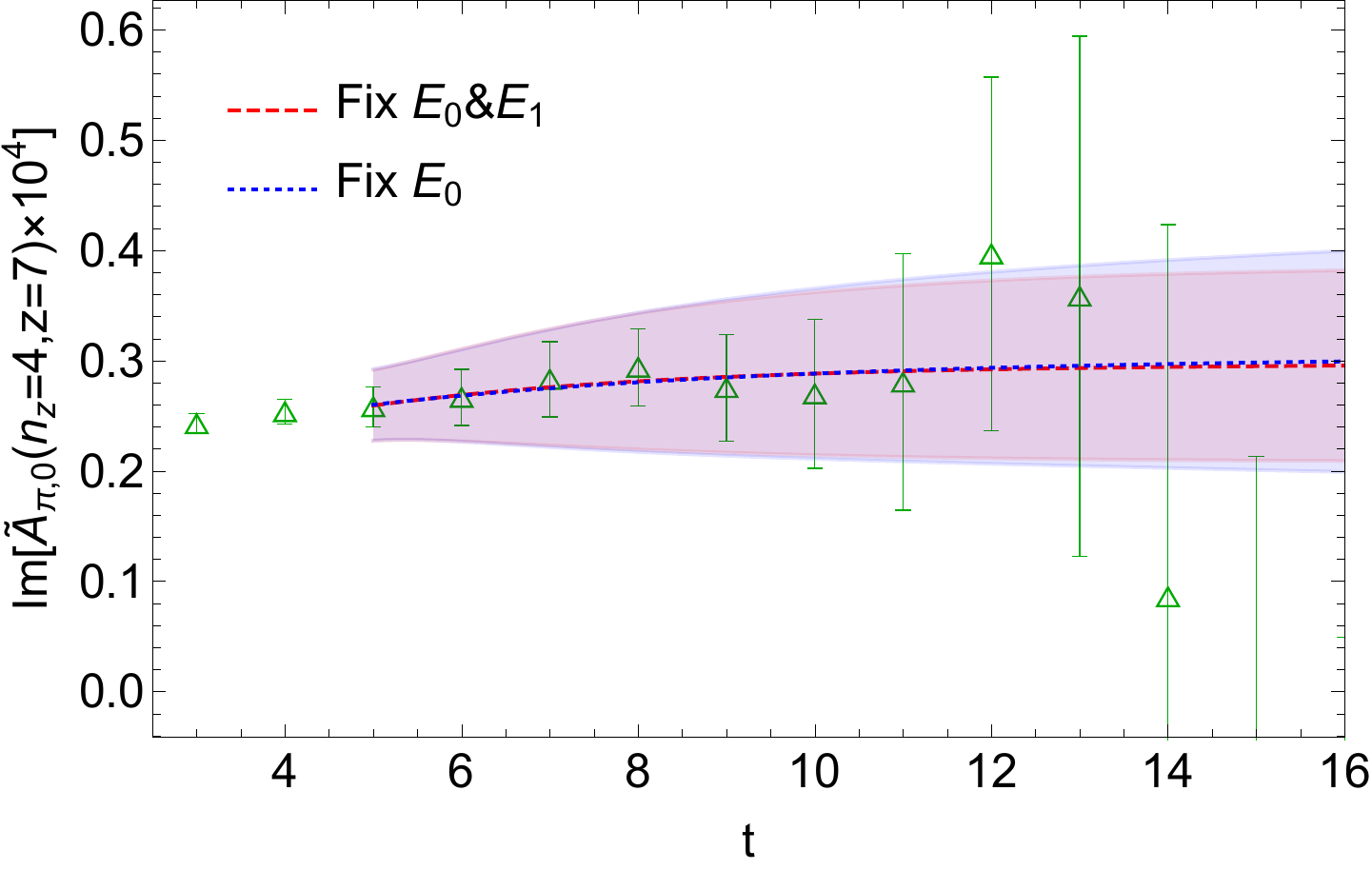}
\includegraphics[width=0.32\textwidth]{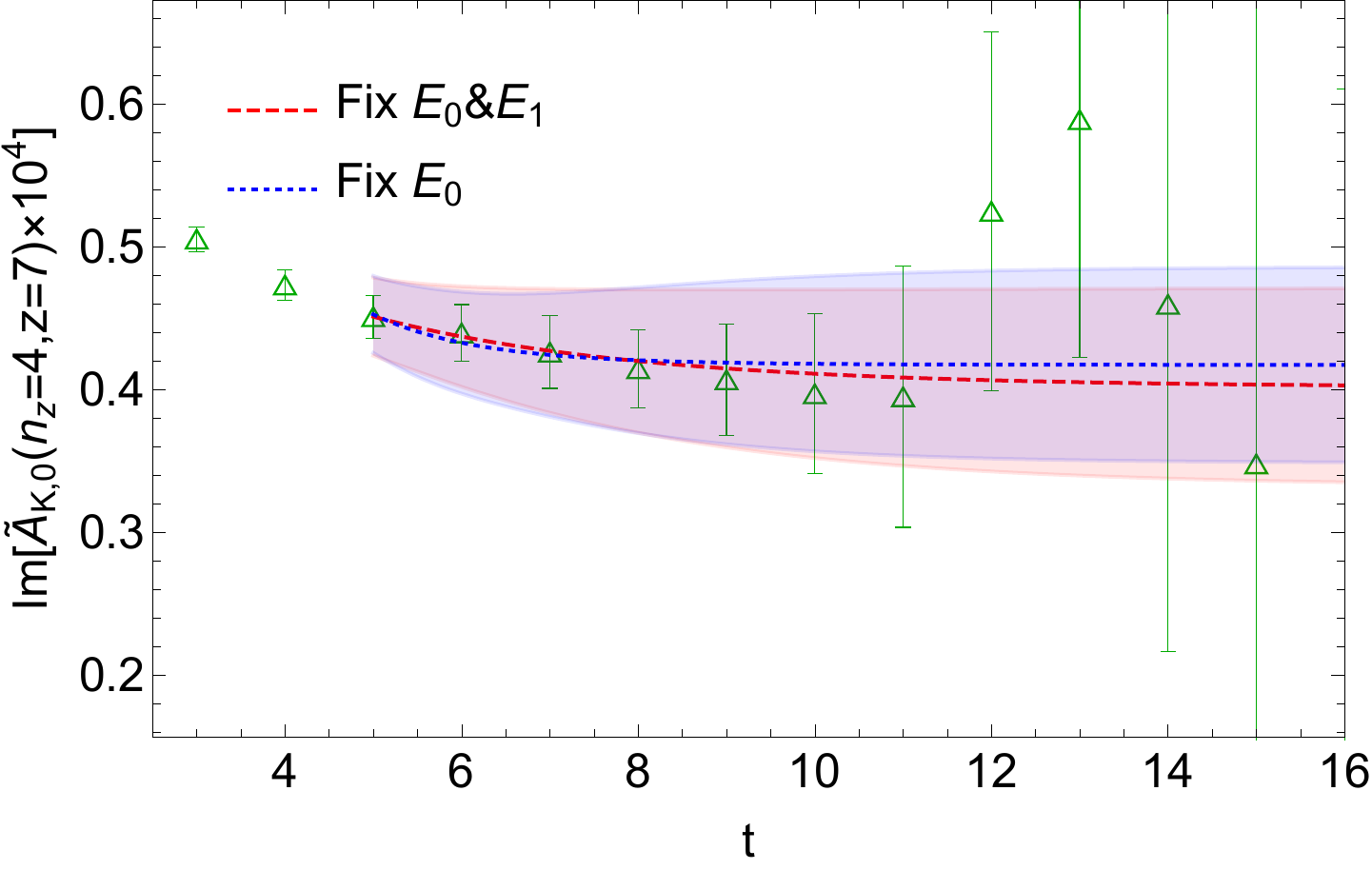}
\includegraphics[width=0.32\textwidth]{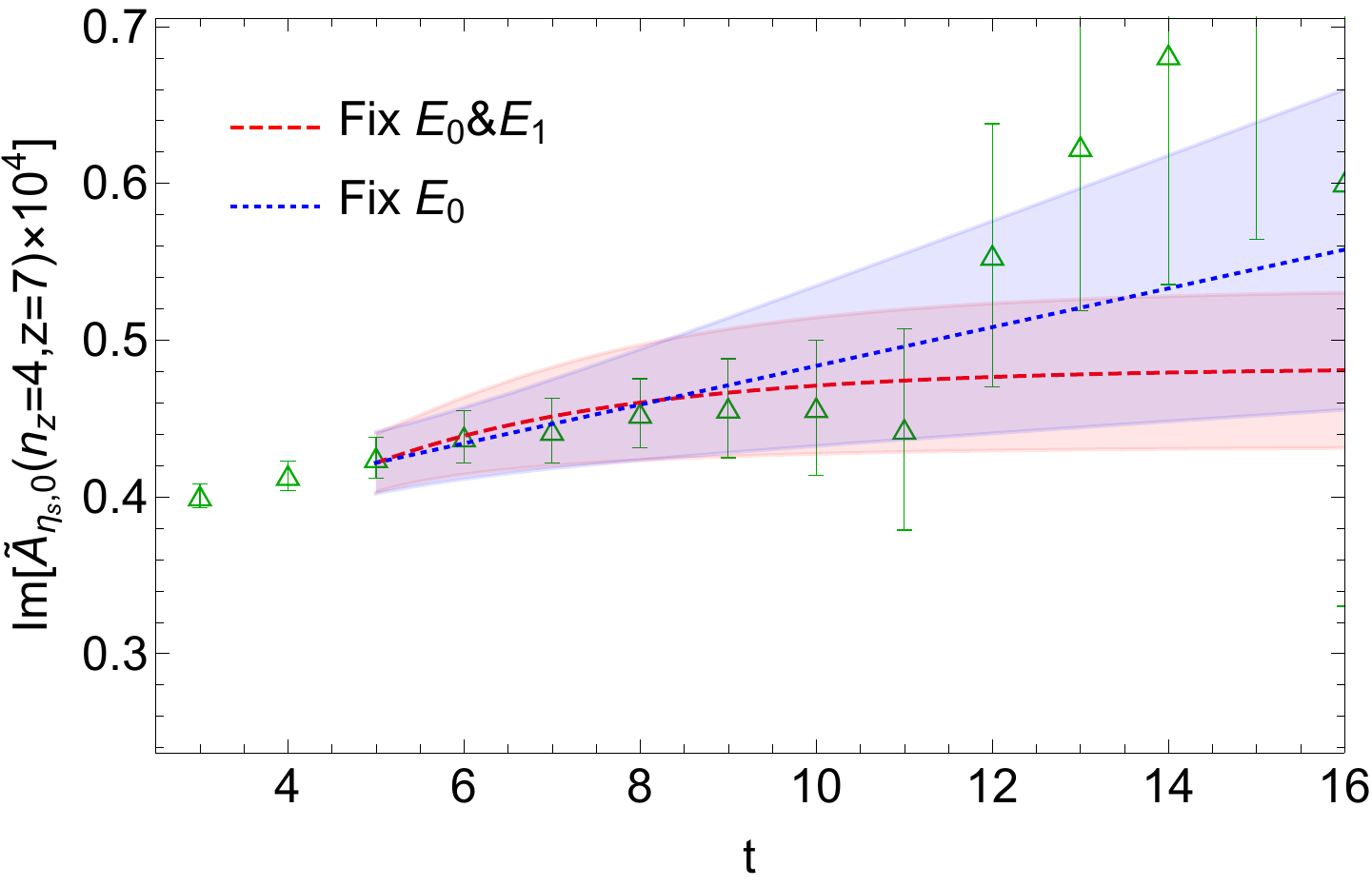}	
\includegraphics[width=0.32\textwidth]{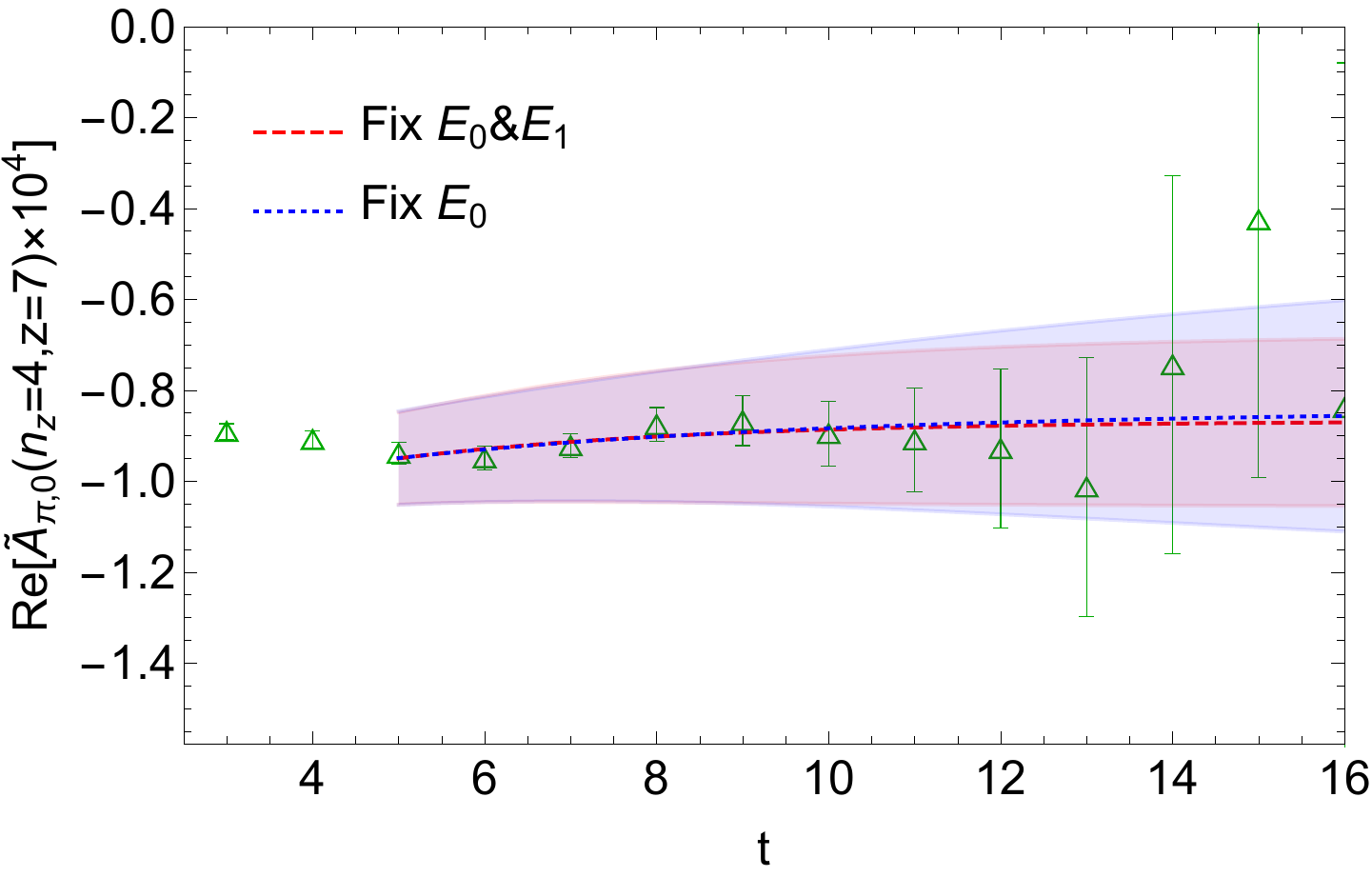}
\includegraphics[width=0.32\textwidth]{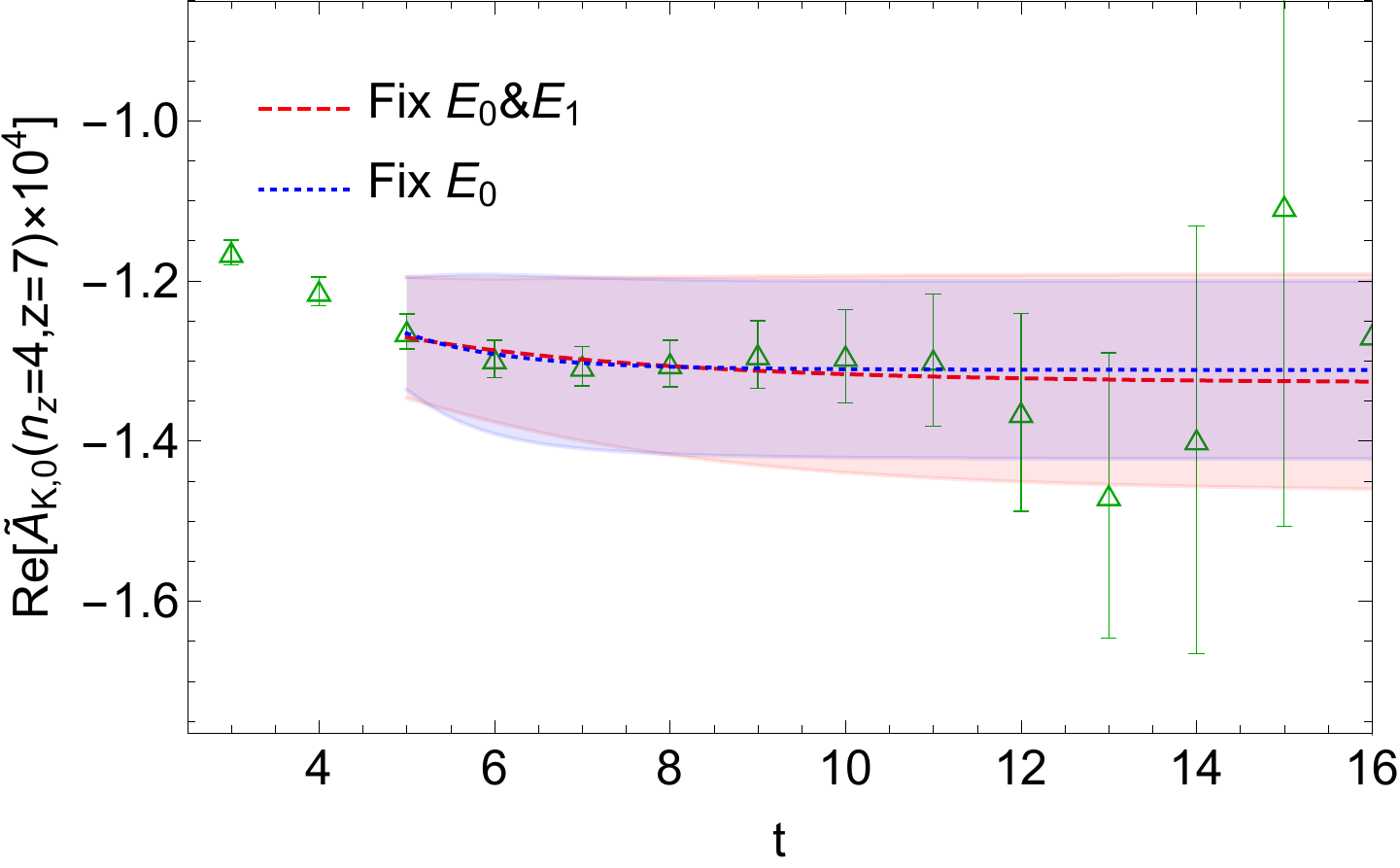}	
\includegraphics[width=0.32\textwidth]{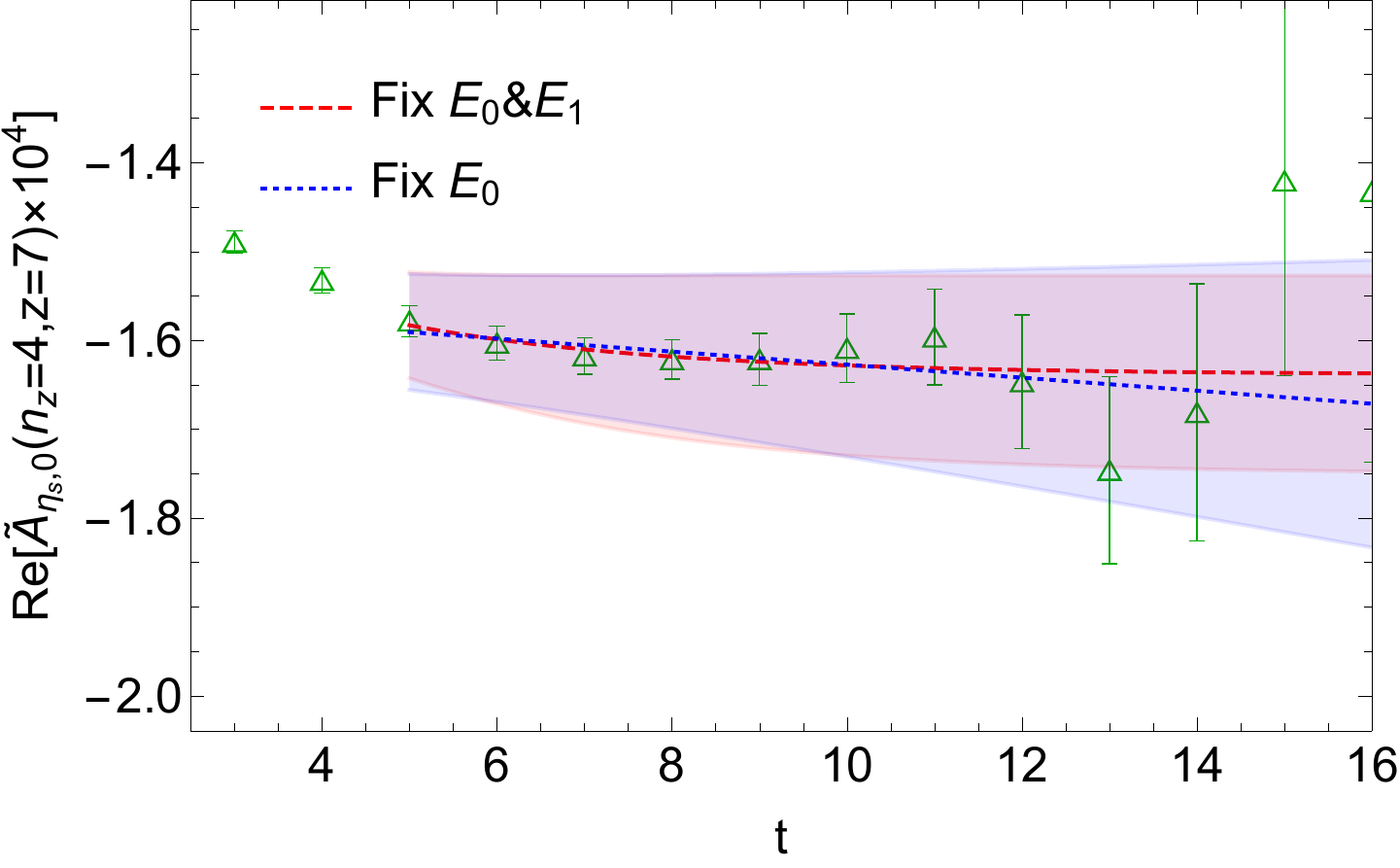}
\caption{The ground-state amplitudes $A_{M,0}$ for the pion (left column), kaon (middle column) and $\eta_s$ (right column) at $z=7$ with boost momentum $P_z=4\frac{2\pi}{L}$ on the a06m310 ensemble.  
Two strategies of two-state fits are used here: fixed $E_0$ (red band) and fixed $E_0$ and $E_1$ (blue band) obtained from the local correlators;
both fits are consistent with each other within uncertainties
The fits with fixed $E_0$ and $E_1$ are more stable in the large-$t$ region;
therefore, we use this fitted strategy for the rest of the analysis.}
\label{fig:A0_comp}
\end{figure*}

Two fit strategies are used to extract the ground-state amplitude $A_{M,0}$ for $z\neq 0$ using the ground-state energy $E_{M,0}$ and excited-state energy $E_{M,1}$ from the simultaneous fit of the HS and DA correlators at $z=0$.
One way of doing this is to fix $E_{M,0}$ at fixed $P$ by simultaneously fitting the HS and $z=0$ DA correlators, and obtain fitting parameters of $A_{M,0}^{\text{DA}}$ and $A_{M,1}^{\text{DA}}$ for the real and imaginary corrector and with a common $E_{M,1}$.
Another way is to fix both $E_{M,0}$ and $E_{M,1}$ from $z=0$ correlator fit of the same boosted momentum, while fitting the imaginary and real parts of DA correlators simultaneously.
To help visualize the resulting ground-state amplitude $A_{M,0}$ from different fit strategies, we multiply the DA two-point correlators by $e^{E_{M,0}t}$.
\begin{align}
\label{eq:CtimesExp}
 \tilde{A}_{M,0}(z,P,t) &=C_M^{\text{DA}}(z,P,t) e^{E_{M,0}t} \nonumber\\
 &= A_{M,0}^{\text{DA}} + A_{M,1}^{\text{DA}}e^{-(E_{M,1}-E_{M,0})t} + ..., 
\end{align}
which should goes to $A_{M,0}$ when $t\to\infty$. 
The reconstructed bands of this quantity are shown in Fig.~\ref{fig:A0_comp} from different fit strategies for the real part and the imaginary part of $A_{M,0}$ at $z=7$, for the largest momentum $P_z=4\frac{2\pi}{L}$ on the a06m310 ensemble.
The fit with fixed $E_{M,0}$, represented by the blue bands, and the fit with fixed $E_{M,0}$ and $E_{M,1}$, represented by the red bands, are consistent with each other within uncertainties.
However, the bands with fixed $E_{M,0}$ and $E_{M,1}$ are more stable in the large-$t$ region.
Thus, the fit with fixed $E_{M,0}$ and $E_{M,1}$ is used in further calculations.

We consider the effects of $t_{\text{min}}$ dependence on the extracted ground-state amplitude $A_{M,0}$ for the three mesons and three ensembles.
The ground-state amplitudes $A_{M,0}$ as functions of $z$ are shown in Fig.~\ref{fig:A0_z_a06} of Appendix~\ref{Appendix B} with multiple $t_{\text{min}}$ choices on ensembles a06m310, a09m310 and a12m310.
The fitted ground-state amplitudes $A_{M,0}$ with smaller $t_{\text{min}}$ tend to have smaller errors.
However, the $\chi^2/\text{dof}$ becomes larger when too small a $t_{\text{min}}$ is chosen, because a two-state fit cannot describe the first few points of $t$ well.
Therefore, $t_{\text{min}}=\{4, 4, 5\}$ are chosen for the a06m310 $\pi$, $K$ and $\eta_s$ fits, $t_{\text{min}}=\{5, 4, 5\}$ are chosen for the a09m310 $\pi$, $K$ and $\eta_s$ fits, and $t_{\text{min}}=\{2, 2, 3\}$ are chosen for the a12m310 $\pi$, $K$ and $\eta_s$ fits.


\section{Results and Discussions}\label{sec:results}
\subsection{Nonperturbative Renormalization}

The Wilson line $\prod_{i=0}^{z-1} U_z(i\hat{z})$ introduces a divergence into the quasi-PDF operator, so the bare matrix elements (ME) cannot be matched directly to physical observables and need to be renormalized.
In contrast to the previous work~\cite{Chen:2017gck} where an effective mass counter-term is used to renormalize the matrix elements, we now follow a standard nonperturbative renormalization (NPR) in regularization-independent momentum-subtraction (RI/MOM) scheme~\cite{Martinelli:1994ty}.
The NPR factors $Z(z,\mu^R,p^R_z,a)$ are calculated by implementing the condition that
\begin{align*}
Z(z,\mu^R,p^R_z,a)& \left\langle S(p|z)\hat{z}) \gamma_z \gamma_5 \left[\prod_n U_z(n\hat{z})\right] S(p|0)\right\rangle_{\substack{p^2=-\mu_R^2,\\p_z=p_z^R}} \nonumber\\
&=
\left\langle\sum_w S(p|z) \gamma_z\gamma_5 \left[\prod_n U_z(n\hat{z})\right] S(p|0)\right\rangle_{\text{tree}}\\
& = e^{-izp_z}\langle S(p)\gamma_z\gamma_5 S(p)\rangle_{p_z=p_z^R},
\end{align*}

\begin{align*}
Z^{-1}&(z,\mu^R,p^R_z,a)=\frac{e^{izp_z}}{12}\Tr\left[\langle S(p)\rangle^{-1}\right.\times\\&\left.\langle S(p|z)\gamma_z\gamma_5(\prod_n U_z(n\hat{z}))S(p|0)\rangle\langle S(p)\rangle^{-1}\gamma_z \gamma_5\right]_{\substack{p^2=-\mu_R^2,\\p_z=p_z^R}}.
\end{align*}
We calculate the NPR factors at $\mu^R=3.8$~GeV, $p_z^R=0$ for all three ensembles.

\begin{figure}
	\centering
	\includegraphics[width=0.9\linewidth]{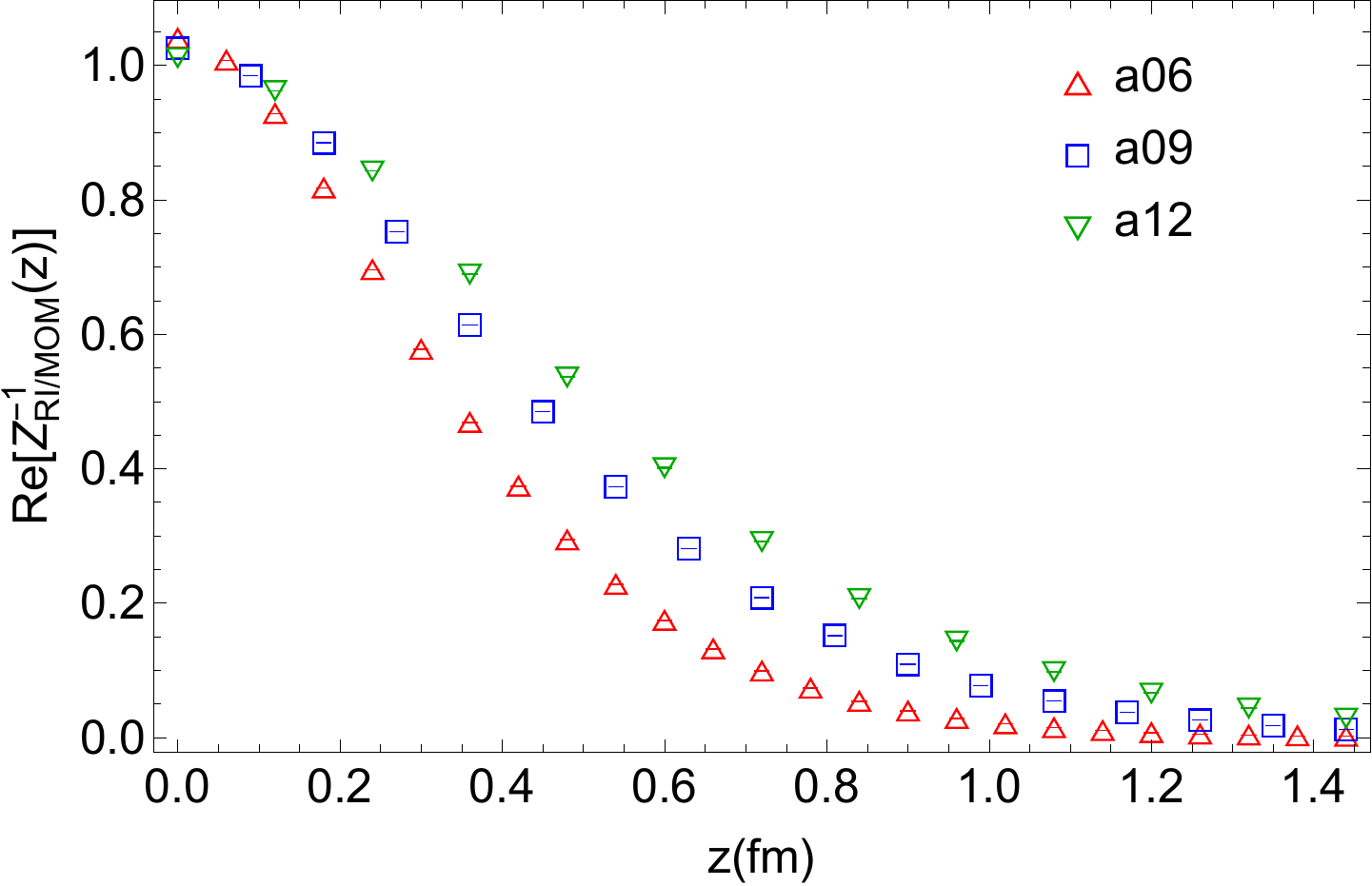}
	\caption{NPR factors for three ensembles in RI/MOM scheme at $\mu^R=3.8$~GeV, $p_z^R=0$. The red triangles, blue squares and green inverted triangles are calculated for $a\in\{0.06,0.09,0.12\}$~fm, respectively. The errors are small and are not visible on the plot.
	}
	\label{fig:zfactor}
\end{figure}

Figure~\ref{fig:zfactor} shows the inverse renormalization factors for the DA on all three lattice ensembles.
The relative errors of these factors are at the percent level and are not visible on the plot. 

\begin{figure*}
	\centering
	\includegraphics[width=0.32\linewidth]{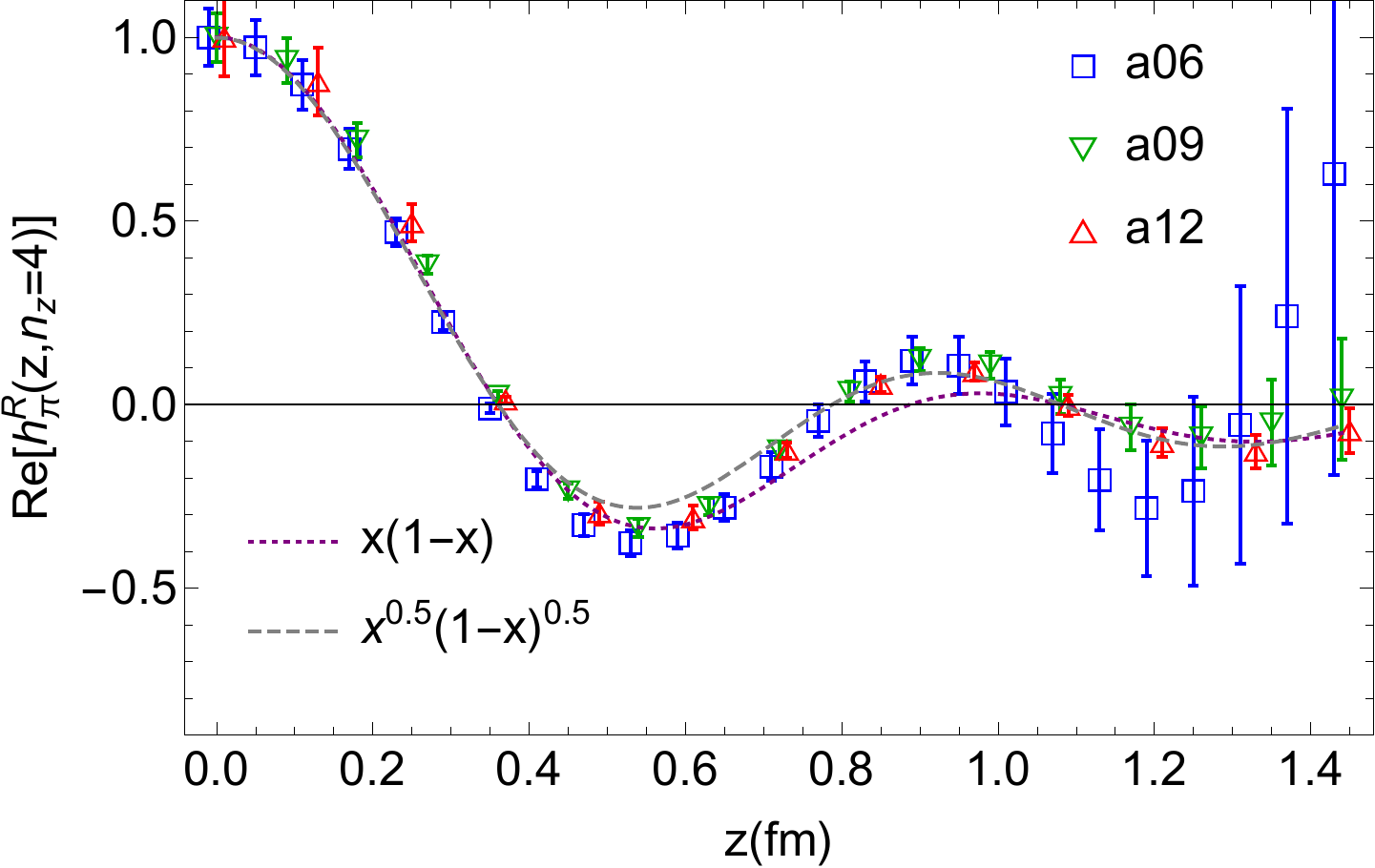}
	\includegraphics[width=0.32\linewidth]{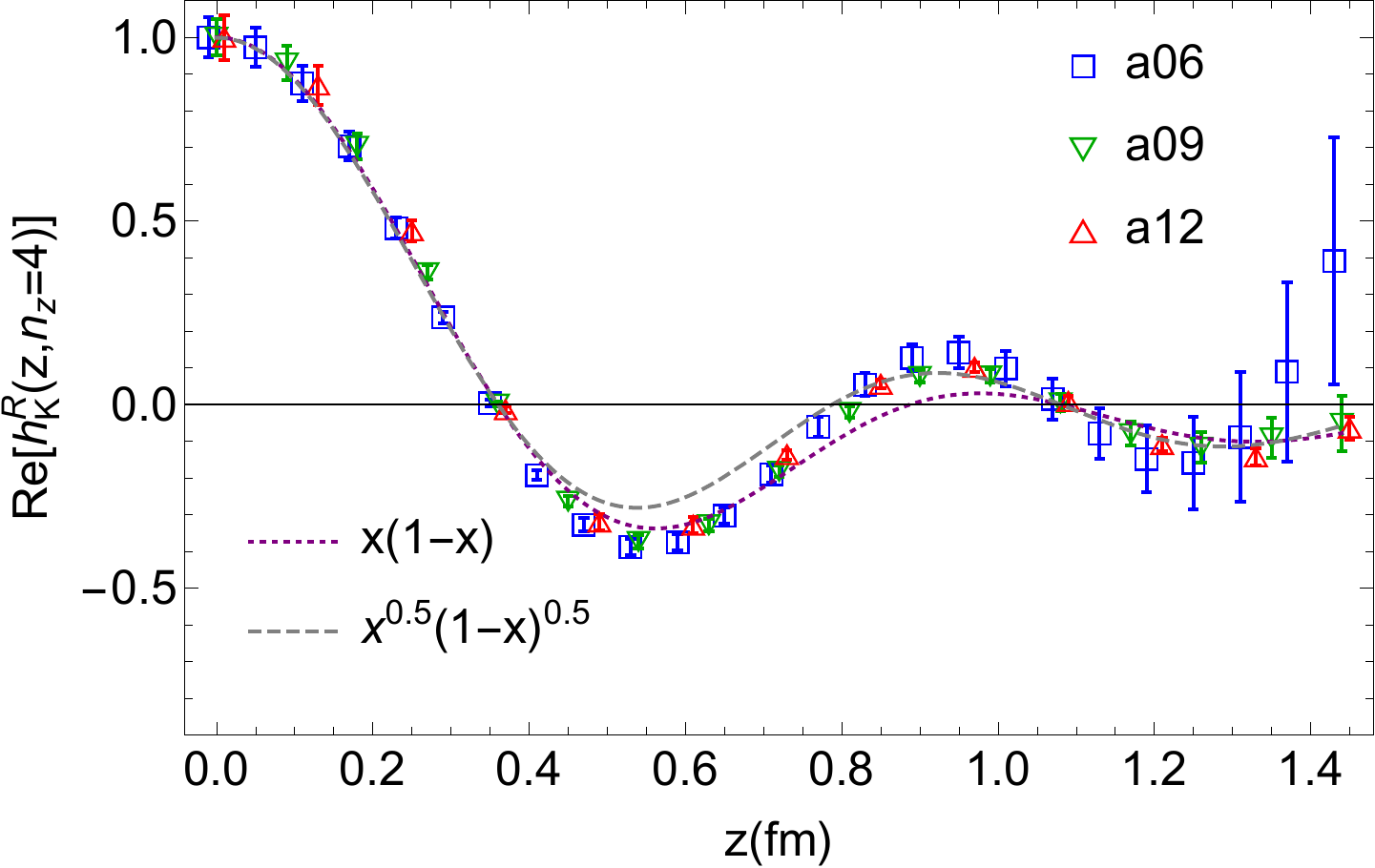}	
	\includegraphics[width=0.32\linewidth]{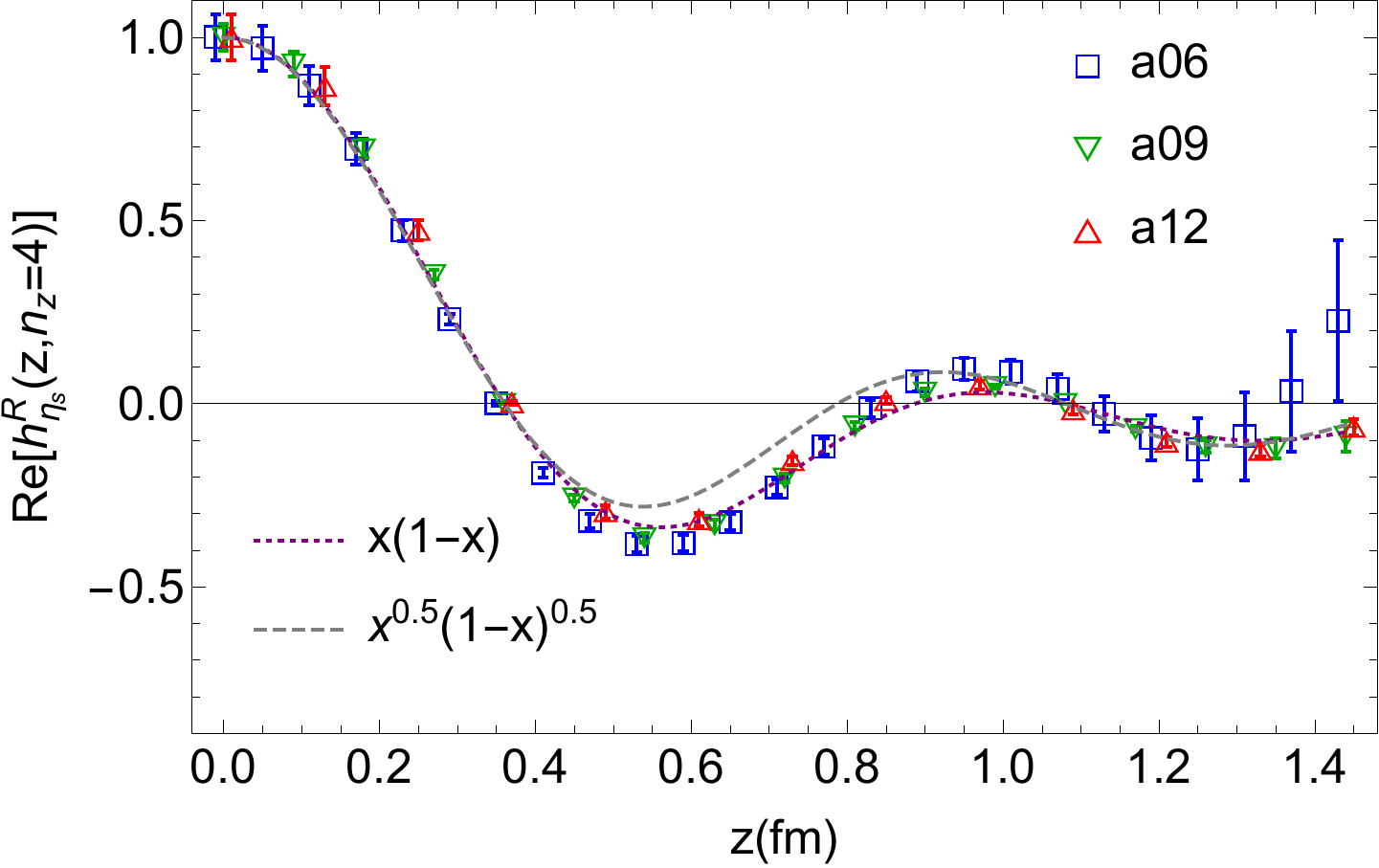}
	\includegraphics[width=0.32\linewidth]{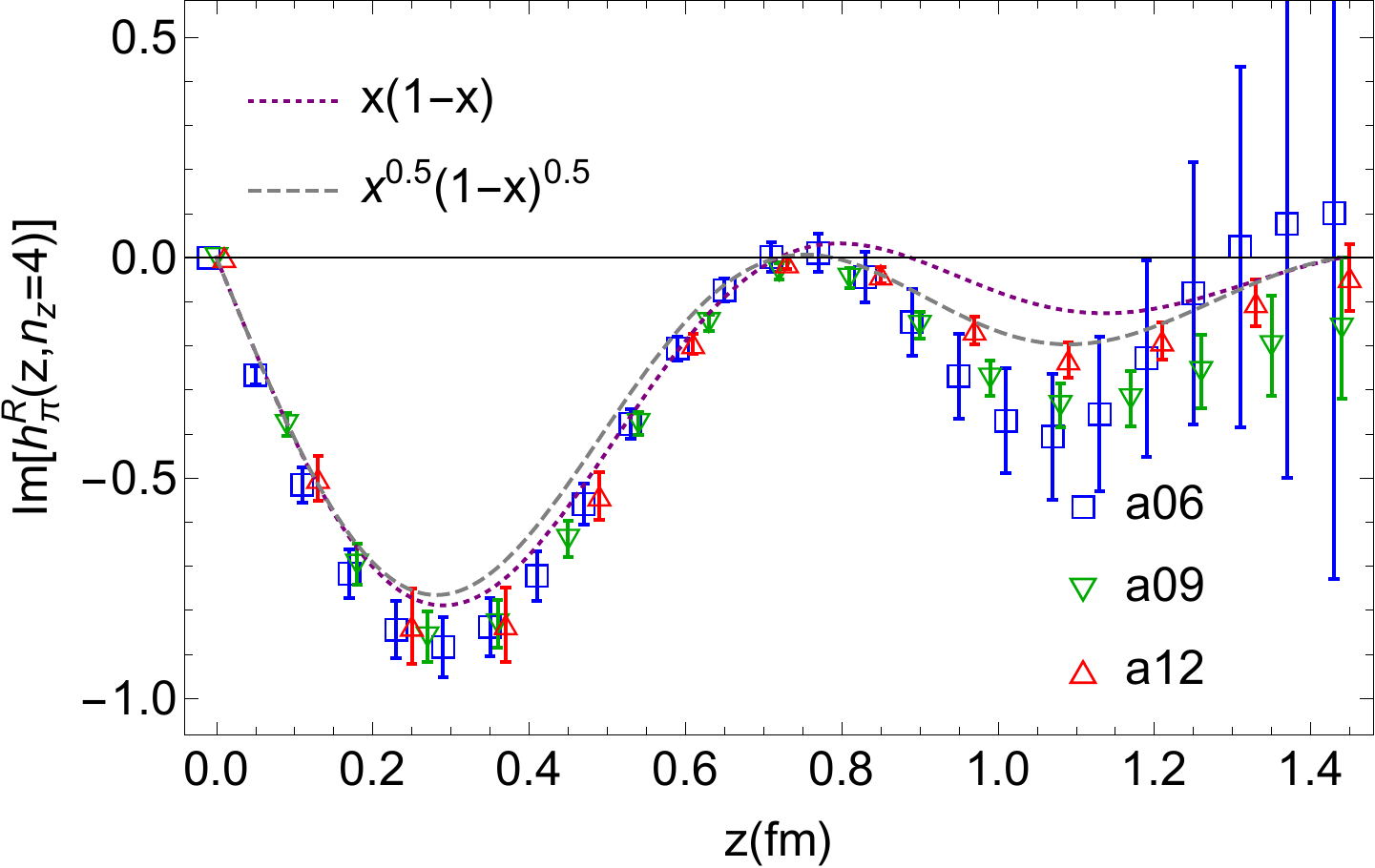}
	\includegraphics[width=0.32\linewidth]{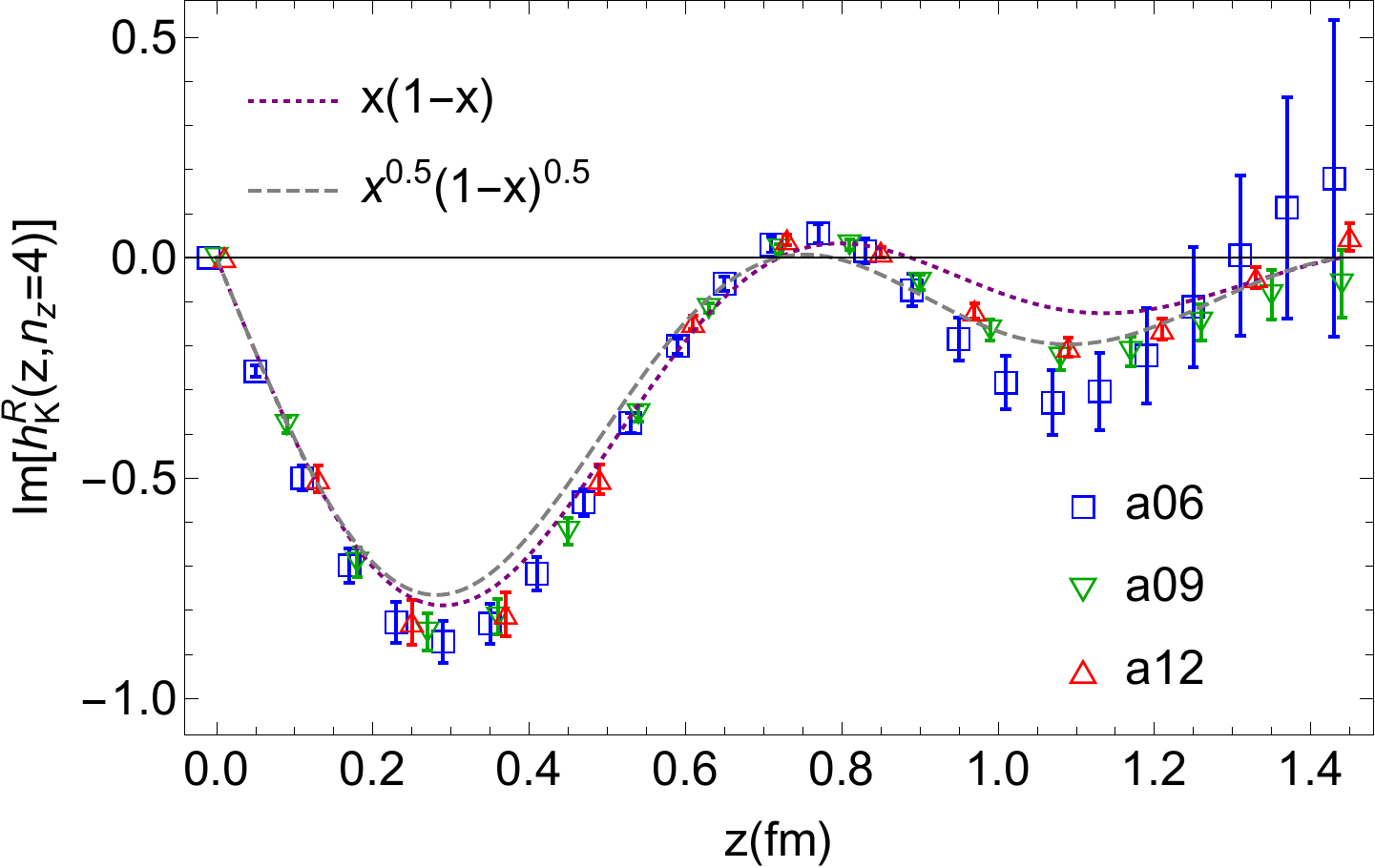}
	\includegraphics[width=0.32\linewidth]{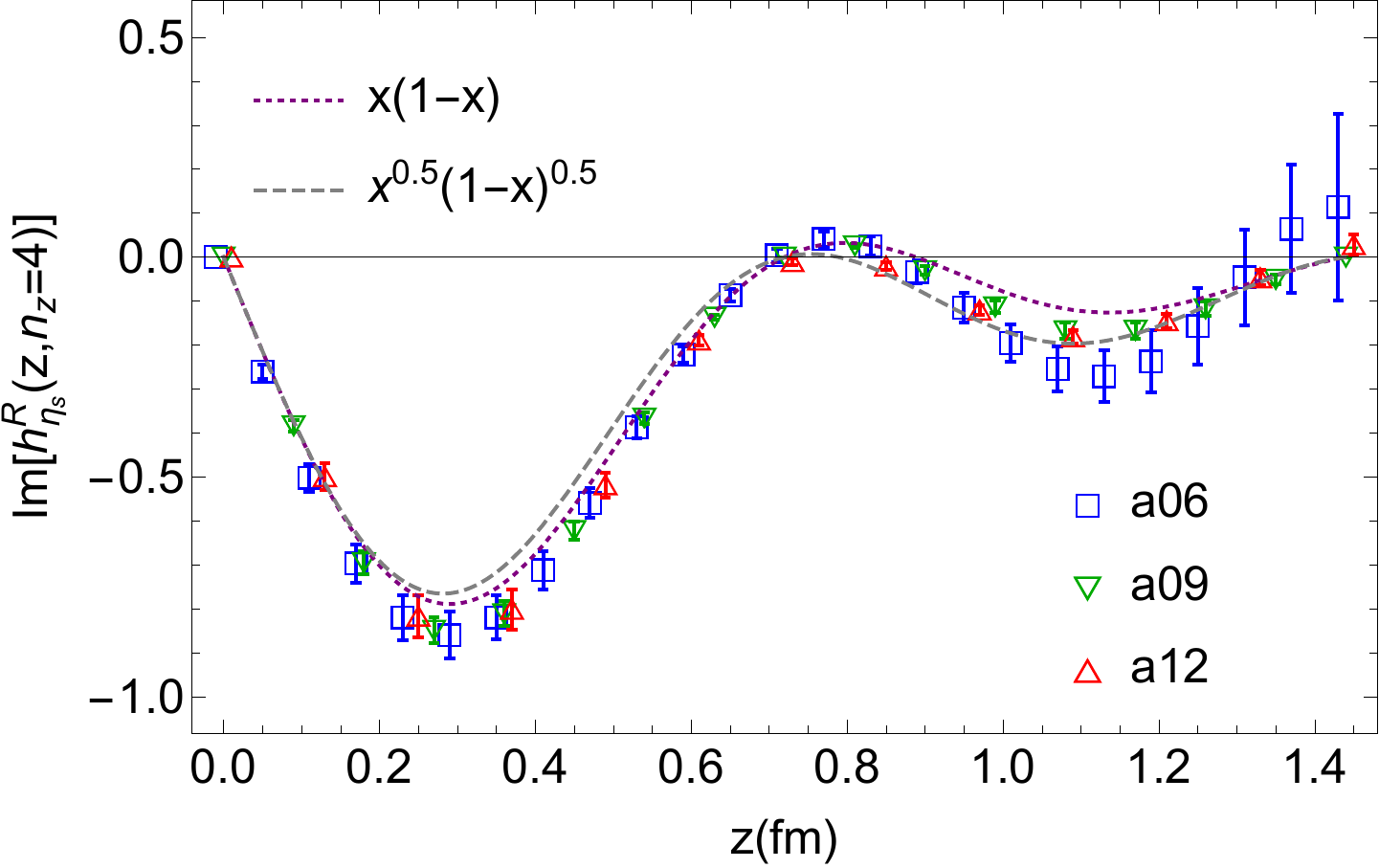}
	\caption{Real (top row) and imaginary (bottom row) renormalized matrix elements at $P_z=4 \frac{2\pi}{L}$ in RI/MOM scheme with $\mu^R=3.8$~GeV, $p_z^R=0$ for $\pi$ (left column), $K$ (middle column) and $\eta_s$ (right column). The dashed lines and dotted lines are the quasi-DA matrix elements matched from the lightcone DA function form $\phi(x)=\frac{8}{\pi}x^{0.5}(1-x)^{0.5}$ and $\phi(x)=6x(1-x)$, respectively. 
	}
	\label{fig:renorm_ME}
\end{figure*}

The renormalized matrix elements are then obtained by
\begin{equation}
\label{RI/MOM}
    h^R_M(z,p_z^R,\mu^R) = h^B_M(z,a) Z(z,\mu^R,p_z^R,a),
\end{equation}
where the bare matrix elements obtained from the ground-state meson amplitude $A_{M,0}$ fit in the previous section via
\begin{align}
h^B_M(z,a)=\frac{A^{\text{DA}}_{M,0}(z,a)}{A^{\text{DA}}_{M,0}(0,a)}.
\label{eq:bare_me_formula}
\end{align}
Figure~\ref{fig:renorm_ME} shows the $n_z=4$ renormalized matrix elements on the three ensembles, along with the quasi-DA matrix elements matched from two lightcone DA function forms, $\phi(x)=x^{\alpha}(1-x)^{\alpha}/\int^1_0 dx\, x^{\alpha}(1-x)^{\alpha}$ with $\alpha=1$ and $\alpha=0.5$, respectively.
The former ($\alpha=1$) is the asymptotic form of the pion lightcone DA~\cite{Lepage:1979zb,Efremov:1979qk}, and the latter ($\alpha=0.5$) has a second moment close to previous lattice computations of the pion DA moments~\cite{Braun:2006dg,Arthur:2010xf,Braun:2015axa,Bali:2017ude,Bali:2019dqc}.
We impose the symmetries to symmetrize the real parts and antisymmetrize the imaginary parts of the matrix elements, and enforce the normalization $\int^1_0 dx\, \phi(x)=1$ so that the central value $h(z=0)=1$.
The matrix elements for lighter mesons are noisier.
We see that the renormalized matrix elements at different lattice spacings are consistent with each other, suggesting that the higher-order discretization effects are small. 
We also note that when $\alpha$ increases, the peaks in $h(z)$ shift toward larger $z$, and the magnitude of the first peak increases while the magnitude of the second peak decreases. 
In our data, the pion result is closer to the form with $\alpha=0.5$.

In Eq.~\eqref{RI/MOM}, the operator that appears in $h_M^B(h,a)$ might mix with other operators.
If it mixes with lower-dimension operators, then subtractions of the lower-dimension operators should be performed first; otherwise, the $Z$ factor in Eq.~\eqref{RI/MOM} will just renormalize the most singular (lowest-dimension) operator in the $a \to 0$ limit rather than the desired operator.
Fortunately, Ref.~\cite{Chen:2017mie} shows that is not the case.
The nonlocal operators used for quasi DAs in this work are the lowest-dimension ones with the same symmetry properties.
This ensures that continuum limit can be taken for Eq.~\eqref{RI/MOM}.
Then, by going to the continuum limit rotational symmetry is restored, so mixing among twist-2 operators of different mass dimensions will not happen.
Also, power-divergent mixing among twist-2 and twist-4 operators was suggested in Ref.~\cite{Rossi:2018zkn}. However, the study in Ref.~\cite{Chen:2016utp} shows that the twist-4 contribution is higher dimension. It can be written as equal-time correlators with two more mass dimensions than the original quasi-distribution operator. Hence, the twist-4 contribution does not cause power-divergent mixing.

Checking these mixings requires a careful analysis of the mixing matrix, which is outside the scope of this work. In general, it is not enough to show that $h_M^R$ of Eq.~\eqref{RI/MOM} has a continuum limit, since, as we argue above, if the operator associated with $h_M^B(h,a)$ mixes with a lower-dimensional operator, then the $Z$ factor can still renormalize this lower-dimensional operator and make $h_M^R$ finite in the continuum limit.
However, the information that the lower-dimension operator provides is different from what we want.
Although the existence of the continuum limit for $h_M^R$ is by itself
a necessary but not sufficient condition for our quasi-DA program, the studies of Ref.~\cite{Chen:2017mie,Chen:2016utp} show that power divergent mixing does not appear in quasi-distributions.

\subsection{Continuum Extrapolation}

Now, we remove the remaining lattice discretization effects by extrapolating the renormalized matrix elements to the continuum by taking the continuum limit $a\to 0$. 
Because the matrix elements with three different lattice spacings do not have data from the same physical $z$'s, we first need to interpolate the points as functions of $z$ for each lattice spacing, then do the extrapolations pointwise on these curves.
For the continuum extrapolation, we use the following functional forms:
\begin{equation}
\label{eq:functional_forms}
    h_M^R(z,a) = h_{i}^R(z) + c_{M,i} a^i + \frac{d_M}{a^2}, 
\end{equation} 
where we use $i=1$ for linear and $2$ for quadratic lattice-spacing dependence.  
We find that the coefficient $d_M$ is consistent with zero within errors, except for kaon and $\eta_s$ at smallest momentum $P_z=0.86\GeV$. Since power divergence should be a short distance property of the Wilson coefficient, the dependence on the long distance properties of $P_z$ and meson flavor suggests that it is due to complications associated with small $P_z$. Hence, we set $d_M = 0$ from now on and focus on $P_z=1.73\GeV$ results in the discussion below.

\begin{figure*}
	\centering
	\includegraphics[width=0.32\linewidth]{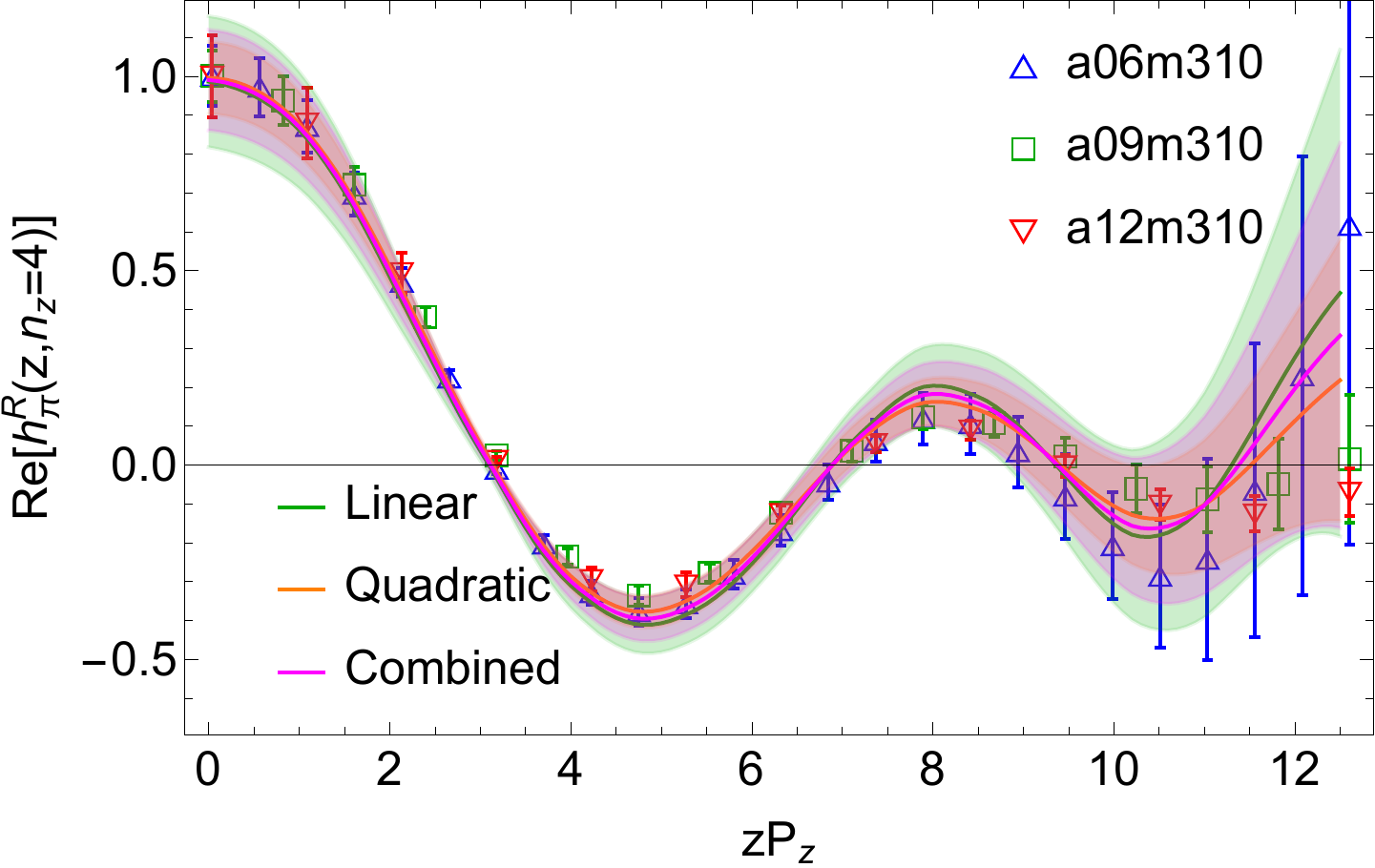}
	\includegraphics[width=0.32\linewidth]{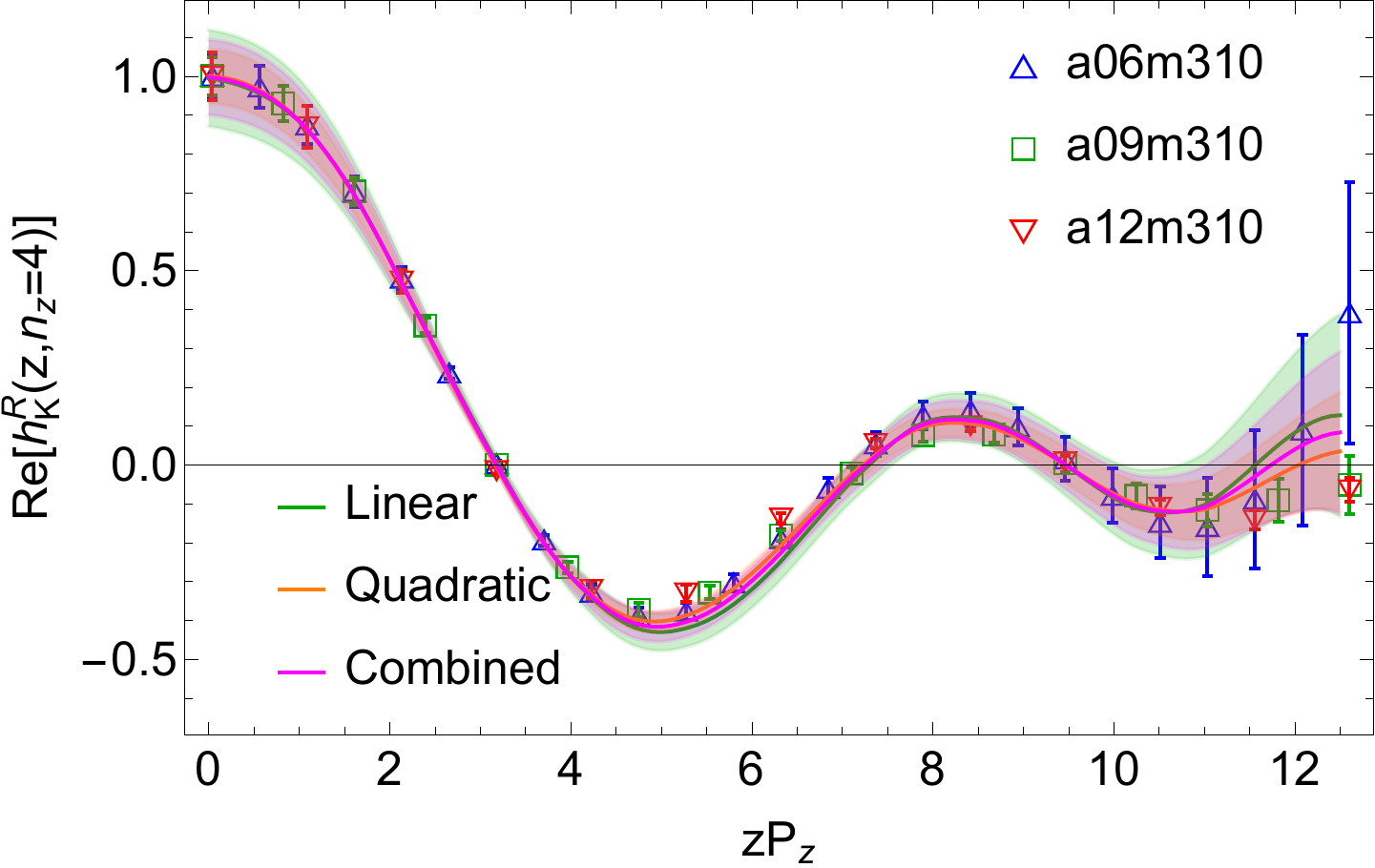}
	\includegraphics[width=0.32\linewidth]{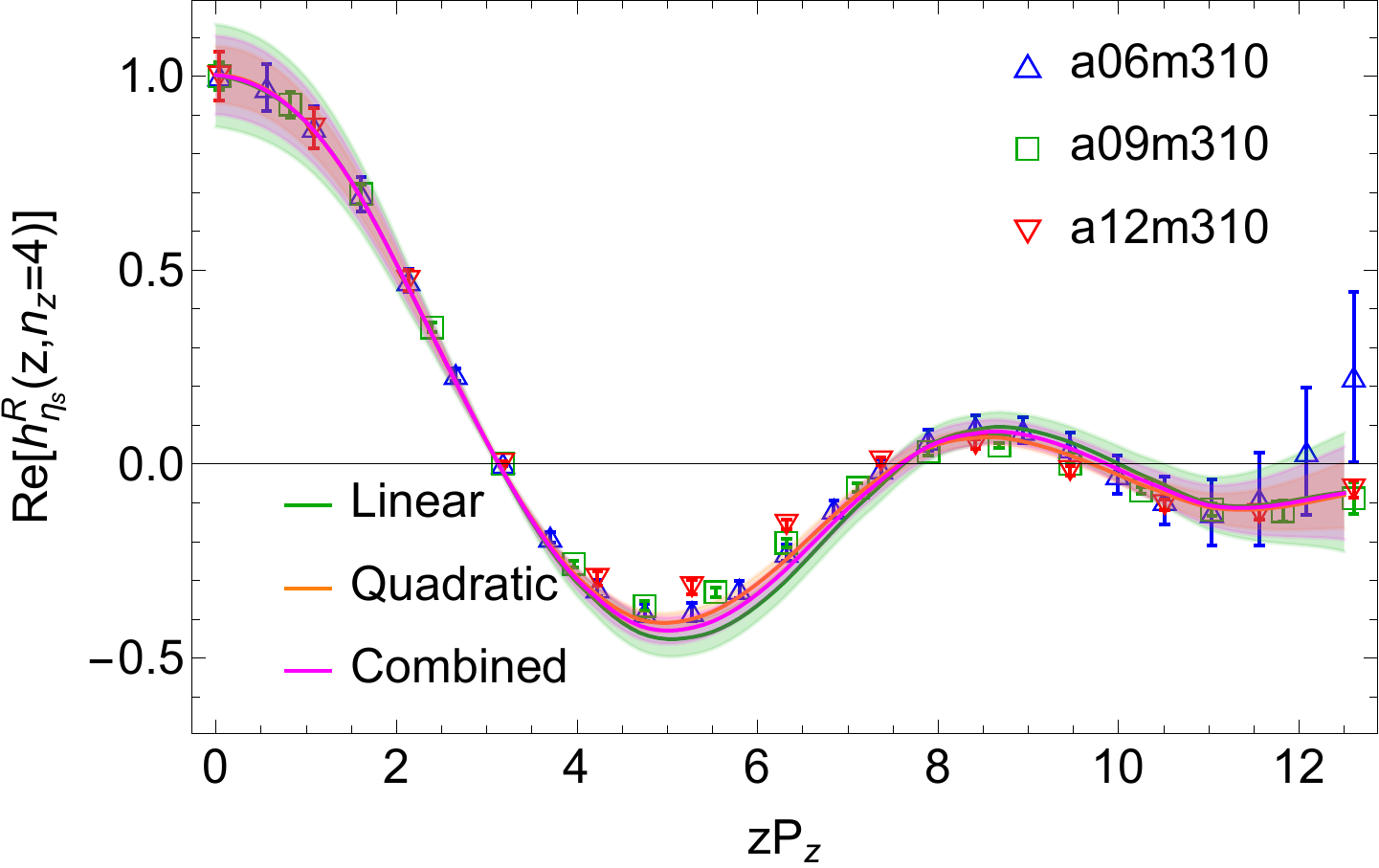}
	\includegraphics[width=0.32\linewidth]{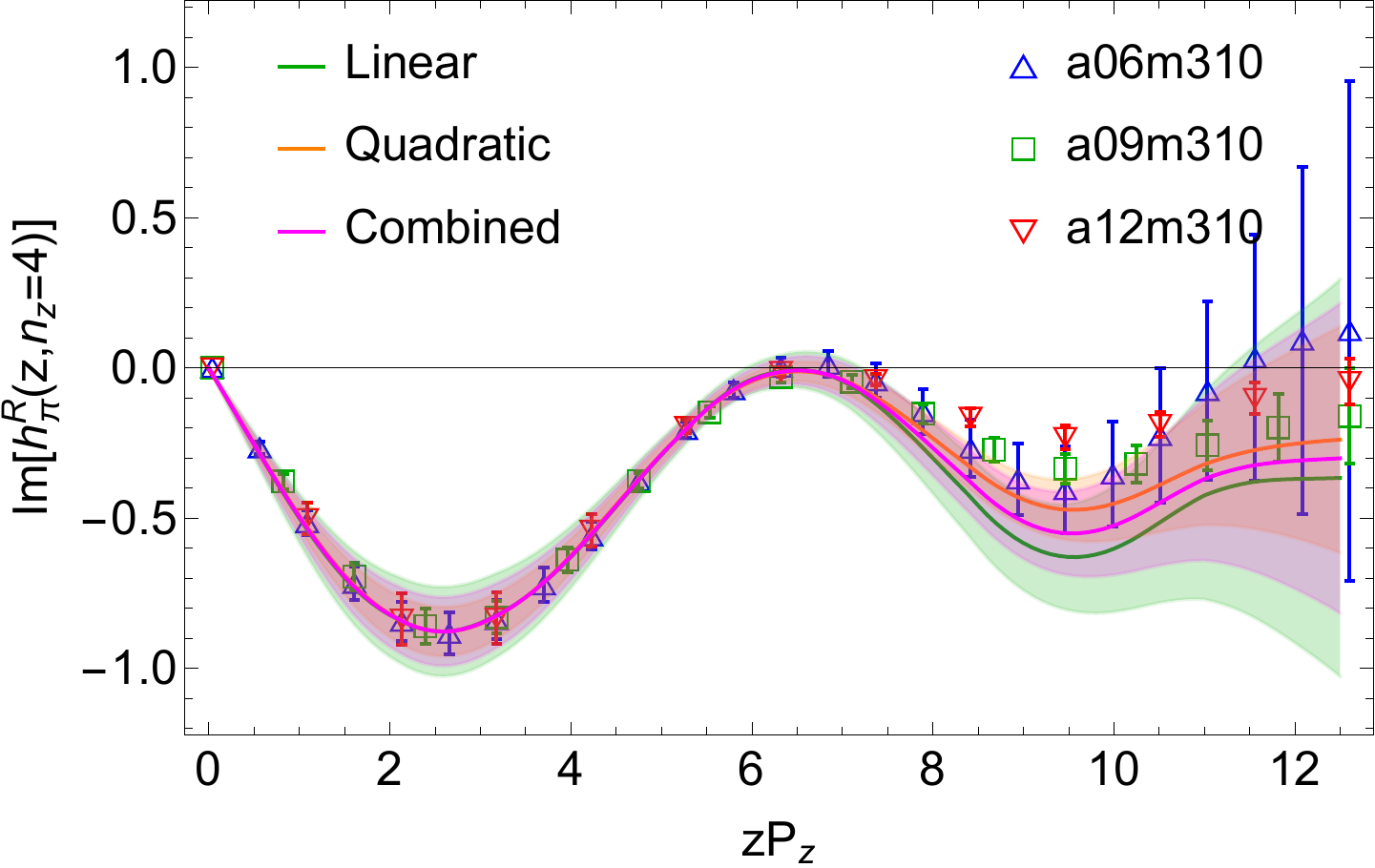}
	\includegraphics[width=0.32\linewidth]{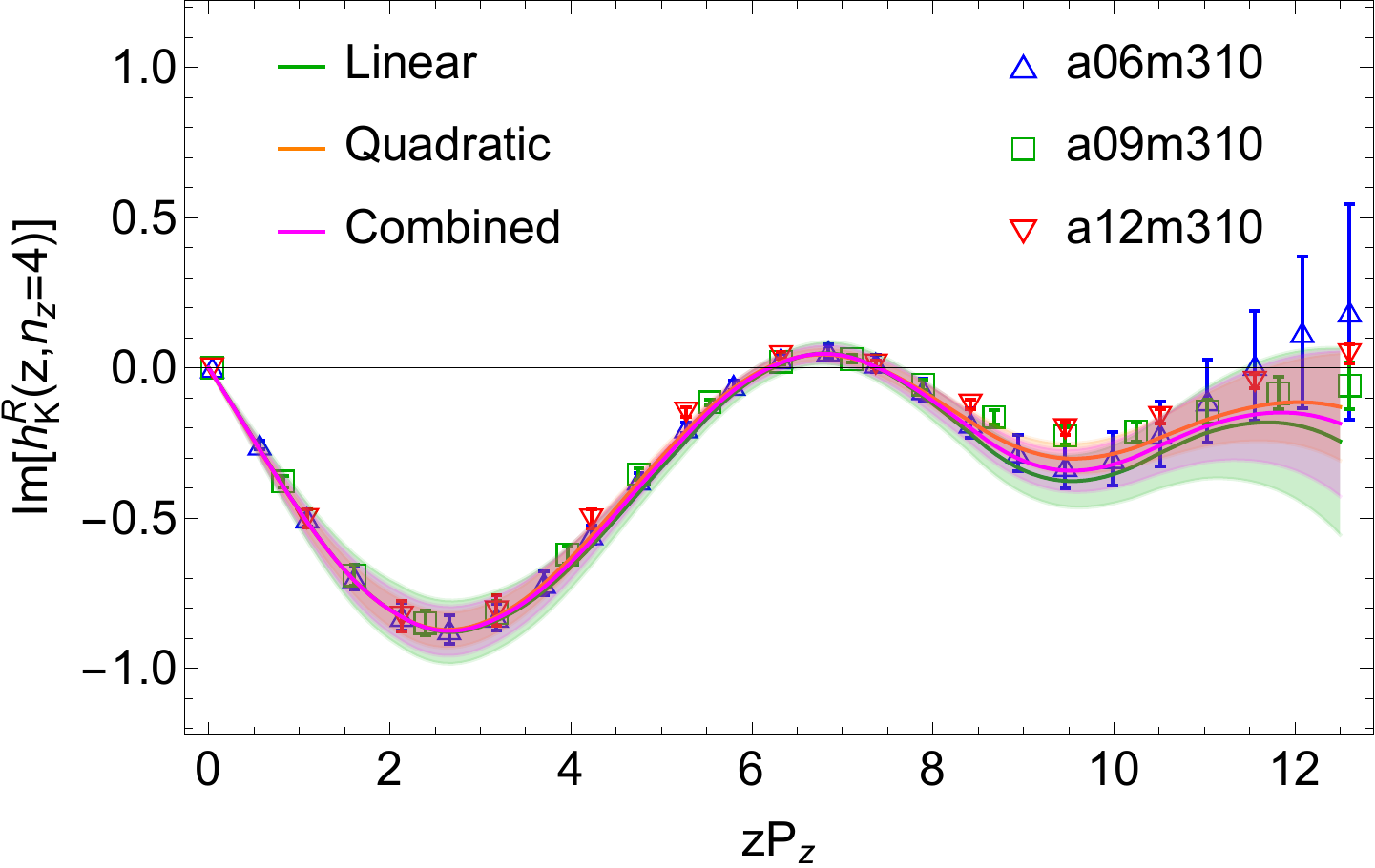}
	\includegraphics[width=0.32\linewidth]{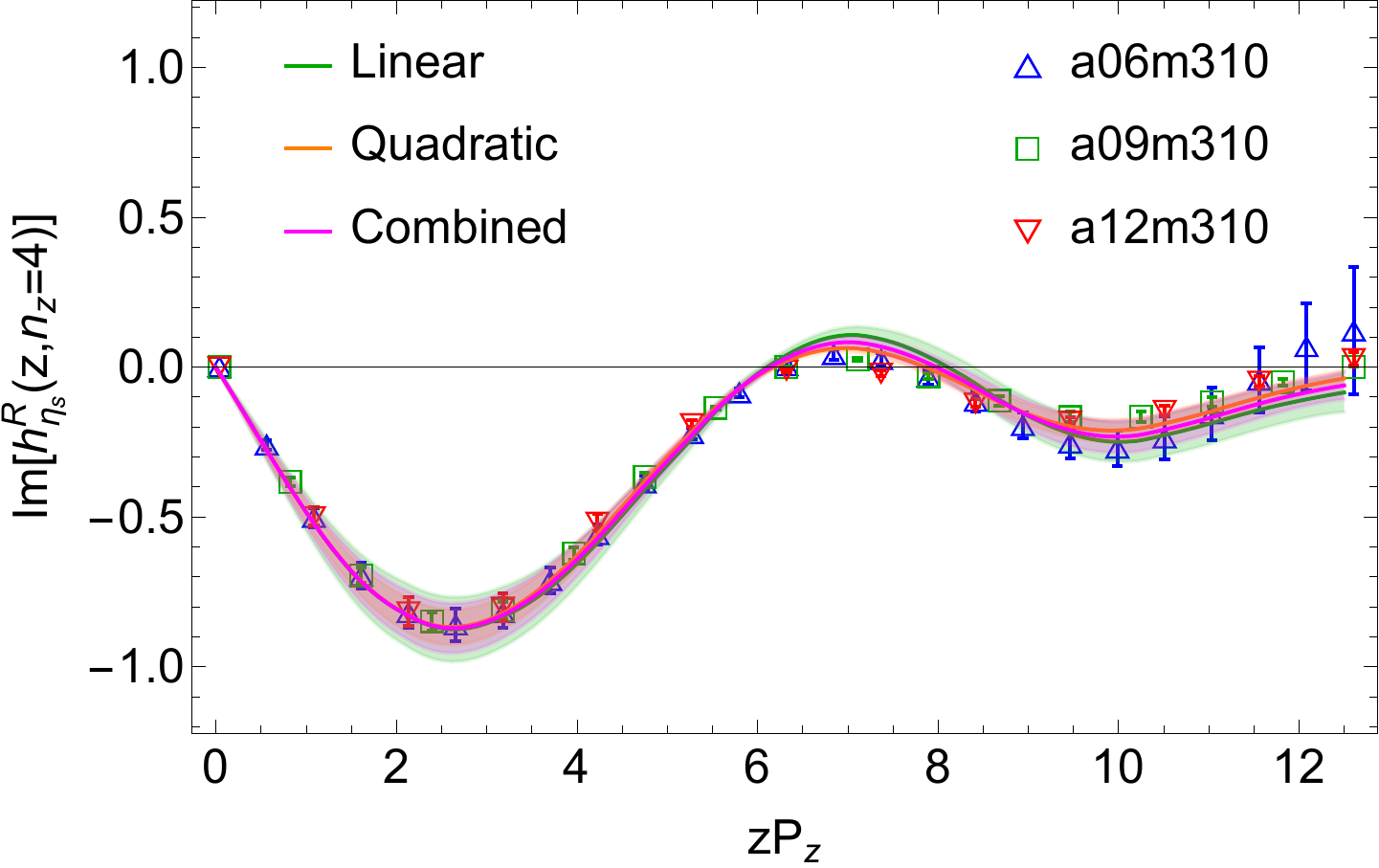}
	\caption{Continuum extrapolation of the real (top row) and imaginary (bottom row) renormalized matrix elements at $\mu^R=3.8$~GeV, $p_z^R=0$ to the continuum from two functional forms and their AIC combination for $\pi$ (left column), $K$ (middle column) and $\eta_s$ (right column). Different extrapolations are consistent with each other.}
	\label{fig:extrapolated_ME}
\end{figure*}

Bootstrap resampling is applied to the three data sets to estimate the error of the continuum extrapolation, since the number of measurements on three ensembles are different.  
The fitted functional forms are consistent with the data points and have average $\chi^2/\text{dof}\approx 1.2$ for $n_z=4$. 
We observe that for the pion the slopes $c_{\pi,1}$ and $c_{\pi,2}$ are consistent with zero for $zP_z<8$. 
Figure~\ref{fig:extrapolated_ME} shows the extrapolated renormalized matrix elements for all mesons at $n_z=4$. We find that at small link lengths $z<0.5$~fm, the  lattice-spacing dependence of the matrix elements is consistent with zero, 
so the extrapolated results are consistent with the data on all ensembles.
At moderate link lengths, $0.5\text{ fm}<z<1\text{ fm}$, near the peaks, the dependence is the most significant and we see $\abs{c_{M,1}} \approx 2\fm^{-1}$ for $K$ and $\eta_s$. 
At large link lengths $z>1$~fm, the lattice-spacing dependence is obscured by the large error, and the extrapolations are mainly constrained by the two cleaner data sets on $a\approx 0.09$~fm and $a\approx 0.12$~fm, where fewer Wilson links are needed at a given physical length of $z$.

To take into account the systematics of using different fitting functions, we used the Akaike information criterion (AIC) technique~\cite{akaike1998bayesian} to combine the linear and quadratic fits: 
\begin{equation}
\label{eq:combined_fit}
    h_M^R(z) = \frac{h_{M,1}^R(z) e^{-(2k_1+\chi_1^2)/2} + h_{M,2}^R(z) e^{-(2k_2+\chi_2^2)/2}} {e^{-(2k_1+\chi_1^2)/2} + e^{-(2k_2+\chi_2^2)/2}},
\end{equation}
where $k_1$ and $k_2$ are the number of free parameters, are both 1 in this case. The quadratic dependence on lattice spacing does not well describe the data; thus, the $\chi^2$ is large in the quadratic extrapolation, and the combined extrapolation is dominated by the linear extrapolation. 
Overall, the extrapolations using the two functional forms are close to each other, so the combined extrapolation is consistent with both results, as shown in Fig.~\ref{fig:extrapolated_ME}. 
Future study using ensembles with different lattice spacing can help resolve any quadratic dependence.

\begin{figure*}
	\centering
	\includegraphics[width=0.45\linewidth]{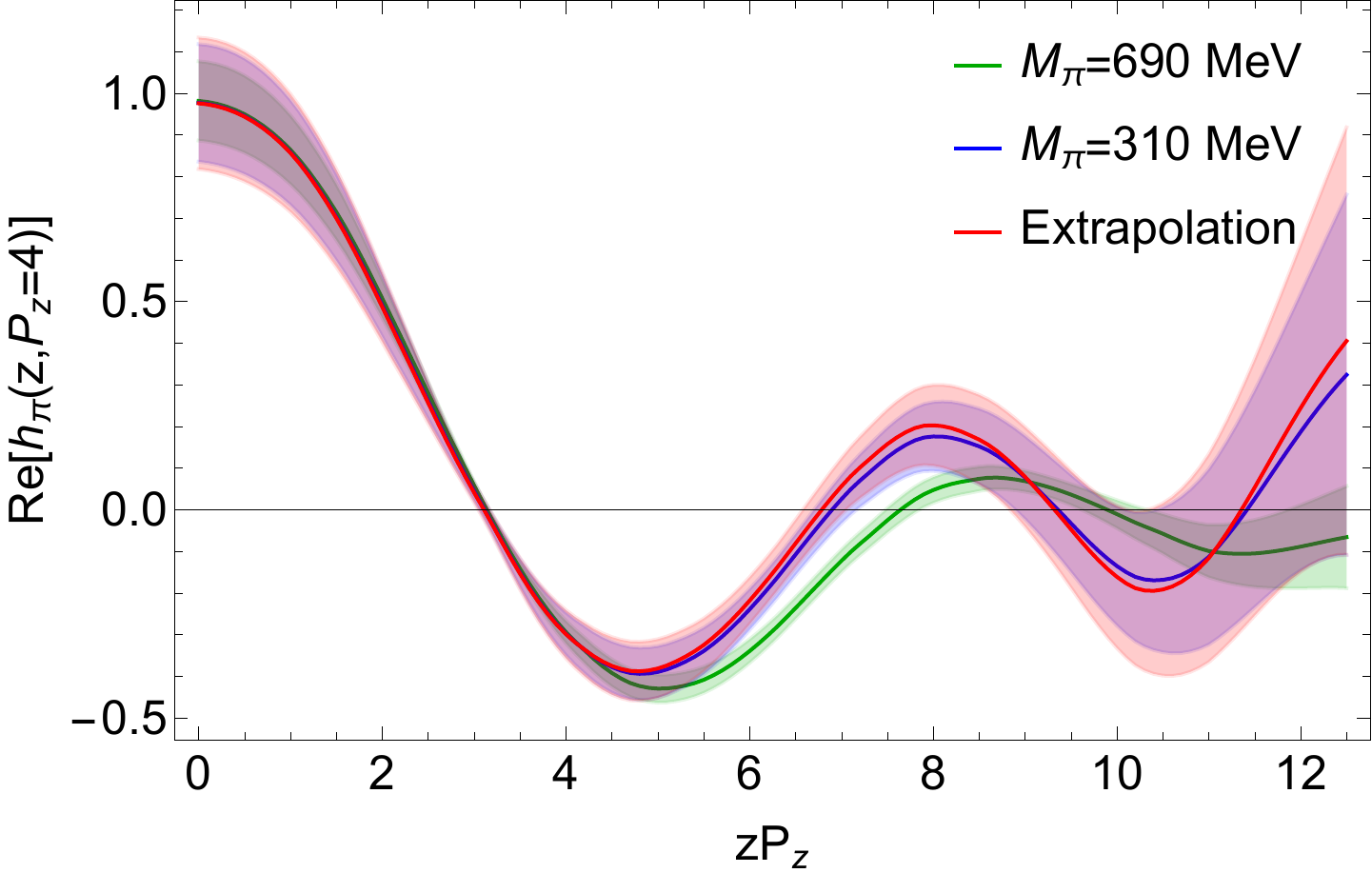}
	\includegraphics[width=0.45\linewidth]{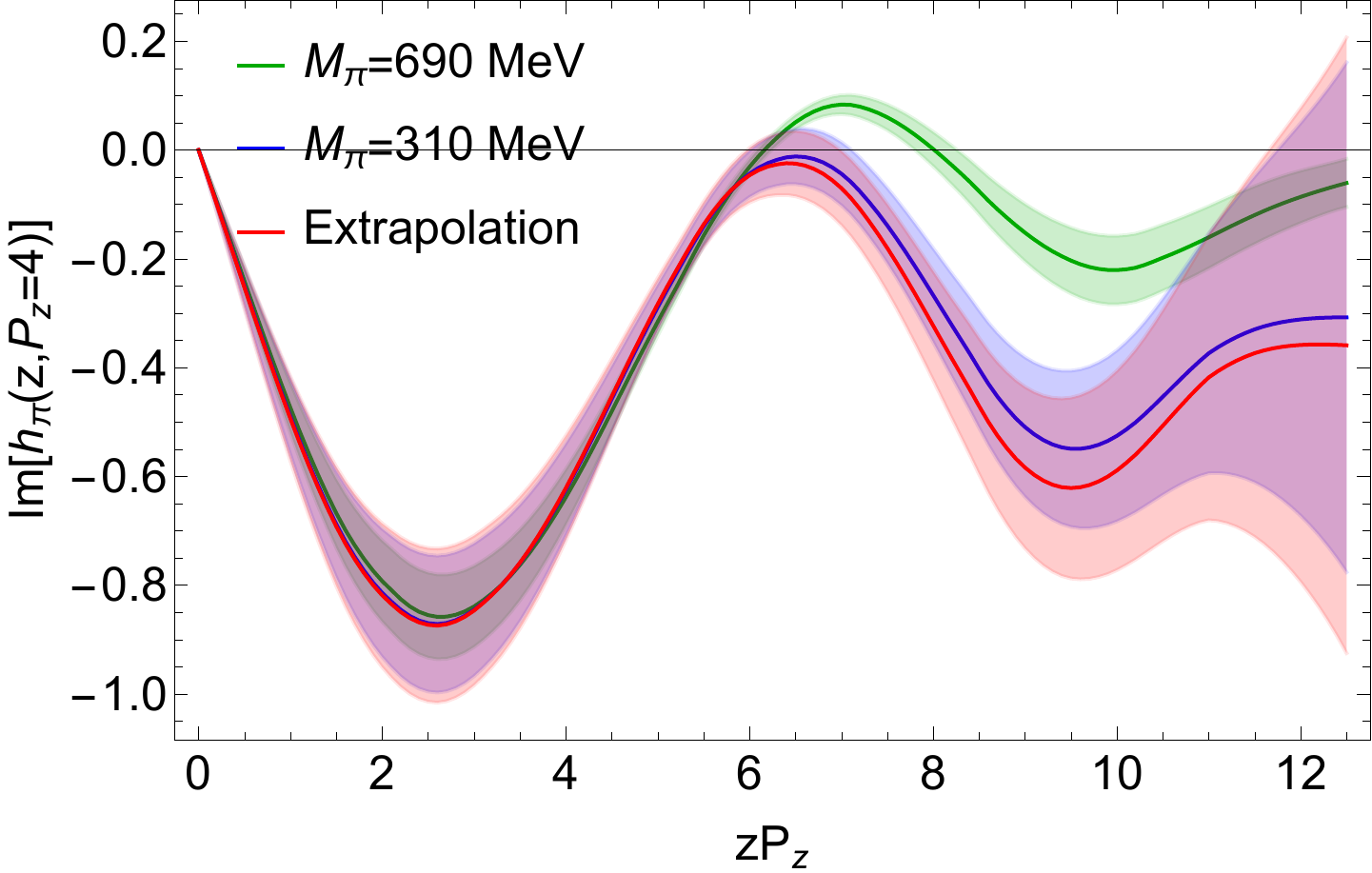}
	\includegraphics[width=0.45\linewidth]{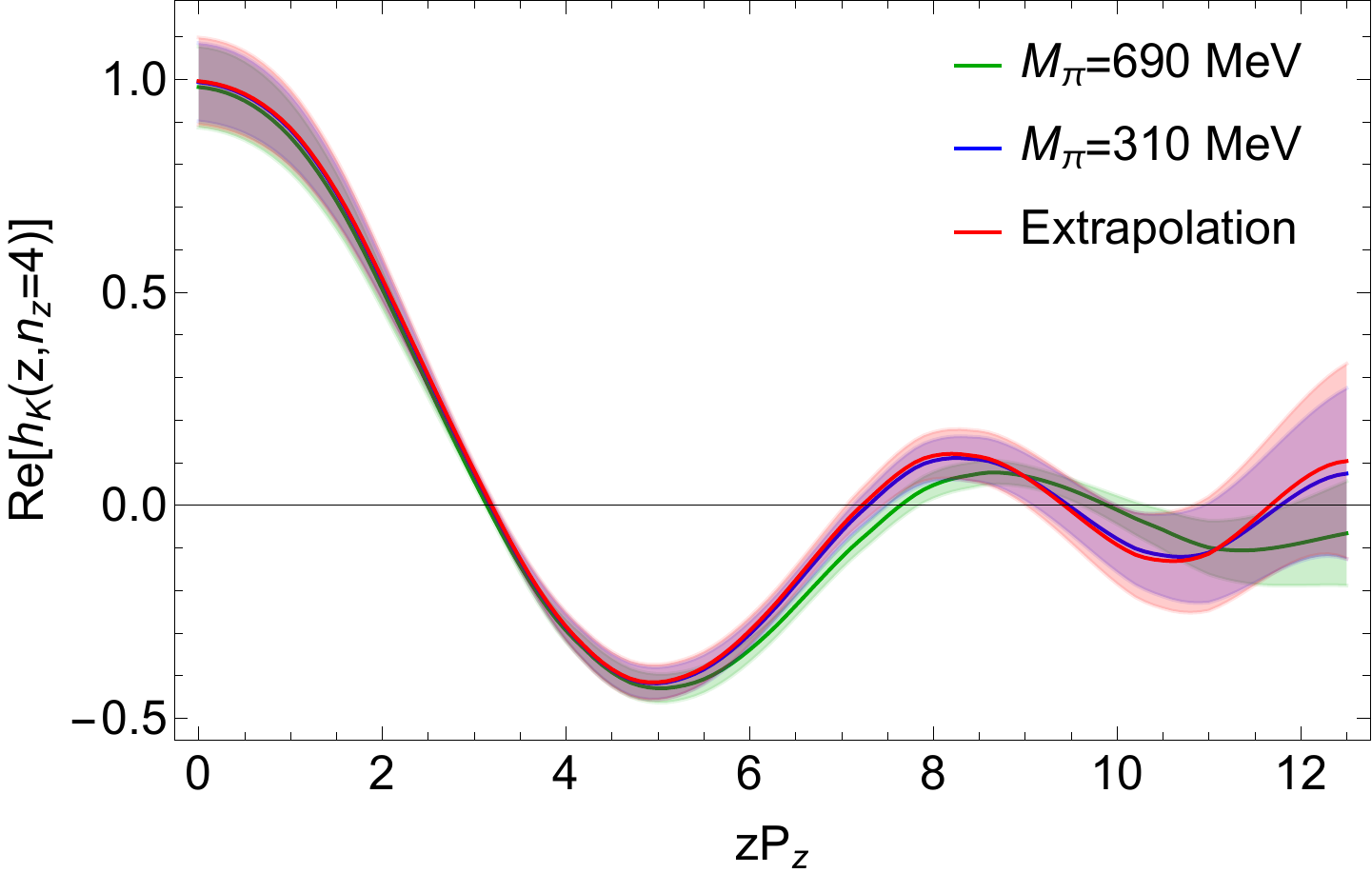}
	\includegraphics[width=0.45\linewidth]{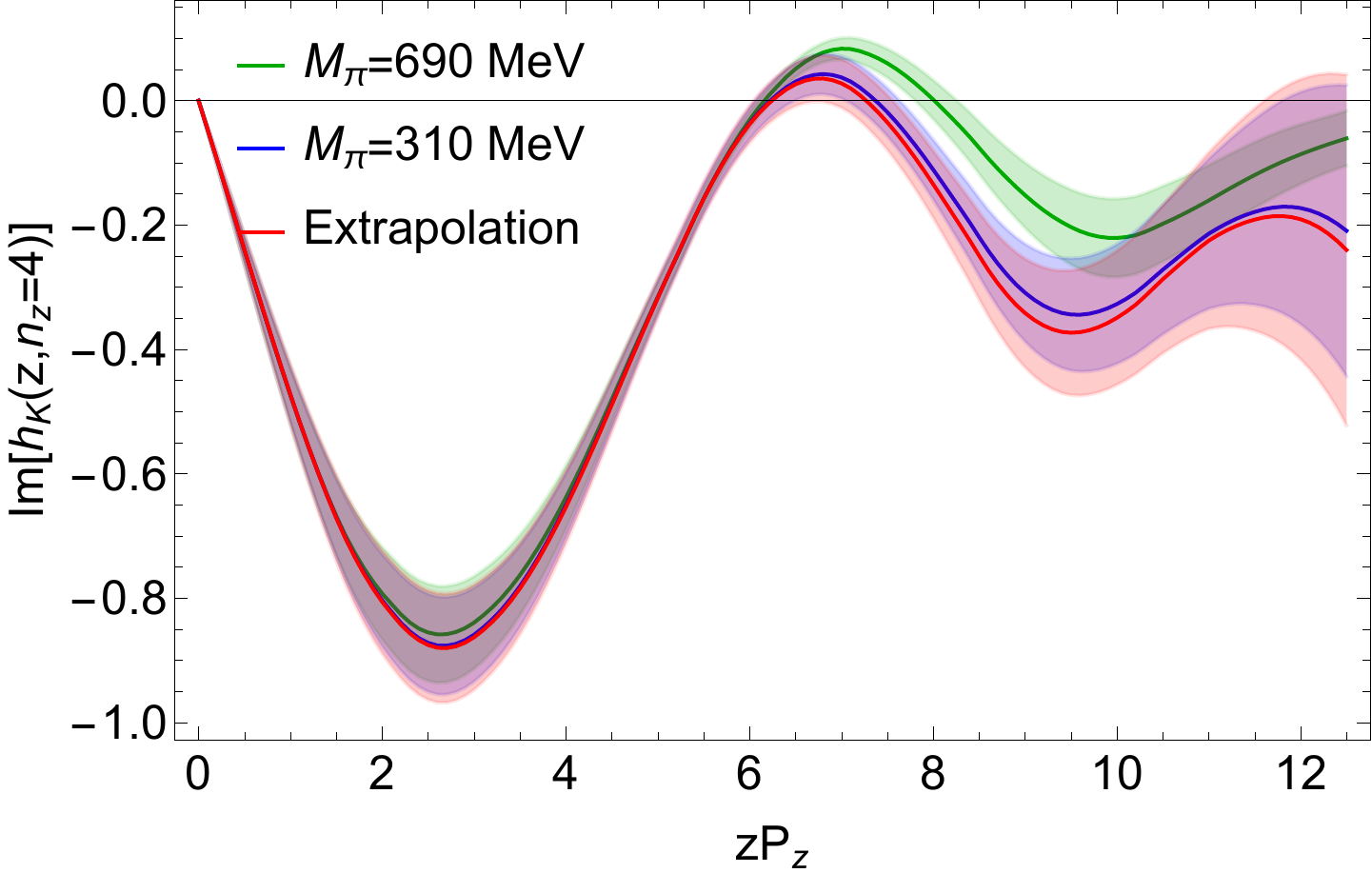}
	\caption{Chiral extrapolation of the $\pi$ (top) and $K$ (bottom) renormalized matrix elements in the continuum to physical pion mass from the $\pi$/$K$ and $\eta$ results for $P_z=n_z\times \frac{2\pi}{L}$ with $n_z=4$. 
	The extrapolated results are close to the $M_\pi=310$~MeV results.}
	\label{fig:mass_extrapolated_ME}
\end{figure*}

The extrapolation formula obtained from one-loop chiral perturbation theory~\cite{Chen:2003fp} is 
\begin{equation}\label{chextr}
    h_M^R(M_\pi,a=0) = s_{M} M_\pi^2 + h(0),
\end{equation}
where the chiral logarithm has been proved to be absent for the DAs of pseudo-Goldstone bosons~\cite{Chen:2003fp}. 
The chiral-extrapolated results are shown in Fig.~\ref{fig:mass_extrapolated_ME}, and they are very close to the ones from calculations at lighter pion mass but with slightly larger error bars due to the extrapolation.
To test whether the higher-loop corrections are significant for $M_\pi=690$~MeV, data at another value of $M_\pi$ is needed. 
In this work, we will use valance pion mass, $M_\pi^\text{val}$, in Eq.~\ref{chextr} for a naive chiral extrapolation to estimate what the DA may be look like at physical pion mass point. 
Future work should include ensembles at lighter pion mass to improve the reliability of the chiral extrapolation and reduce the uncertainty due to such extrapolation.

\subsection{Quasi-DA matrix elements to lightcone DA}

\begin{figure}
\centering
\includegraphics[width=0.9\linewidth]{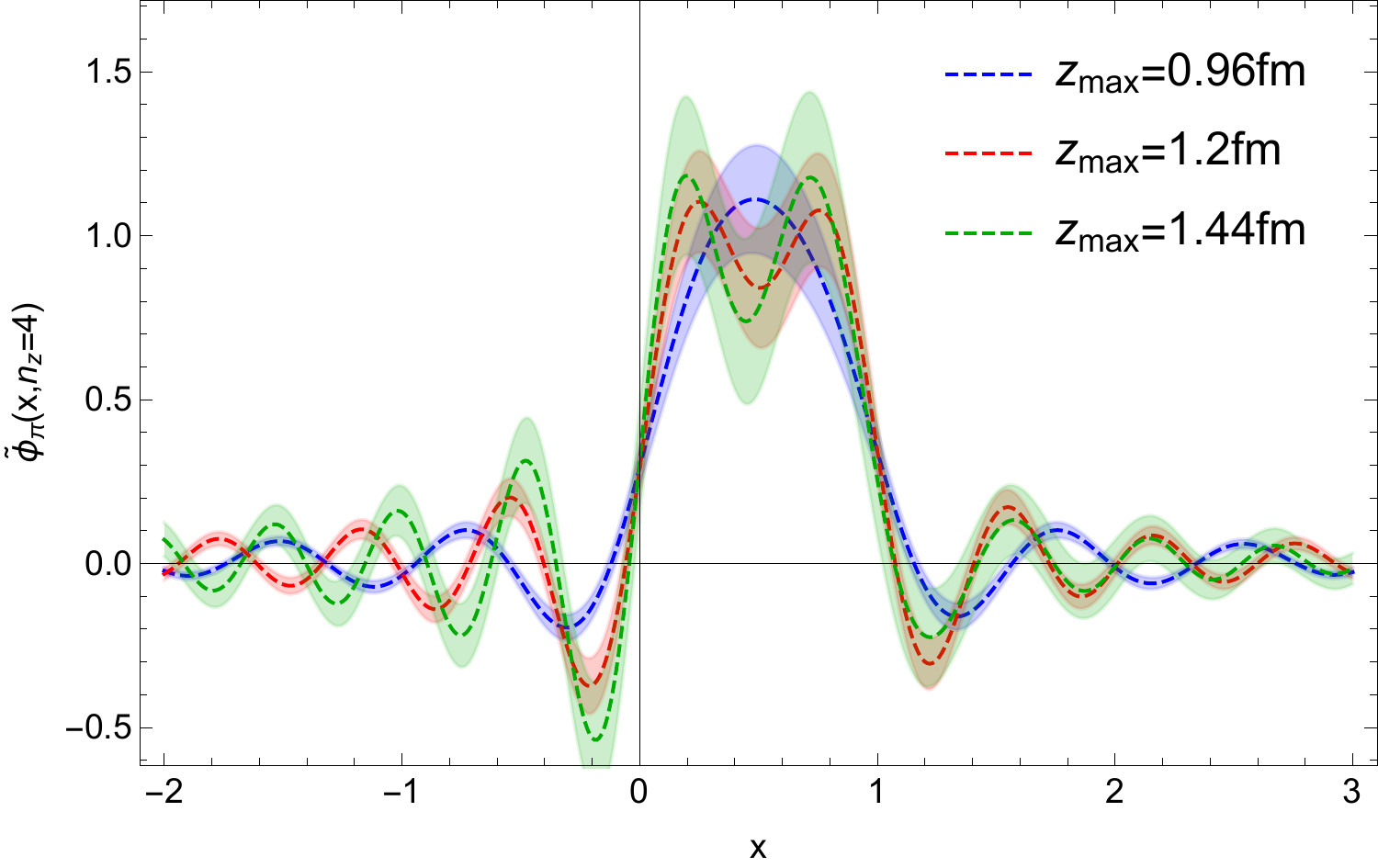}
\caption{
The pion quasi-DA obtained from Fourier transformation of the RI/MOM renormalized matrix elements at $P_z=n_z \frac{2\pi}{L}$ with $n_z=4$, $\mu^R=3.8$~GeV, $p_z^R=0$.
The shape of the peak is sensitive to the $z_\text{max}$ used in the Fourier transformation, and the distribution unphysically oscillates in the large positive and negative $x$ regions. 
}
\label{fig:ft_DA}
\end{figure}

The standard procedure to obtain the lightcone DA via quasi-DA is to first Fourier transform the chiral and continuum extrapolated matrix elements from the coordinate space to the momentum space (i.e. $x$ space), then to apply the inverse matching kernel to obtain the lightcone DA.
The quasi-DA is obtained through
\begin{equation}
 \label{eq:ft}
     \tilde{\phi}_M(x,\mu^R,p_z^R,P_z) = \int \frac{dz}{2 \pi}\, e^{-i(1-x)zP_z} h_M^R(zP_z,p_z^R,\mu^R).
\end{equation}
Because our matrix elements in coordinate space are discretized and bounded in the range $|z|<1.44$~fm, we can only do a truncated Fourier transformation with $|z|\le z_\text{max} \le 1.44$~fm after interpolating the data.
This truncation will introduce a step function into the Fourier transformation and lead to oscillations in the quasi-DA in momentum space.
This was first observed in the nucleon PDF studies~\cite{Chen:2017mzz,Green:2017xeu}, and multiple solutions have been proposed to help resolve or minimize the issue~\cite{Lin:2017ani,Ishikawa:2019flg,Karpie:2019eiq}.
A similar problem is also observed in our meson-DA study; an example from the pion quasi-DA is shown in Fig.~\ref{fig:ft_DA}.
Not only does the pion distribution have similar oscillations, but it is worse than those observed in the nucleon PDF distribution in Ref.~\cite{Chen:2017mzz,Green:2017xeu}.
In addition, the shape of the peak at $x=\frac{1}{2}$ is sensitive to the choice of $z_\text{max}$ used in the Fourier transformation, causing large uncertainty in the DA determination.

To better constrain the DAs such that they vanish outside the physical region, $x=[0,1]$, we adopt the fitting approach by parametrizing the distribution amplitude using the commonly used meson PDF global-fitting form
\begin{align}
\label{eq:DAform}
    f_{m,n}(x) &= \frac{1}{B(m+1,n+1)} x^m (1-x)^n,\\
    B(m+1,n+1) &= \int_0^1 dx\,x^m (1-x)^n,
\end{align}
where $B(m+1,n+1)$ is the beta function, which normalizes the lightcone DA such that the area under the curve is unity.
We then obtain the parameters $m$ and $n$ for the meson lightcone DAs by fitting to the lattice matrix elements $h$
\begin{align}\label{eq:match_to_me}
    h(z,\mu^R,&p_z^R,P_z) = \int_{-\infty}^{\infty} dx \int_0^1 dy\,\nonumber\\ &C\left(x,y,\left(\frac{\mu^R}{p_z^R}\right)^2,\frac{P_z}{\mu},\frac{P_z}{p_z^R}\right) f_{m,n}(y) e^{i(1-x)zP_z} ,
\end{align}
where $C$ is the matching kernel for the DA~\cite{Liu:2018tox} with $\mu=2\GeV$ (the $\overline{\text{MS}}$ renormalization scale), $\mu^R=3.8\GeV$, and $p_z^R=0$. 
This approach was originally proposed for the pion valence PDF~\cite{Izubuchi:2019lyk}.

\begin{figure*}
	\centering
	\includegraphics[width=0.43\linewidth]{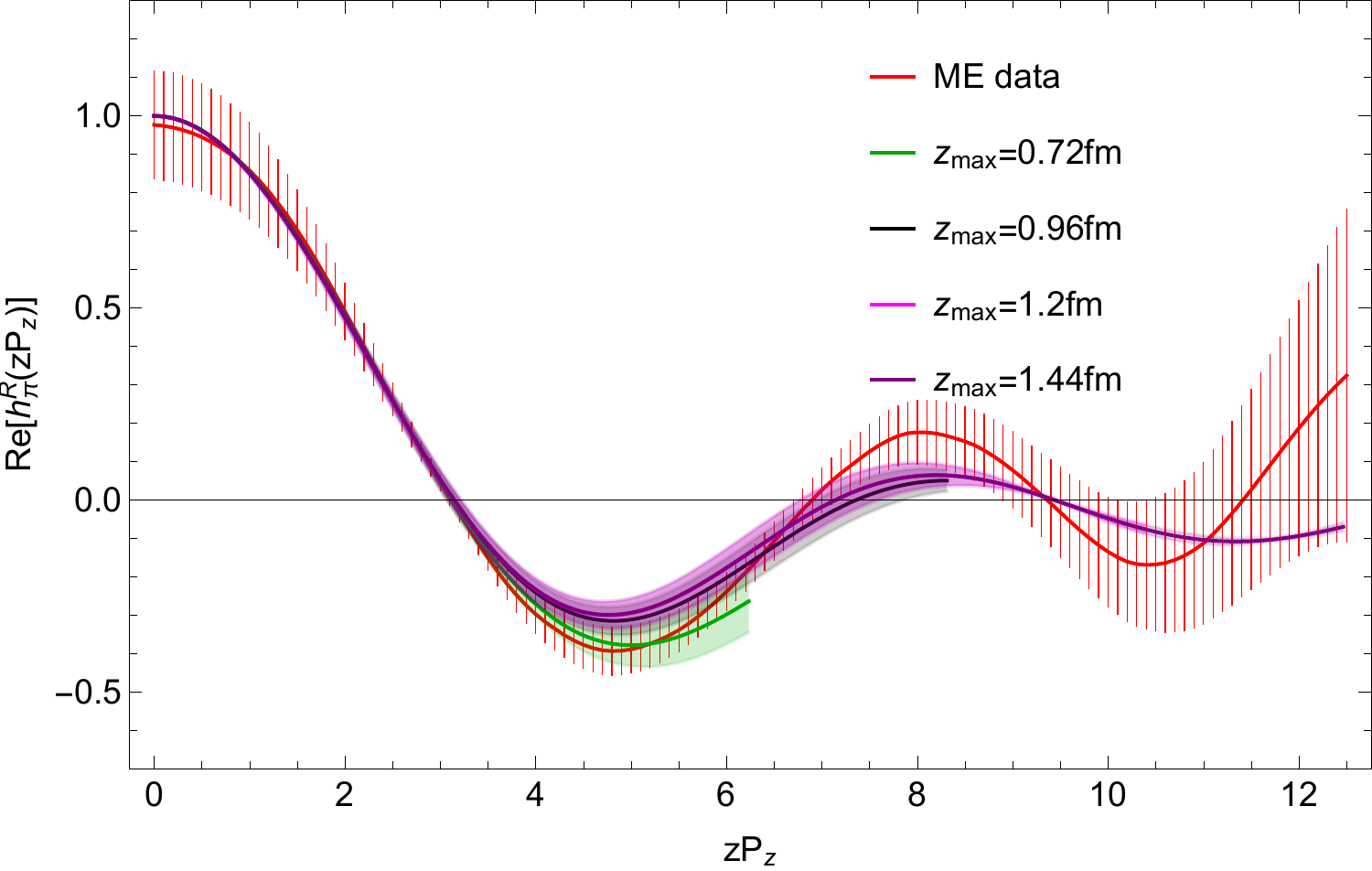}
	\includegraphics[width=0.43\linewidth]{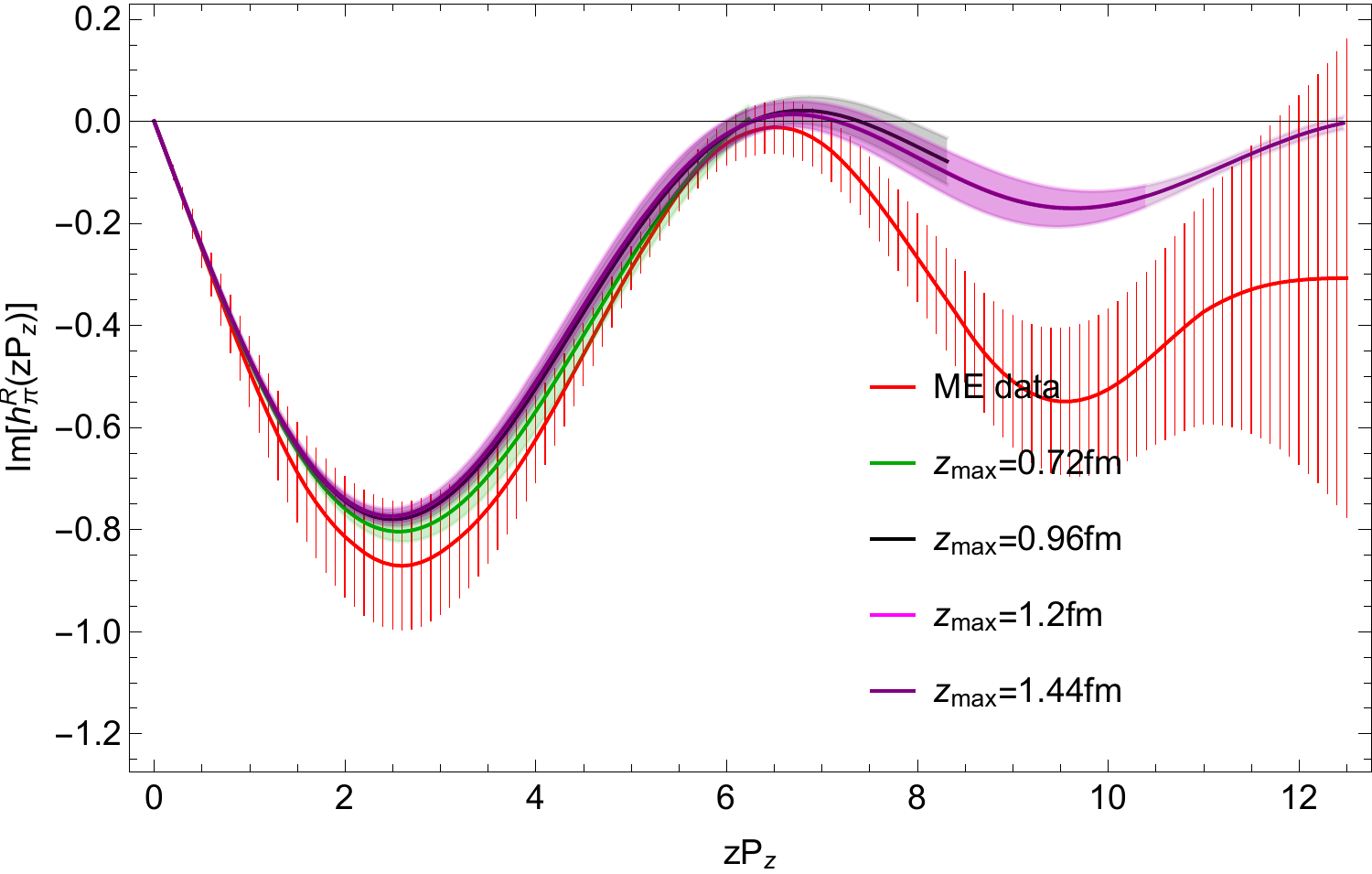}
	\includegraphics[width=0.43\linewidth]{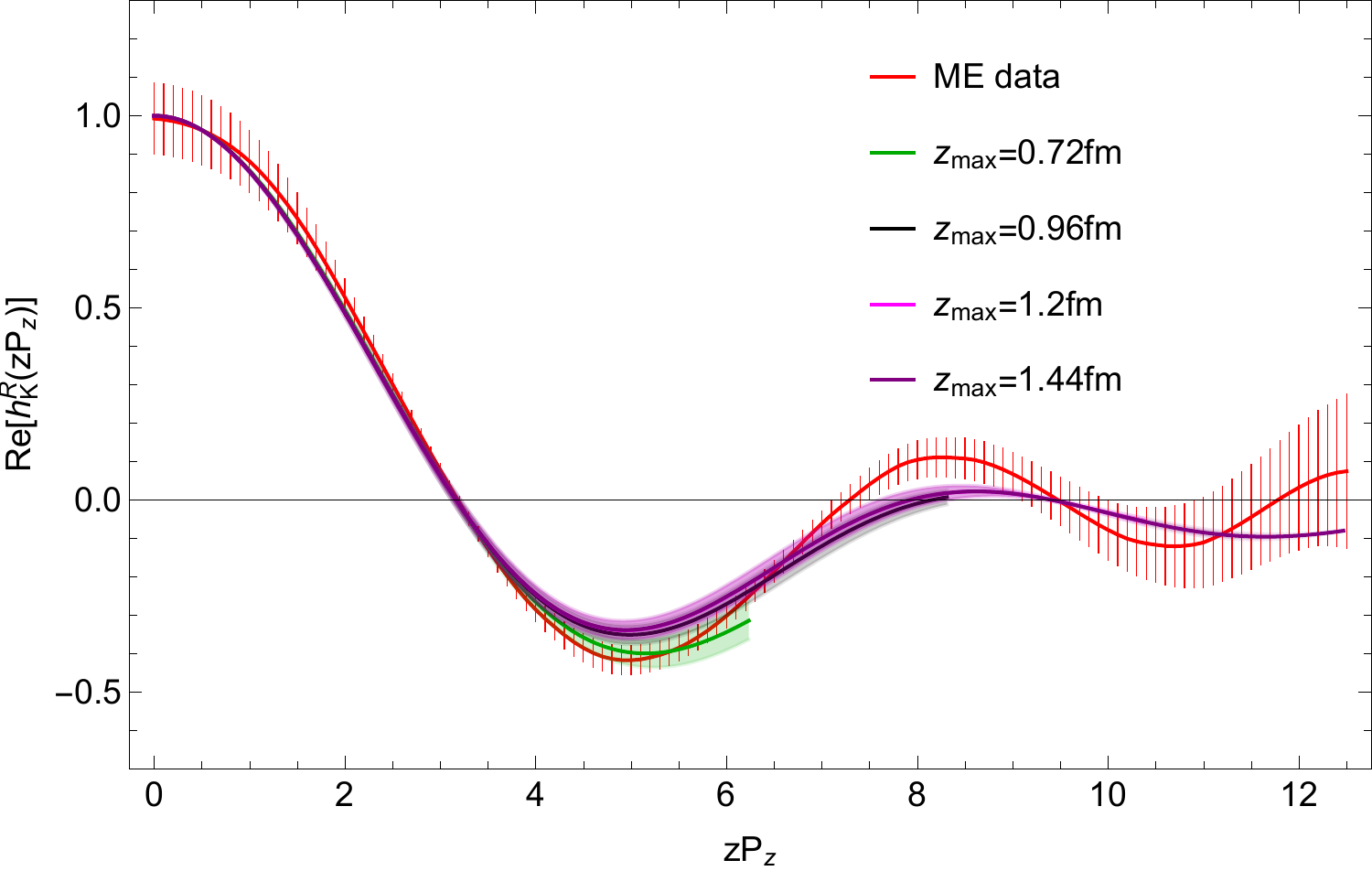}
	\includegraphics[width=0.43\linewidth]{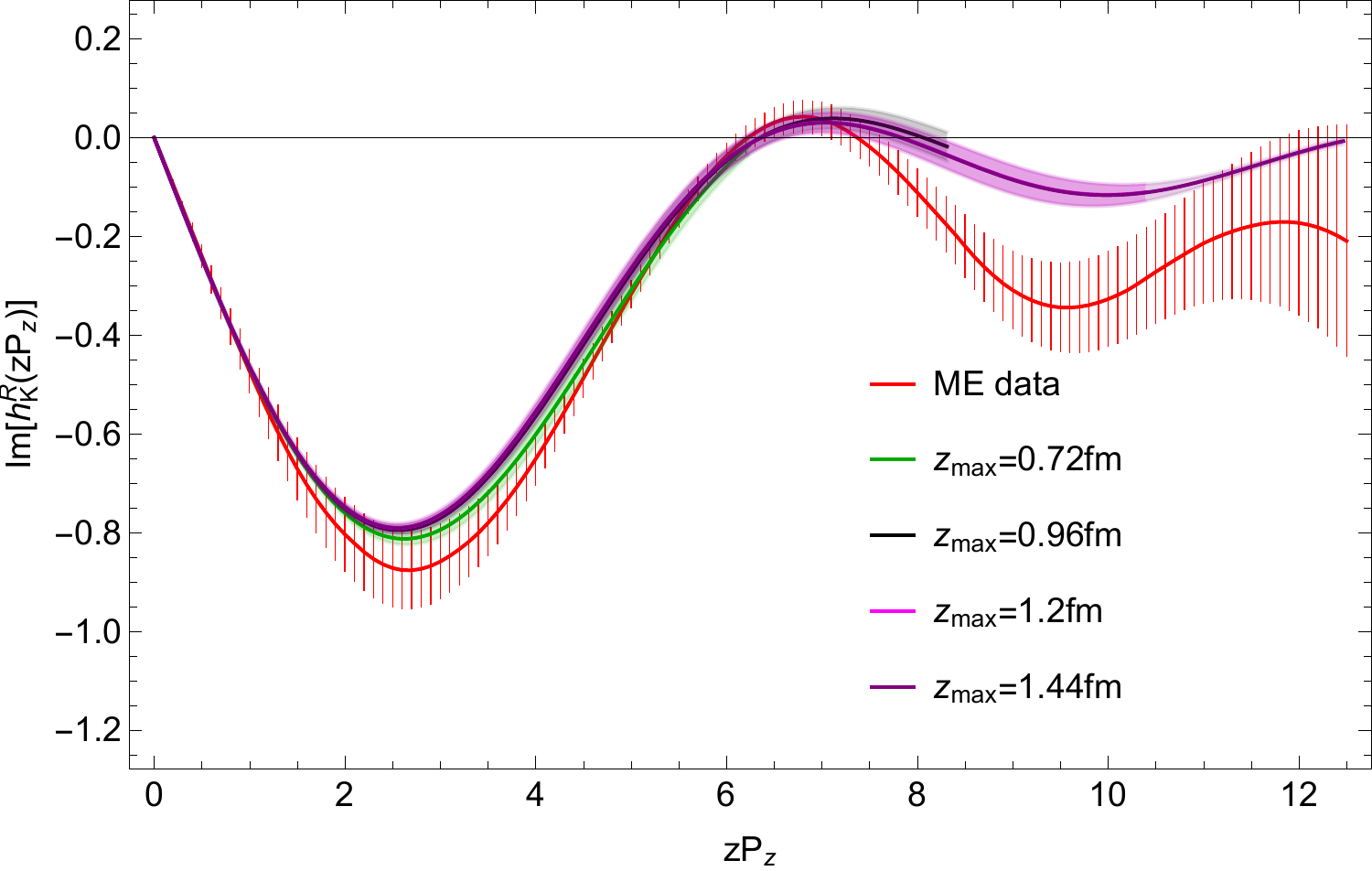}
	\caption{Fit of the matched function form to $P_z=4 \frac{2\pi}{L}$ pion (upper) and kaon (lower) renormalized matrix elements at $\mu^R=3.8$~GeV, $p_z^R=0$ in range $|z|<z_{\text{max}}$. 
	We note that the fit with $z_{\text{max}}=1.08$~fm is the same as the fit with $z_{\text{max}}=1.44$~fm.
	If we go to smaller $z_{\text{max}}$, the exponential $a,b$ will become larger to recover the large amplitude of the first peaks.}
	\label{fig:fit_ME_zmax}
\end{figure*}

Figure~\ref{fig:fit_ME_zmax} shows the reconstructed matrix elements from Eqs.~\eqref{eq:DAform} and \eqref{eq:match_to_me} using the fitted parameters, $m$ and $n$, for all three mesons, along with the input chiral-continuum--extrapolated ones at $P_z=1.73$~GeV. 
Results using different values of $z_{\text{max}}$, ranging from 0.72~fm to 1.44~fm, as input data $h(z,\mu^R,p_z^R,P_z)$ are also shown. 
The $\chi^2/\text{dof} = 1.02(58)$ is small for the fit of full-range pion data $z_{\text{max}}=1.44\fm$, and it reproduces the peak locations.
However, we can see from the plot that the fitted function cannot reproduce the large amplitude of the secondary peaks. 
This indicates that more complicated forms need to be used.
The fit results at the two largest $z_{\text{max}}$ are consistent with each other, because $z_{\text{max}}=1.08$~fm already covers the secondary peaks, and the fit is trying to recover the large amplitude there, resulting in a small ${m,n}$. 
When we truncate the data at smaller $z_{\text{max}}$, then the fit is trying to recover the large amplitude at the first peaks, resulting in a larger ${m,n}$.
For the remaining part of the paper, we only show the fit results for the full-range data with $z_{\text{max}}=1.44\fm$.

\begin{figure}
	\centering
	\includegraphics[width=0.9\linewidth]{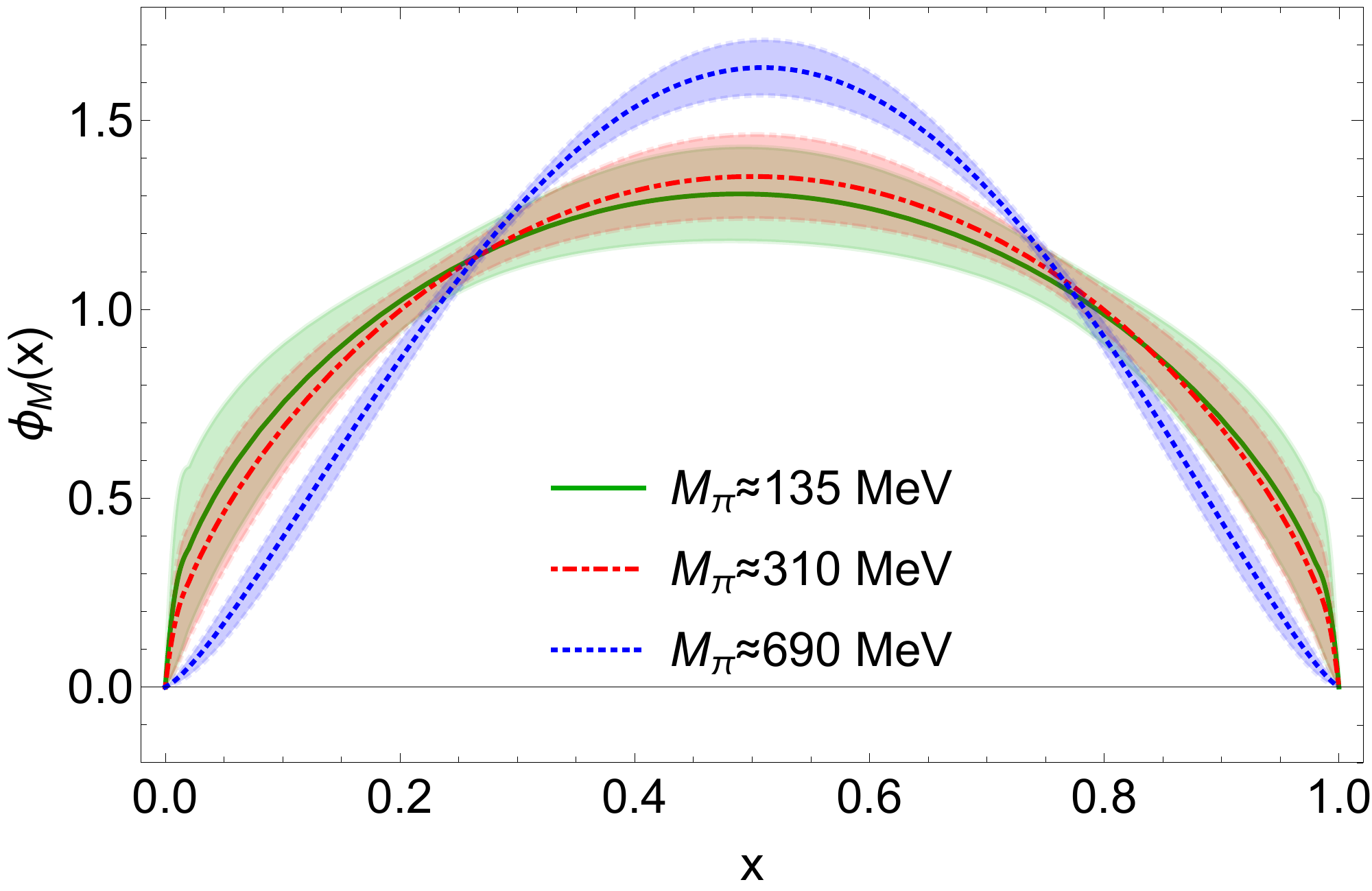}
	\caption{
	Pion distribution amplitude dependence on pion mass at $P_z=1.73\GeV$ as a function of Bjorken-$x$.
	The lighter mesons have a broader distribution.
	}
	\label{fig:fit_x_all}
\end{figure}

We first study the pion-mass dependence of the pion distribution amplitude in the continuum limit.
Figure~\ref{fig:fit_x_all} shows pion DA results using pion masses of 690, 310 and extrapolated to 135~MeV.
We remind the reader that our chiral extrapolation is dominated by 310-MeV results.
Nevertheless, the DAs for the heavier mesons, at strange point, have a narrower distribution, showing a similar trend as suggested in Ref.~\cite{Roberts:2019ngp}.
Mapping out how the DA shapes change as a function of quark masses helps us understand the origin of mass~\cite{Aguilar:2019teb}, which is a priority research direction for a future EIC and other facilities.
We will leave a more complete study of the quark-mass dependence of the DAs to the future. 

\begin{figure*}
	\centering
	\includegraphics[width=0.46\linewidth]{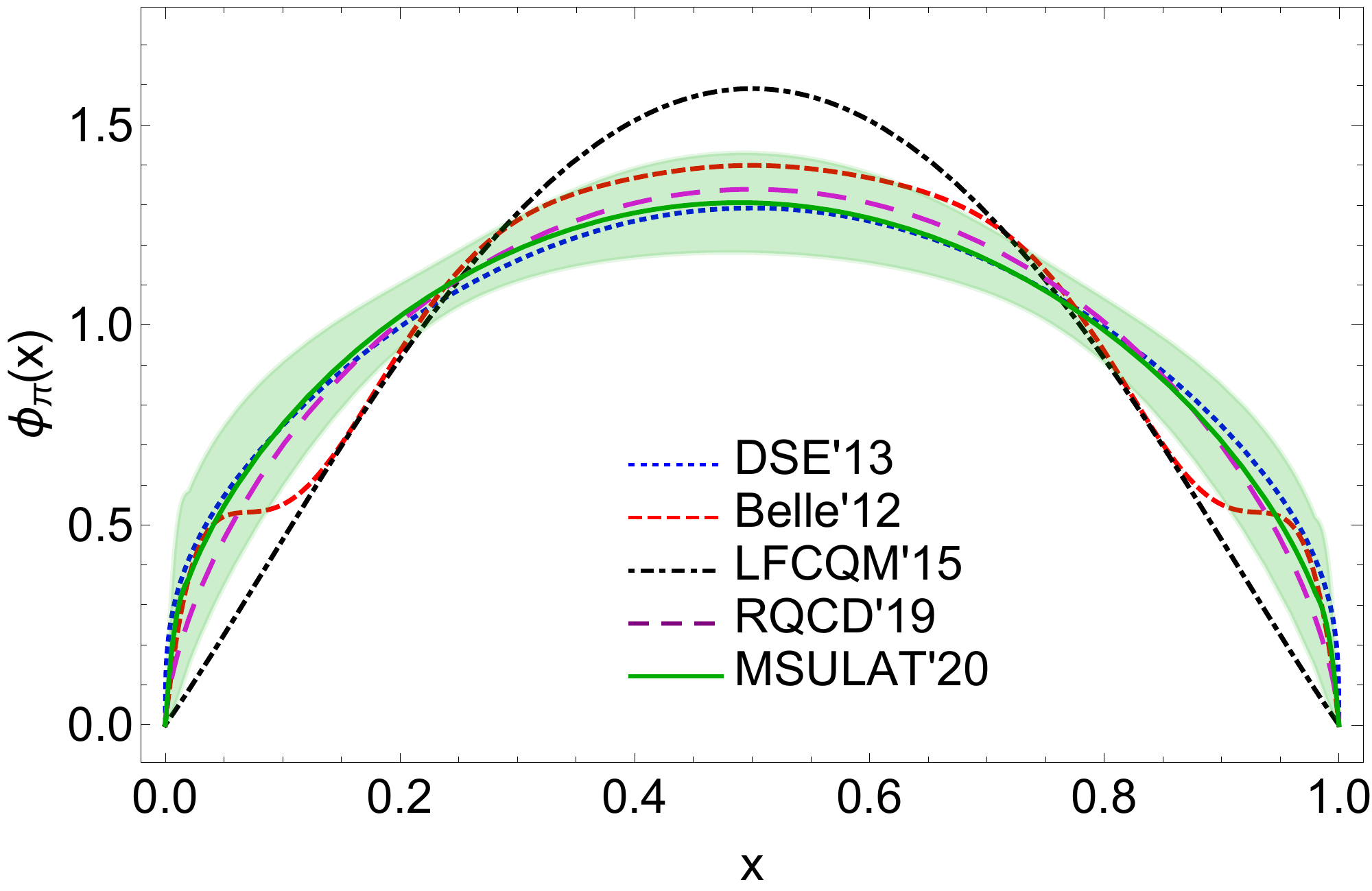}
	\includegraphics[width=0.46\linewidth]{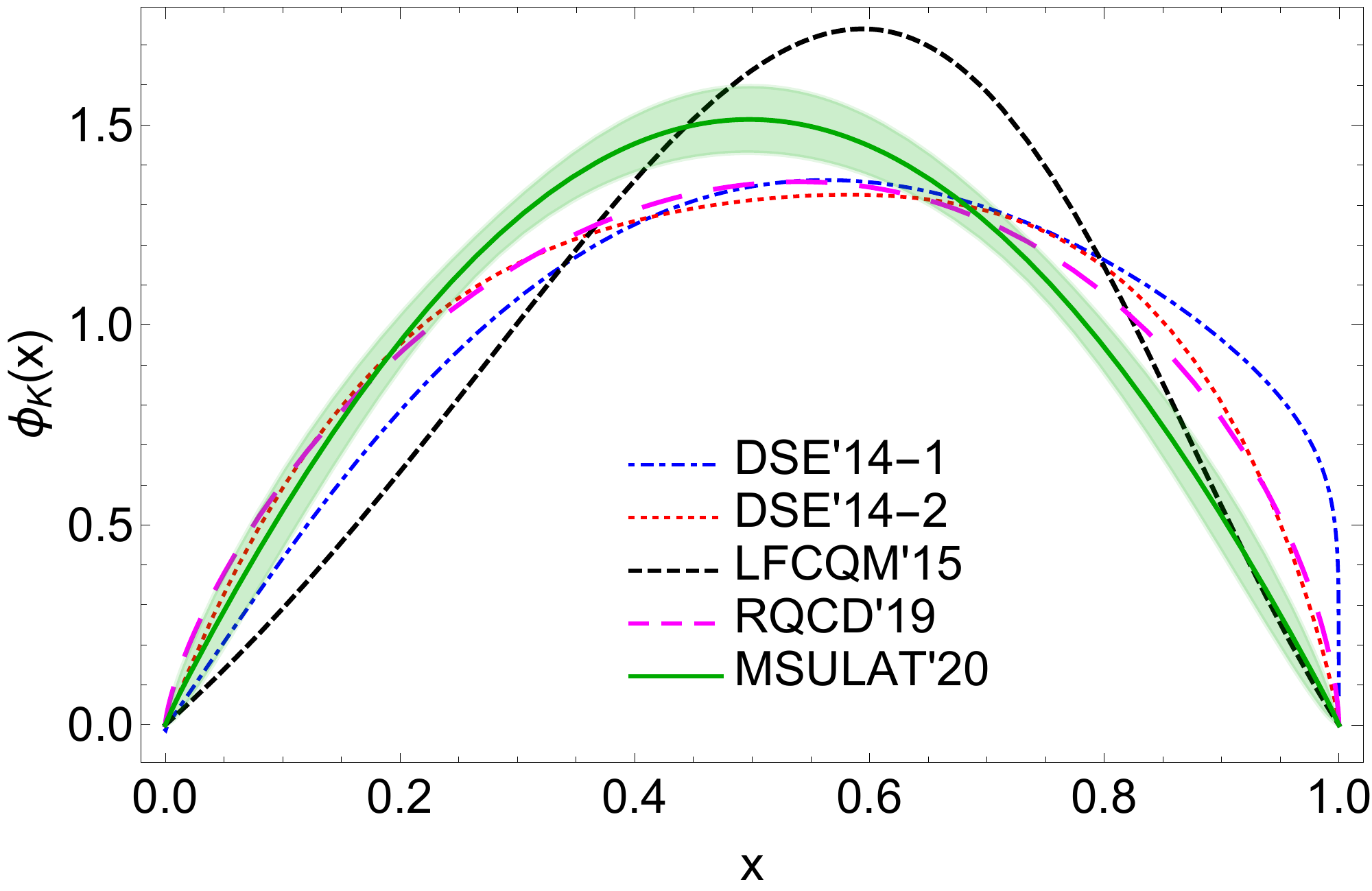}
	\caption{Fit of the $P_z=4 \frac{2\pi}{L}$ pion (left) and kaon (right) data to the analytical form in Bjorken-$x$ space, compared with previous calculations (with only central values shown).
	Although we do not impose the symmetric condition $m=n$, both results for the pion and kaon are symmetric around $x=1/2$ within error.
	\label{fig:fit_x}
	}
\end{figure*}

Our pion distribution amplitude extrapolated to the physical pion mass is shown on the left-hand side of Fig.~\ref{fig:fit_x} with the fitted parameters $m=0.57(27)$ and $n=0.60(26)$.
We also show results from
the Dyson-Schwinger equation (DSE) prediction (DSE'13) with the form $\phi_\pi(x)=1.81[x(1-x)]^{0.31}[1-0.12C_2^{0.81}(2x-1)]$~\cite{Chang:2013pq}, 
the data from Belle experiments~\cite{Agaev:2012tm}, 
the prediction of the light-front constituent-quark model (LFCQM'15)~\cite{deMelo:2015yxk}, 
and the fit to the form Eq.~\eqref{eq:DAform}, of the second moment~\cite{Bali:2019dqc} (labeled as ``RQCD'19''). 
Our pion result is consistent with the DSE and ``RQCD'19'' moment reconstructed results, showing a broader distribution than the LFCQM result. 
Our pion amplitude obtained through the parametrization is constrained to physical region $0<x<1$ by definition, and, therefore, has a higher peak 
compared with the results in our previous work~\cite{Chen:2017gck}. 
RQCD also calculated the $x$-dependent pion distribution amplitude using multiple Euclidean correlation functions~\cite{Bali:2018spj} on a $N_f=2$ 295-MeV pion mass, $a \approx 0.071$~fm lattice-spacing ensemble. 
They found a much broader distribution than our results.  
Using the parameters $m=1.04(20)$ and $n=1.05(20)$ obtained from fitting the kaon matrix elements, we obtain the kaon lightcone DA, as shown on the right-hand side of Fig.~\ref{fig:fit_x}. 
We compare the kaon result with
DSE predictions~\cite{Shi:2014uwa} (labeled as ``DSE'14-1'' and ``DSE'14-2''), 
the LFCQM result (LFCQM'15)~\cite{deMelo:2015yxk}, 
and the fit to the form Eq.~\eqref{eq:DAform} of the first and second moments~\cite{Bali:2019dqc} (labeled as ``RQCD'19''). 
Again, the kaon DA has higher peak compared with the one in our previous work~\cite{Chen:2017gck}, but no observed asymmetric around $x=1/2$. Our kaon distribution is narrower than the DSE and RQCD moment-reconstructed results. 

With the fitted DA, we can calculate their second moments by integration
\begin{equation}
    \langle\xi_M^2\rangle=\int_0^1 dx\, \phi_M(x)(2x-1)^2.
\end{equation}
We find $\langle\xi_\pi^2\rangle=0.244(30)$ for pion and $\langle\xi_K^2\rangle=0.198(16)$ for kaon.
A comparison with previous moment calculations on lattice is shown in Table~\ref{tab:PionMoments}.
The pion moments calculated from our $x$-dependent distributions suffer from larger error due to the usage of larger momentum in the hadron states, while the traditional moment calculations rely on hadrons at rest to obtain better signal.
Our pion results are generally consistent with earlier lattice determinations using the moment approach.
However, our kaon second moment is about 20\%  smaller;
this is anticipated since our kaon $m$, $n$ in Eq.~\eqref{eq:DAform} are larger.
The kaon distribution is narrower than pion one and almost symmetric around $x=1/2$; therefore, we have a smaller kaon moment.

\begin{table*}[tbp]
\begin{center}
\begin{tabular}{|c|c|c|c|c|c|c|c|c|}
\hline
References & Sea quarks & Valence quarks & $\langle \xi^2 \rangle_\pi$ & $\langle \xi^2 \rangle_K$ & Renormalization & a (fm) & $M_\pi$ (MeV) & $M_\pi L$  \\
\hline
MSULat'20 (this work) & 2+1+1f HISQ & clover & 0.244(30) & 0.198(16) & RI-MOM & 0.06--0.012 & 310--690 & 4.4--10 \\
\hline
RQCD'19~\cite{Bali:2019dqc} & 2+1f clover &  clover & 0.234(6)(6) & 0.231(4)(6) &   RI'-SMOM & 0.039--0.086 & 130--420 & 3.6--6.4 \\
\hline
RQCD'17~\cite{Bali:2017ude} & 2+1f clover & clover & 0.2077(43) & N/A & RI'-SMOM &  0.086 &   222--420 & 3.9--5.8 \\
\hline
RQCD'15~\cite{Braun:2015axa} & 2f clover & clover & 0.236(4)(4) & N/A & RI'-SMOM &  0.06--0.08 &   150--260 & 3.4--4.8 \\
\hline
RBC/UKQCD'10~\cite{Arthur:2010xf} & 2+1f DWF & DWF & 0.28(1)(2) &  0.26(1)(2) &  RI'/MOM & 0.11 &  330--670 & 4.5--9.2\\
\hline
QCDSF'07~\cite{Braun:2006dg} & 2f clover & clover &0.260(39) &0.260(6) & RI/MOM   & 0.06--0.085 & 580--1170 & 4.6--9.6 \\
 \hline 
\end{tabular}
\end{center}
\caption{\label{tab:PionMoments} Summary of past dynamical calculations of the second moment of the pion DA. All results listed here are renormalized in $\overline{\text{MS}}$ scheme at 2~GeV, except for QCDSF'07 at 2.69~GeV.
}
\end{table*}

\subsection{Machine Learning Predictions for Lightcone DAs}

Another approach to obtain lightcone DAs from the spatial matrix elements is to apply machine learning.
The idea here is to train a supervised machine-learning model with randomly generated pseudo-data which have similar properties to the DAs and are constrained by the same physical requirements.
The model is then applied to real lattice matrix elements in coordinate space to predict the lightcone DAs.
A similar application to PDFs was studied in Ref.~\cite{Karpie:2019eiq}, where instead of real lattice data, a set of pseudo-data generated from global-fitting results was used to test the method. Note that Ref.~\cite{Karpie:2019eiq} attempted to reconstruct nucleon PDFs using pseudo lattice data but did not finish by using actual lattice data to obtain PDFs.

In this work, we use the multilayer perceptron (MLP) regressor~\cite{hinton1990connectionist,glorot2010understanding,he2015delving}, a machine-learning algorithm implemented in the Python scikit-learn package~\cite{pedregosa2011scikit}.
Since this is a first attempt to use purely lattice data to reconstruct the distribution functions, we use the same parametrization formula as shown in Eq.~\eqref{eq:DAform}, and their linear combinations with 100,000 randomly generated ${m,n}$ pairs in Eq.~\eqref{eq:DAform}, evaluated at 99 points $x\in(0,1)$ as outputs of the model.
Random relative noise at each point is added to these samples.
Then, we apply Eq.~\eqref{eq:match_to_me} at renormalization scale $\mu^R=3.8$~GeV, $p_z^R=0$ to obtain the corresponding matrix elements at $z\in[0,24]\times 0.06\fm$ in coordinate space as inputs of the model.
We train and test the MLP regressor on these labelled pseudo-data.
The model optimizes the squared-loss $L=\sum_i (y_i^\text{pred}-y_i)^2$, where a large relative deviation near the boundary $x\in\{0,1\}$ will not contribute much to the loss because of its small amplitude.
We tune the hyperparameters of the model, i.e., the geometry of the hidden layer and the activation function, with GridSearchCV in scikit-learn.
The optimized model is a MLP regressor of three hidden layers with 100 perceptrons and the activation function $f(x)=\max(0,x)$ on each layer.

\begin{figure*}
	\centering
	\includegraphics[width=0.32\linewidth]{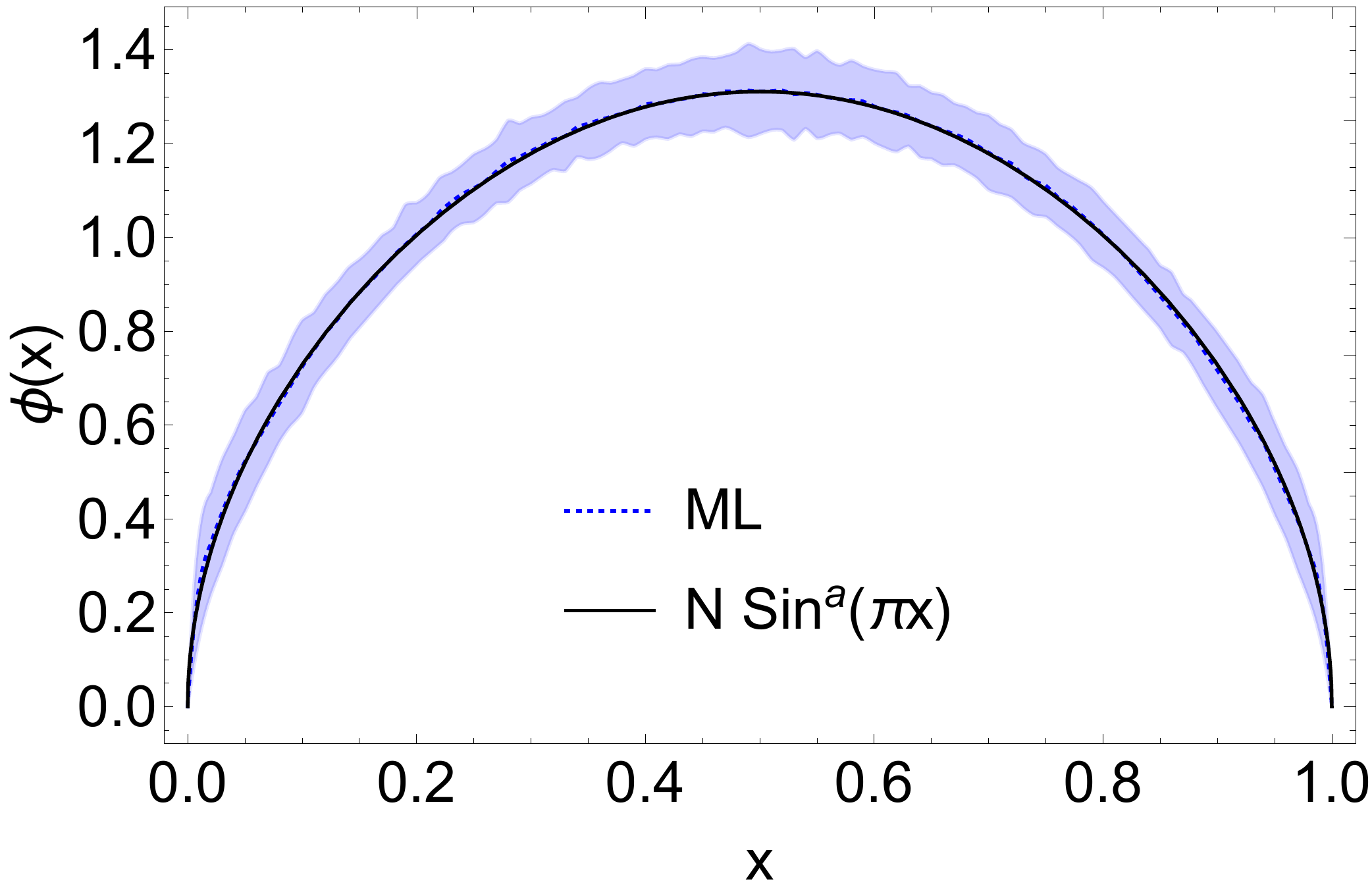}
	\includegraphics[width=0.32\linewidth]{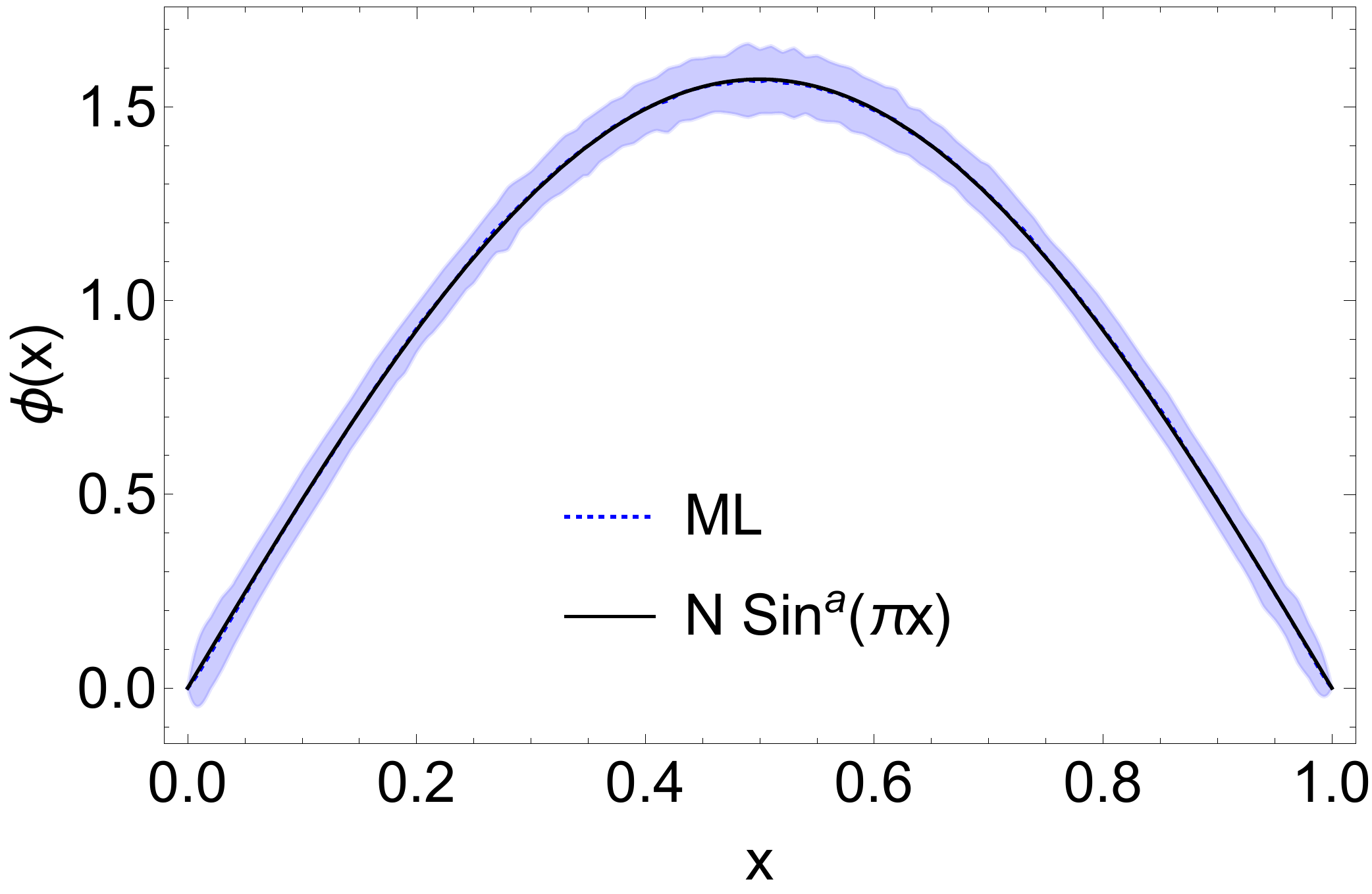}
	\includegraphics[width=0.32\linewidth]{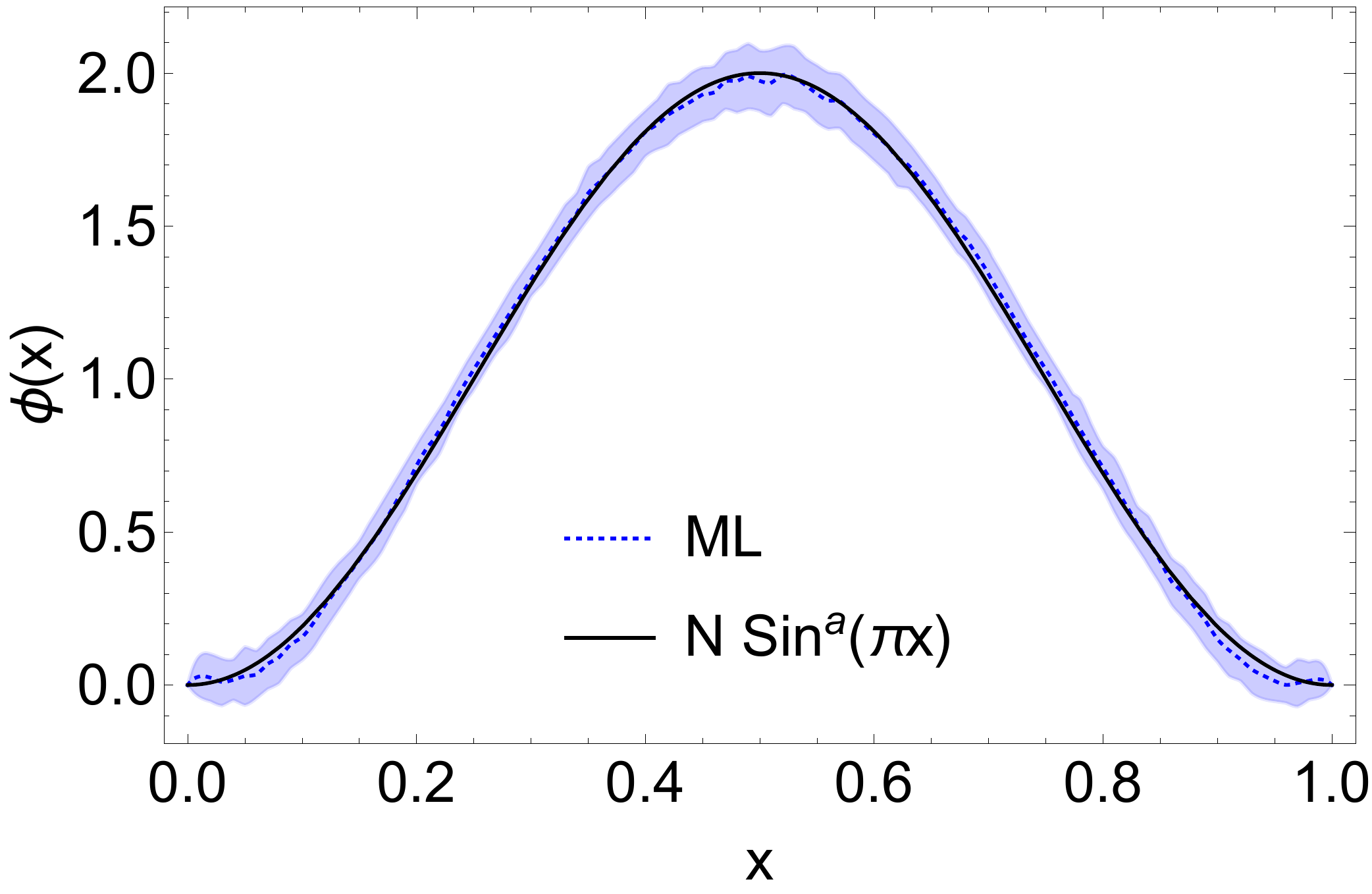}
	\caption{The machine-learning (ML) predictions test on a toy-model distribution with the form $f(x)=N \sin^\beta(\pi x)$ using noisy input pseudo-data generated from $\beta=0.5$ (left), $1$ (middle), $2$ (right). The blue band indicates the uncertainty calculated using bootstrap sampling, and the black curve is the exact function we use to generate those inputs. 	
	The consistency with the true value indicates that extrapolation to unknown form is promising.} 
	\label{fig:ml_sin}
\end{figure*}

To make sure that the above procedure works, we test our procedure on a simpler formula.  
We generate a test set of data with the same constraints but from different form, $f(x)=N \sin^\beta(\pi x)$, to check the stability of the model when extrapolating to unknown functions.
We generate the test data for $\beta\in\{0.5, 1, 1.5, 2\}$.
After transforming to coordinate space, we generate 1000 samples for each $\beta$, following a Gaussian distribution $N(\mu,\sigma^2)$ with $\mu=h(z), \sigma=h(z) \times 0.1 \exp[0.1z]$ to simulate the noise from data on lattice.
We test the model on these sets of noisy data.
It turns out that the the model works fairly well, as shown in Fig.~\ref{fig:ml_sin}, indicating that even if the lattice results do not follow the functional form we used to train the data, the model is able to give close predictions.

\begin{figure*}
	\centering
	\includegraphics[width=0.32\linewidth]{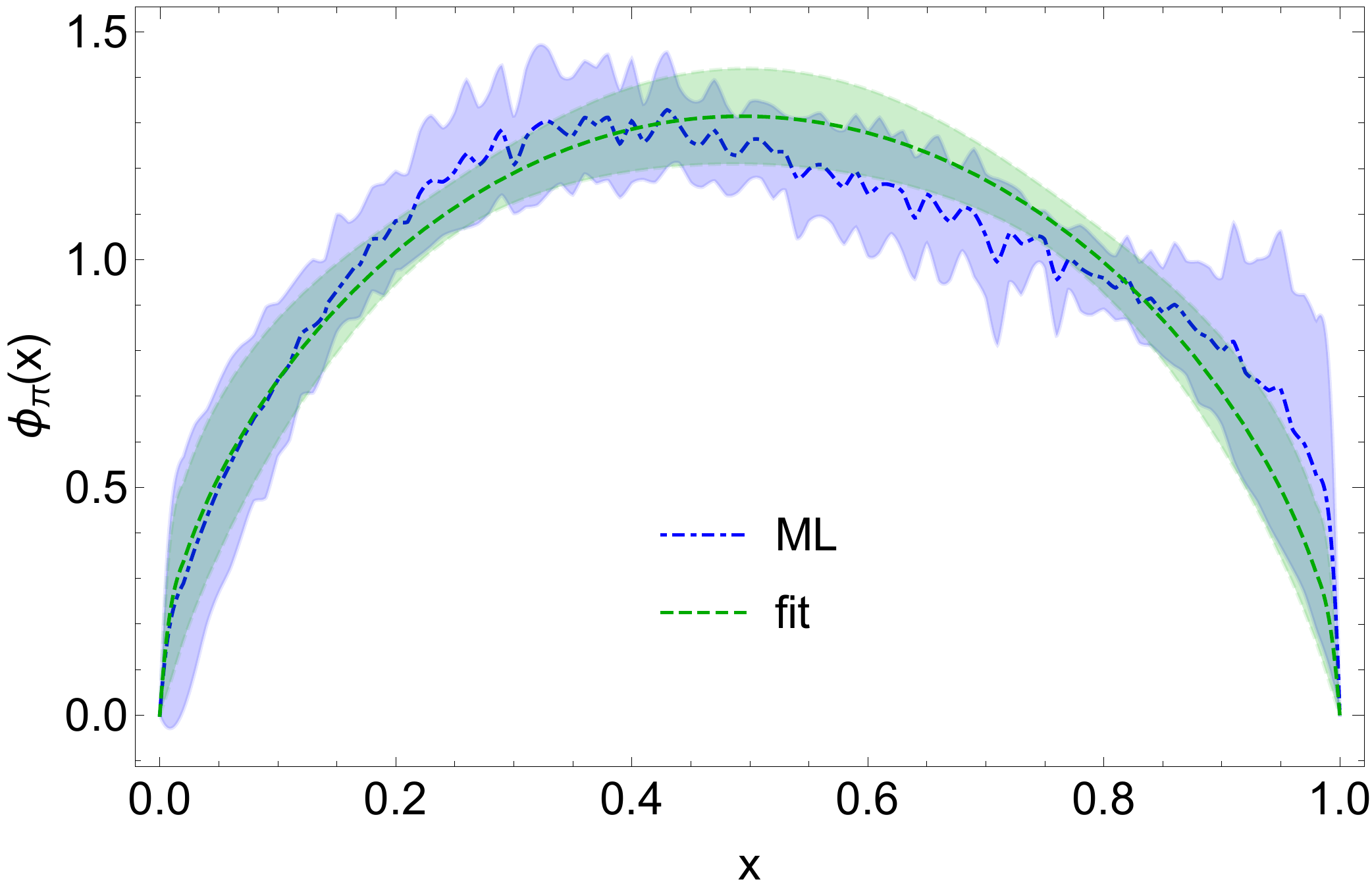}
	\includegraphics[width=0.32\linewidth]{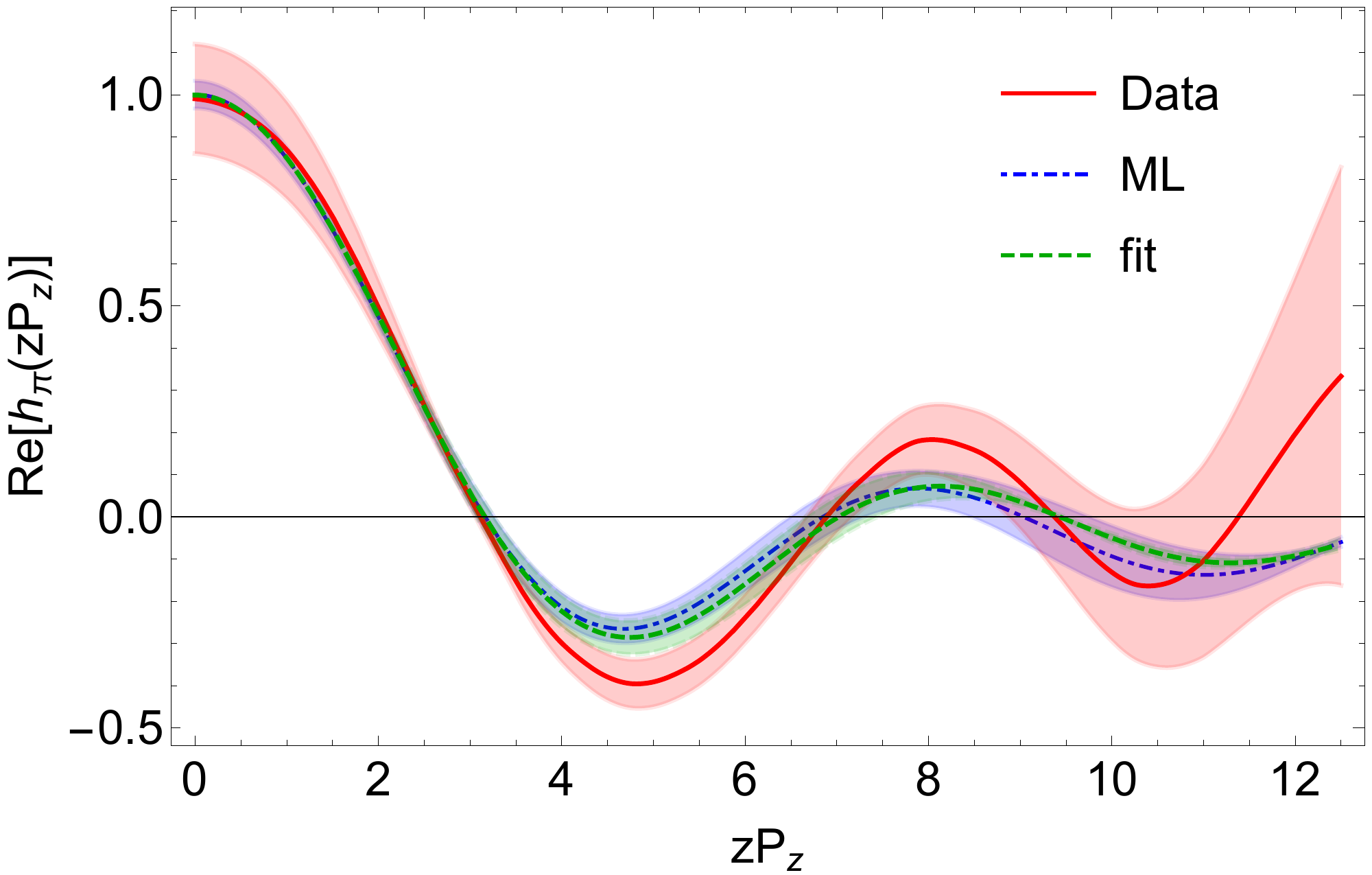}
	\includegraphics[width=0.32\linewidth]{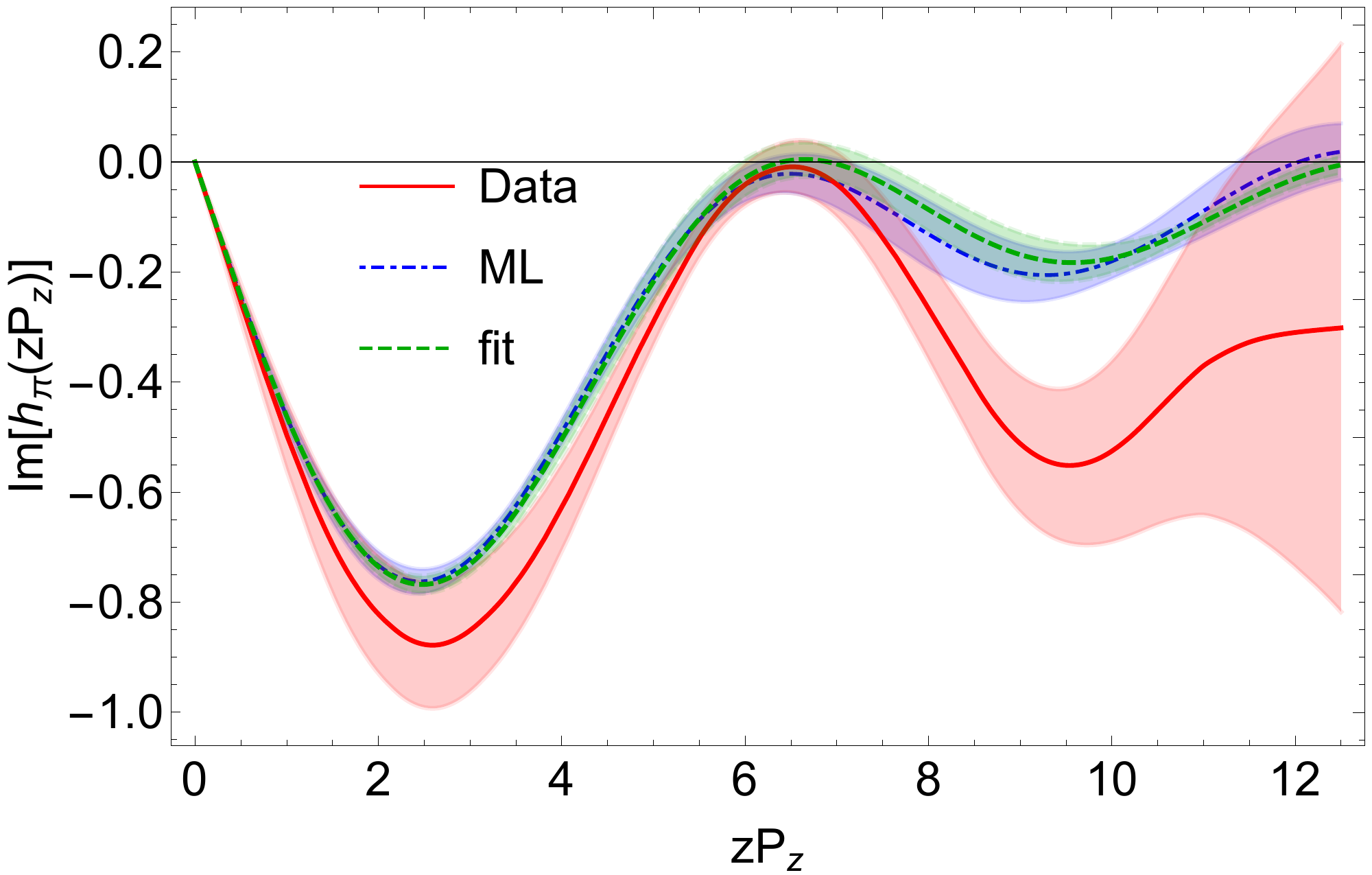}
	\includegraphics[width=0.32\linewidth]{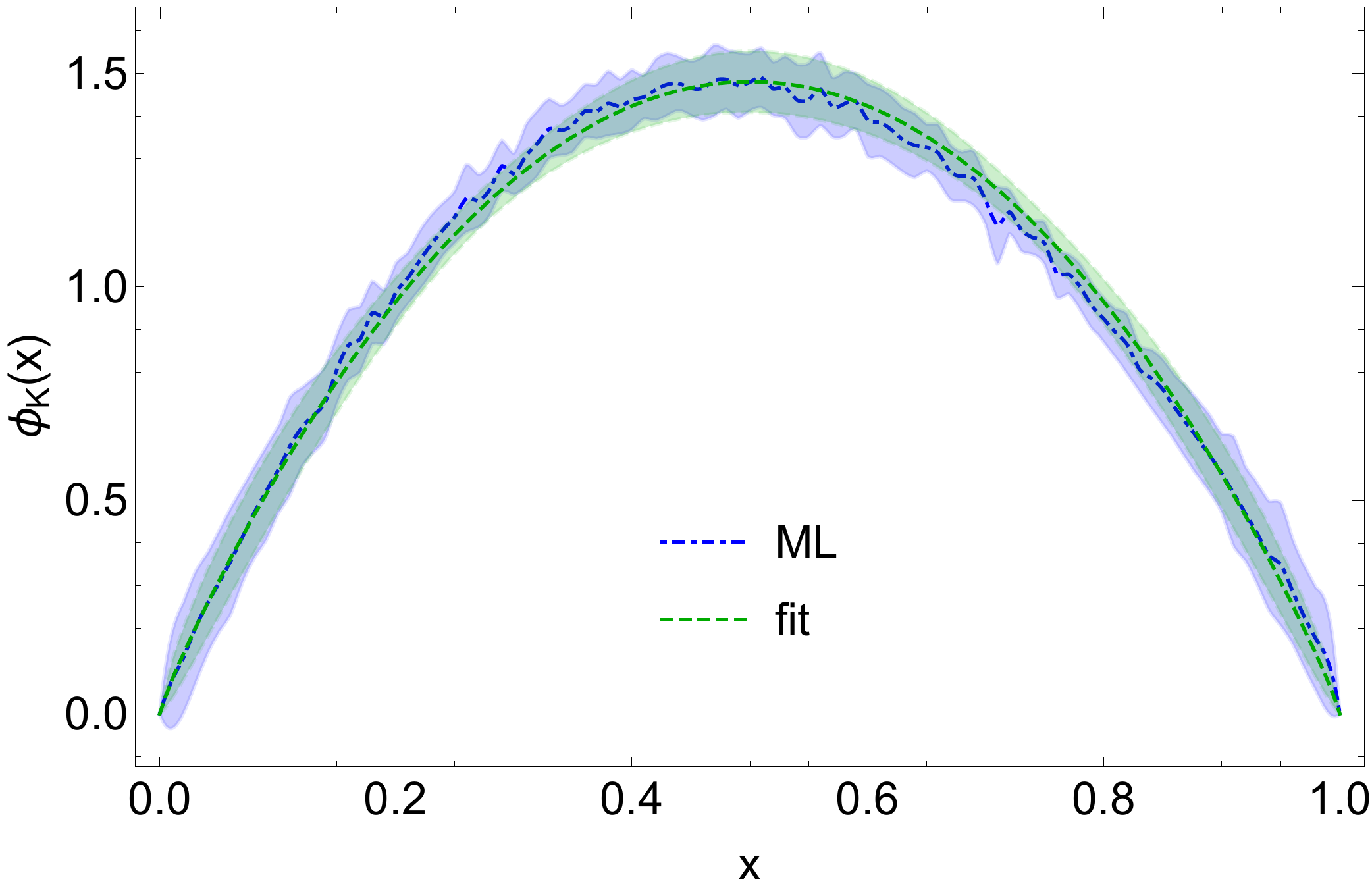}
	\includegraphics[width=0.32\linewidth]{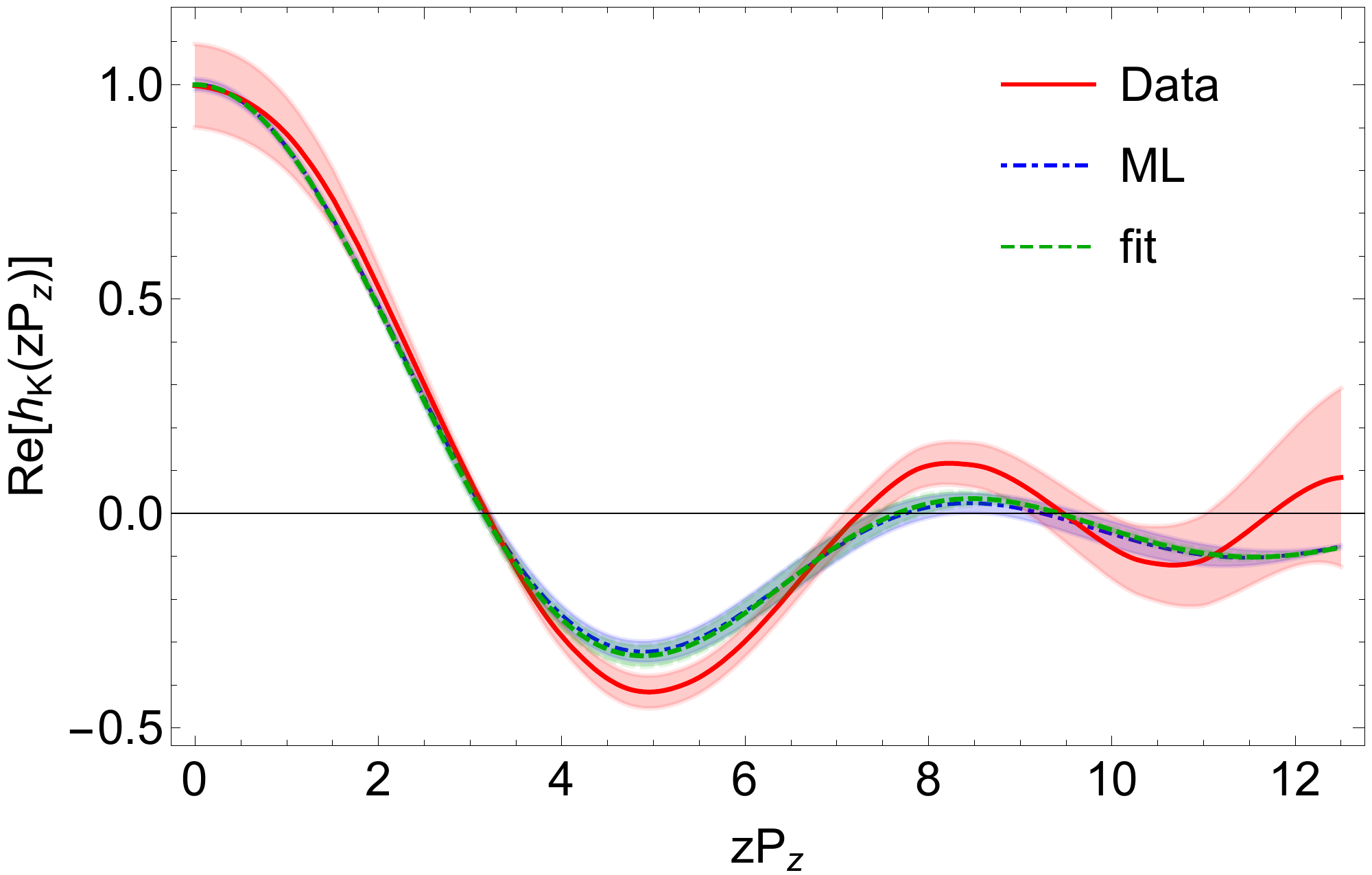}
	\includegraphics[width=0.32\linewidth]{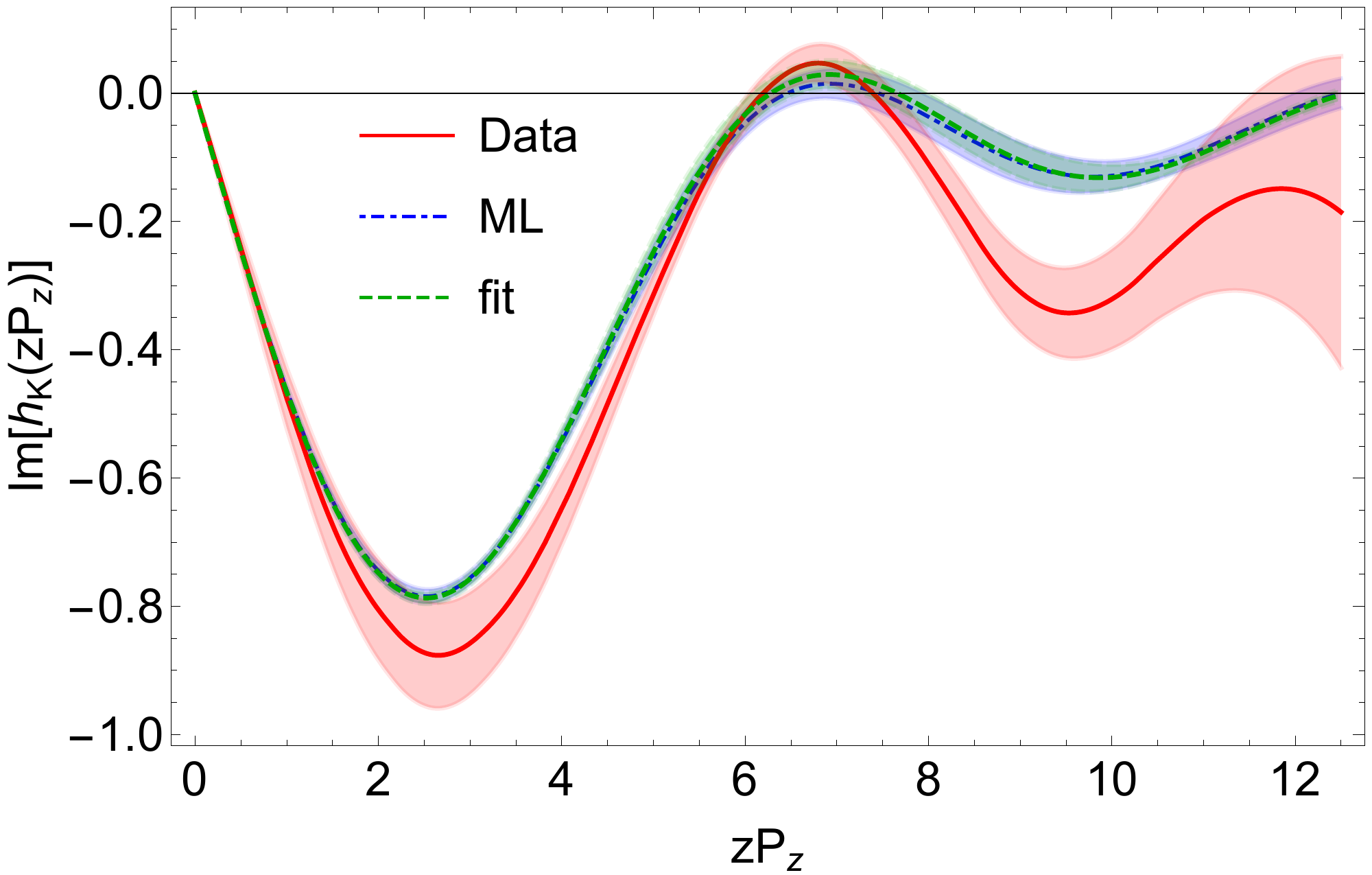}
	\includegraphics[width=0.32\linewidth]{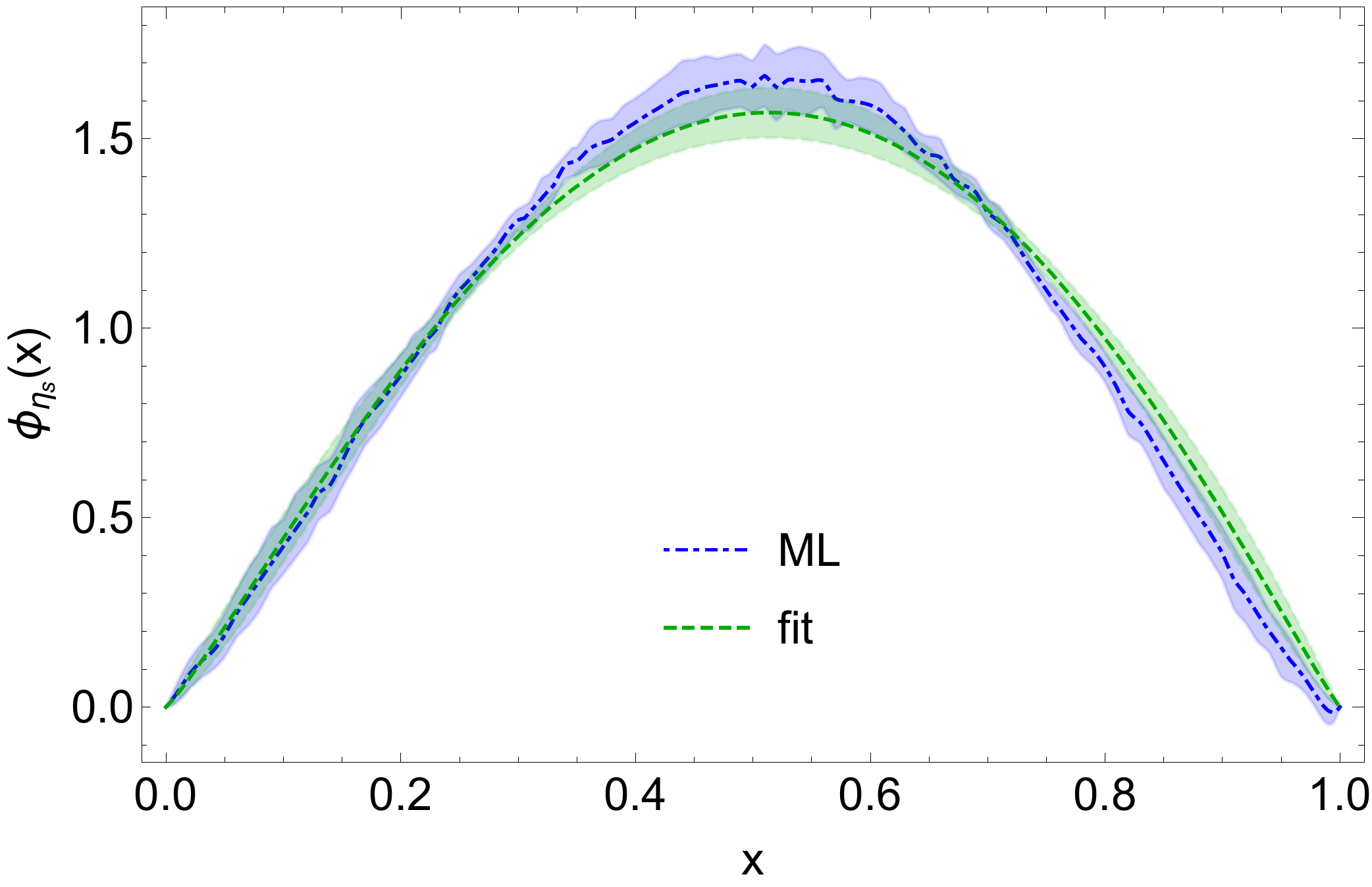}
	\includegraphics[width=0.32\linewidth]{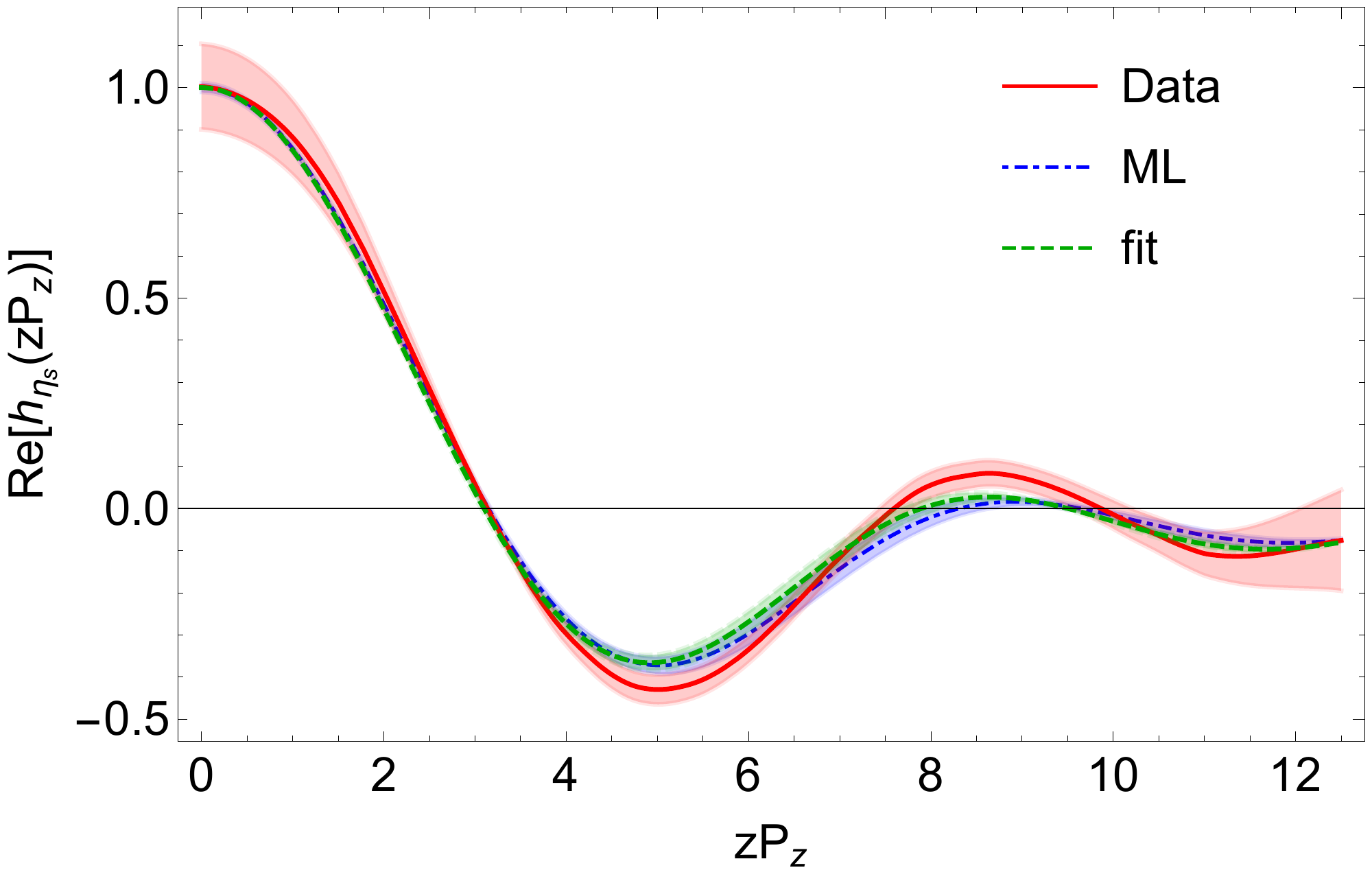}
	\includegraphics[width=0.32\linewidth]{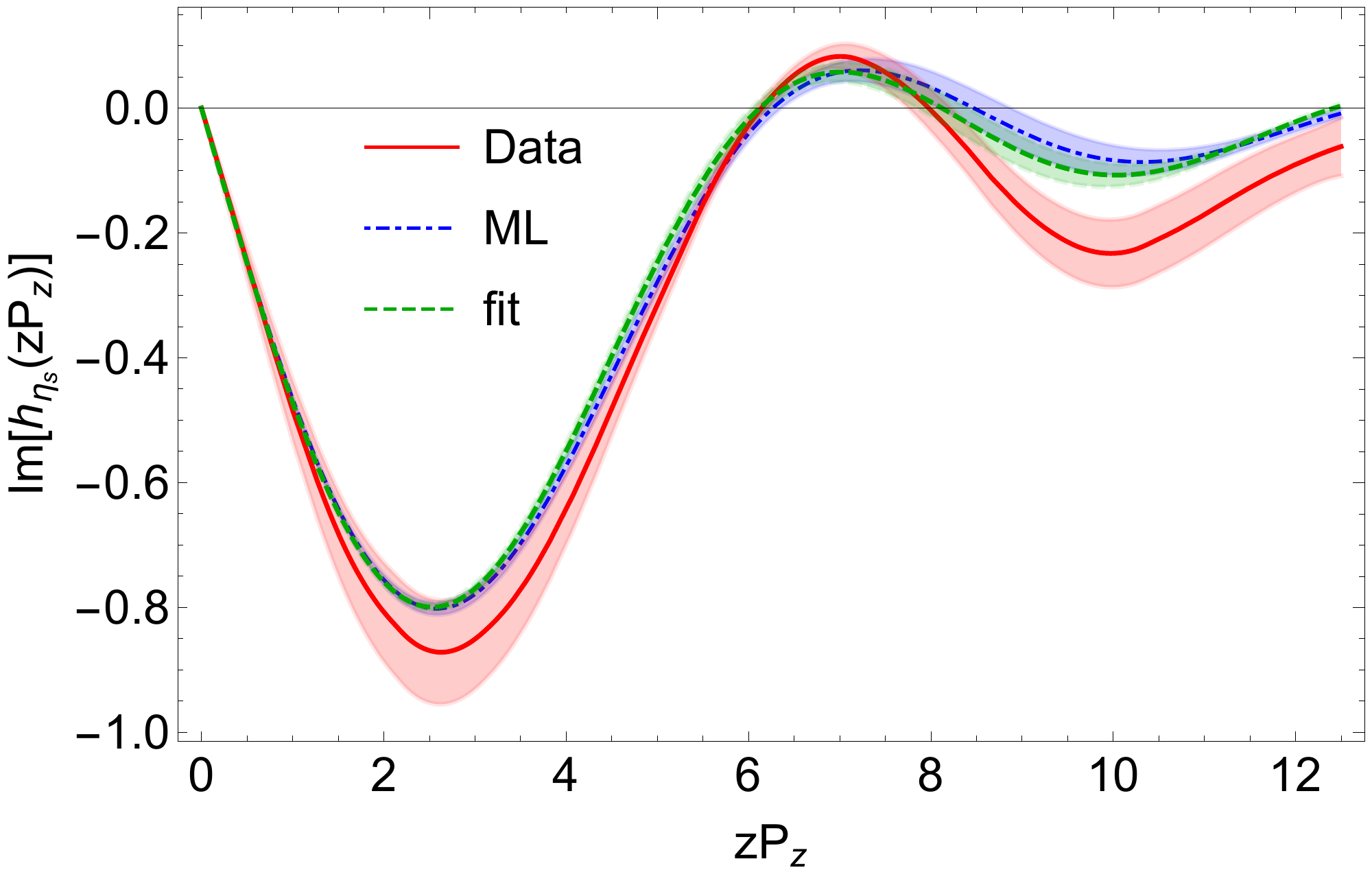}
\caption{\label{fig:ml_pred}
The machine-learning predictions on meson distribution amplitudes (leftmost column) of $\pi$ (top row), $K$ (middle row), and $\eta_s$ (bottom row) at $P_z=4$. The right two columns show the ML reconstructed of matrix elements as a function of $zP_z$ along with the input lattice data (shown in pink).}
\end{figure*}

With the success of the simple sine-function tests, we apply our procedure to the chiral-continuum extrapolated pion, kaon and $\eta_s$ lattice data.
However, simply applying our procedure to real lattice data gives a very noisy distribution.
This is mainly due to the fact that the trained network knows nothing about the physics, especially around $x=0$ and 1, which sometimes causes unstable distributions. 
To solve the problem, we divide the target lightcone DA pseudo data by a factor of $x^d(1-x)^d$ to increase the weight near the boundary, which stabilizes the prediction while keeping these points finite.
We found $d=0.3$ gives the most stable results, as shown in the left-most column of Fig.~\ref{fig:ml_pred}.
For the less noisy data sets from $\eta_s$ and $K$ mesons, the output distributions are more stable and have smaller uncertainty.
However, for the noisier pion data, the prediction becomes much worse.
This is not a surprise, as most ML training networks require high-statistics data to work well.
We also show a comparison with the fit method described in the previous subsection for both DAs and how the ML reproduces the coordinate-space matrix elements (the left two columns of Fig.~\ref{fig:ml_pred}).
Note that in this study, the machine-learning results are very close to the fit ones.
This is likely due to the fact that we set up the training data with the same form as the fitting approach.
In future work with higher-statistics data, more function forms should be included in the training process to remove the parametrization dependence.

\section{Summary and Outlook}\label{sec:summary}

In this work, we presented an updated lattice calculation of the pion, kaon and $\eta_s$  distribution amplitudes using the LaMET/quasi-distribution approach. We not only improved our previous single--lattice-spacing calculations~\cite{Chen:2017gck} with smaller statistical errors for all mesons, but also extended the calculations to two smaller lattice spacings, 0.09 and 0.06~fm. 
This allowed us to perform a continuum extrapolation using the lattice data and address issues relating to power-divergent mixing among twist-2 operators and among twist-4 operators~\cite{Rossi:2018zkn}.  
Our analysis confirmed that the coefficient of the leading $1/a^2$ power divergence is consistent with zero within errors for $P_z=1.29$ and $1.72\GeV$. 
This power divergence is not seen in our extrapolation (keeping $P_z$ constant while taking $a \to 0$), together with the absence of mixing to lower dimensional non-local operators~\cite{Chen:2017mie,Chen:2016utp}, suggests the power divergent mixing problem does not happen.

We attempted a naive chiral extrapolation to the physical pion mass $M_\pi=135$~MeV using 690-MeV and 310-MeV renormalized matrix elements. 
We used two strategies to extract the lightcone DAs. First, we fit the continuum-chiral--extrapolated matrix elements in coordinate space using Eq.~\ref{eq:match_to_me} with the distribution form used by global fit, Eq.~\ref{eq:DAform}. 
Our results in $\overline{\text{MS}}$ at 2-GeV show a pion distribution symmetric around $x=1/2$ and having broader distribution than the asymptotic prediction, consistent with prior DSE results. 
The second moment, taking the integral of our pion DA, gives 0.244(30), which is consistent with past direct lattice-QCD moment calculations. 
Our kaon DA has a narrower distribution than the pion one, but we do not observe asymmetric behavior after the continuum-chiral extrapolation.
This is likely due to the fact that our light-quark mass is not far enough away from the strange-quark mass, and thus the milder asymmetric distribution that washed out in the  increase uncertainties of continuum-chiral extrapolation.
As a result, our second moment of the kaon DA, 0.198(16), is about 20\% smaller than the previous direct calculation.
Future calculations with improved statistics and lighter quark mass will be crucial to resolve this question.

Our second strategy used a machine-learning algorithm to make predictions of the meson DAs. Our procedure has been tested with a simpler sine function that mimics the lattice data statistical distribution, modified for stable outputs.
The same setup is trained using pseudo-lattice data with Eq.~\ref{eq:DAform}, before being applied to the continuum-chiral--extrapolated lattice data to predict the meson DAs.
Further tuning is needed to obtain a stable output from the network.
We found that the ML can give stable predictions on the more precise dataset in the cases of $K$ and $\eta_s$ with the predicted result to similar the fitting one.
This is likely due to the fact that pseudo-data generated to train the model is limited to Eq.~\ref{eq:DAform} so far, but getting nonzero results is quite exciting for a first result.
Future work with even higher precision data would allow us to explore wider range of the training models, remove the model dependence, and see the impacts on the real lattice data.

\section*{Acknowledgments}
We thank the MILC Collaboration for sharing the lattices used to perform this study. The LQCD calculations were performed using the Chroma software suite~\cite{Edwards:2004sx}  with the multigrid solver algorithm~\cite{Babich:2010qb,Osborn:2010mb}. 
This research used resources of 
the National Energy Research Scientific Computing Center, a DOE Office of Science User Facility supported by the Office of Science of the U.S. Department of Energy under Contract No. DE-AC02-05CH11231 through ERCAP; 
the Extreme Science and Engineering Discovery Environment (XSEDE), which is supported by National Science Foundation grant number ACI-1548562;
facilities of the USQCD Collaboration, which are funded by the Office of Science of the U.S. Department of Energy, 
Extreme Science and Engineering Discovery Environment (XSEDE), which is supported by National Science Foundation grant number ACI-1548562;
and supported in part by Michigan State University through computational resources provided by the Institute for Cyber-Enabled Research (iCER).  
RZ, and HL are supported by the US National Science Foundation under grant PHY 1653405 ``CAREER: Constraining Parton Distribution Functions for New-Physics Searches''. The work of HL is also partly supported by the  Research  Corporation  for  Science  Advancement through the Cottrell Scholar Award.
JWC is partly supported by the Ministry of Science and Technology, Taiwan, under Grant No. 108- 2112-M-002-003-MY3 and the Kenda Foundation.

\appendix
\section{Kaon asymmetry}
\begin{figure*}
	\centering
	\includegraphics[width=0.32\linewidth]{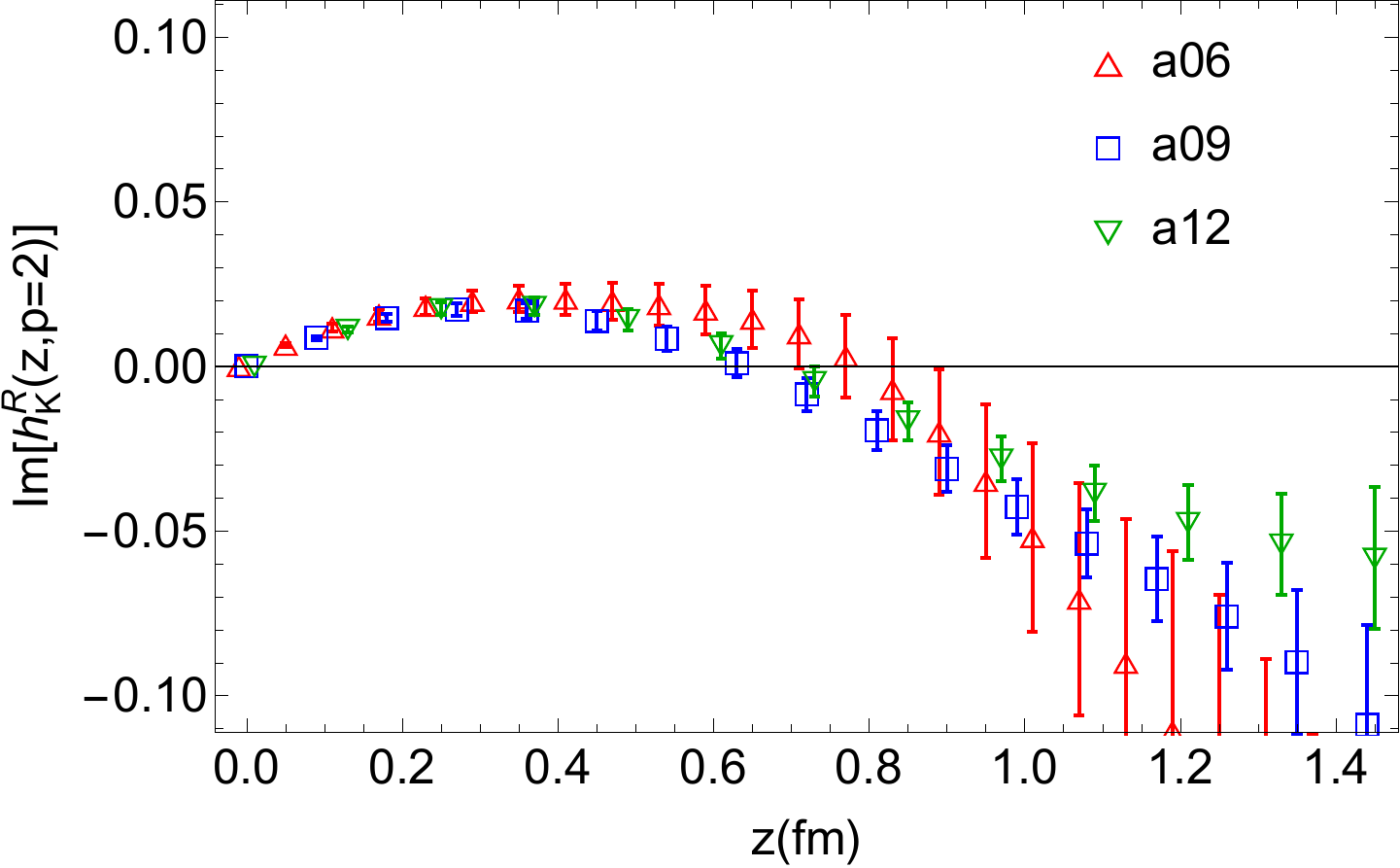}
	\includegraphics[width=0.32\linewidth]{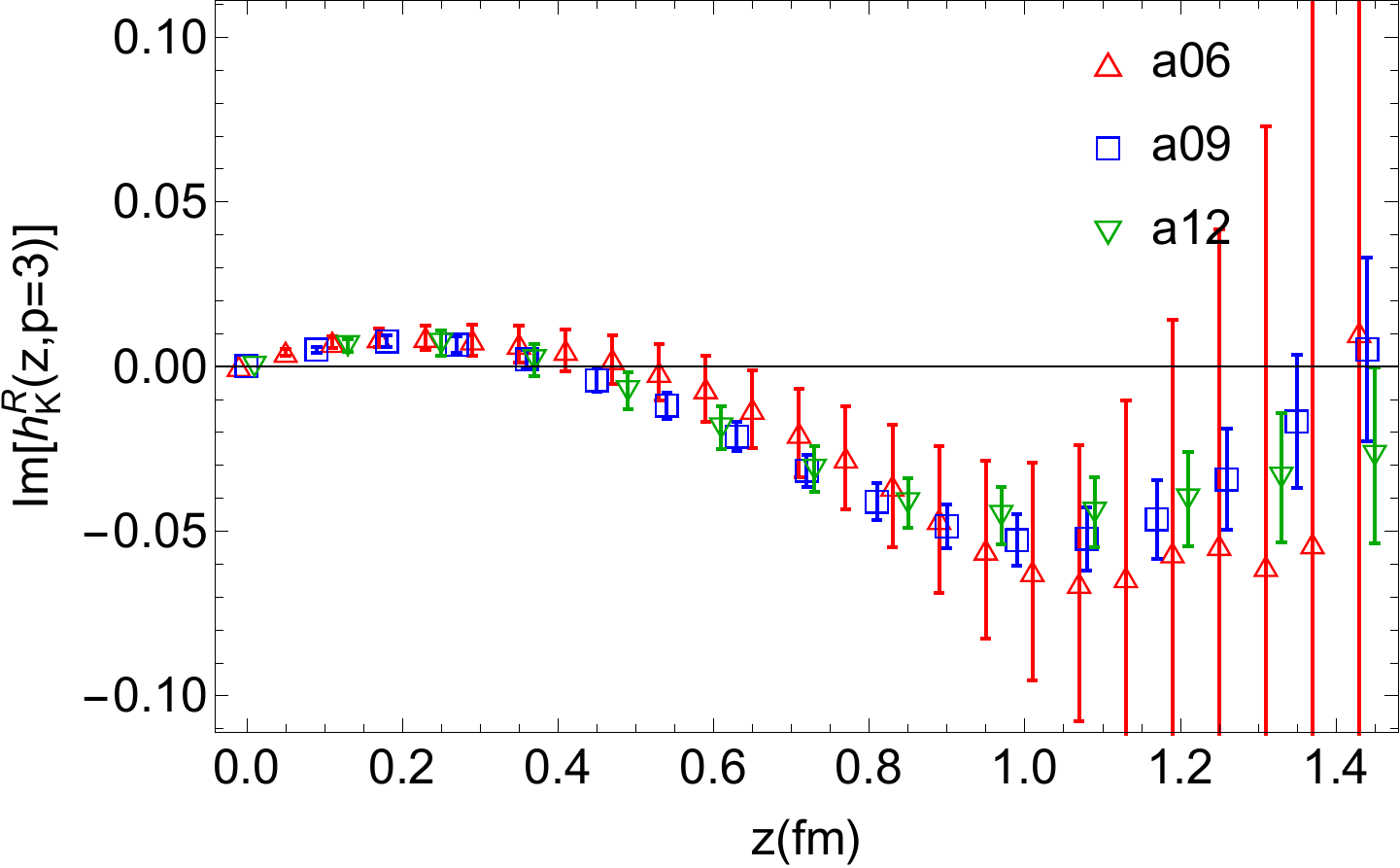}
	\includegraphics[width=0.32\linewidth]{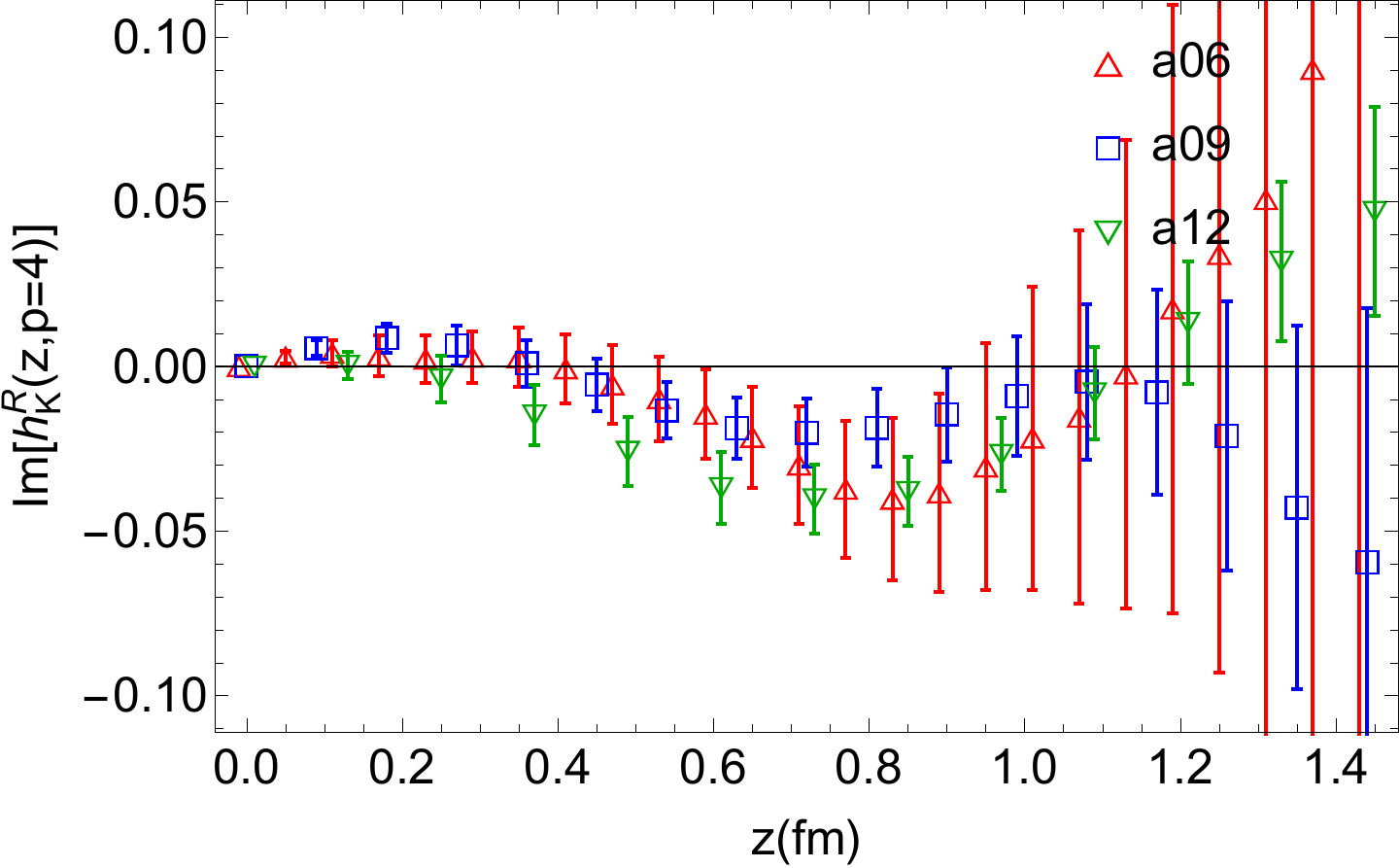}
	\caption{Imaginary part of rotated matrix elements for $K$ at $P_z=n_z \frac{2\pi}{L}$ with 
	$n_z=2/3/4$. From the $a\approx 0.12$~fm data we see that there is  an asymmetry in $K$. However, this asymmetry becomes consistent with zero when extrapolated to the continuum.
	}
	\label{fig:ME_rot}
\end{figure*}

We note that the kaon DA we obtain in this approach is symmetric around $x=\frac{1}{2}$, inconsistent with the kaon asymmetry found in the previous work~\cite{Chen:2017gck}. We note that the matching kernel preserves the symmetry in quasi-DA. Because of the unsolved issues in the FT and matching procedure, we check the asymmetry directly in the coordinate space of quasi-DA. As described in Ref.~\cite{Chen:2017gck}, the asymmetry comes from the nonzero imaginary part after a phase rotation of the quasi-DA matrix elements,
\begin{equation}
    \tilde{H}^R(zP_z,p_z^R,\mu^R) = e^{-izP_z/2} h^R(zP_z,p_z^R,\mu^R)
\end{equation}
then the FT formula Eq.~\eqref{eq:ft} will become
\begin{equation}
\label{eq:ft_tilde}
    \tilde{\phi}(x,\mu^R,p_z^R,P_z) = \int dz\, e^{-i(1/2-x)zP_z} \tilde{H}^R(zP_z,p_z^R,\mu^R).
\end{equation}
We can see from Eq.~\eqref{eq:ft_tilde} that if $\tilde{H}^R$ is real, $\tilde{\phi}(x)=\tilde{\phi}(1-x)$ will hold.
The phase-rotated matrix elements for $K$ at $n_z=2/3/4$ are shown in Fig.~\ref{fig:ME_rot}. From the data on $a\approx 0.12$~fm lattice, we see a clear nonzero imaginary part for the kaon. Yet, when we extrapolate to the continuum, the imaginary part of $n_z=4$ becomes consistent with zero. Thus our kaon result in continuum at $n_z=4$ is close to a symmetric distribution.

\section{Additional Figures}
\label{Appendix B}
\begin{figure*}
	\centering
    \includegraphics[width=0.32\linewidth]{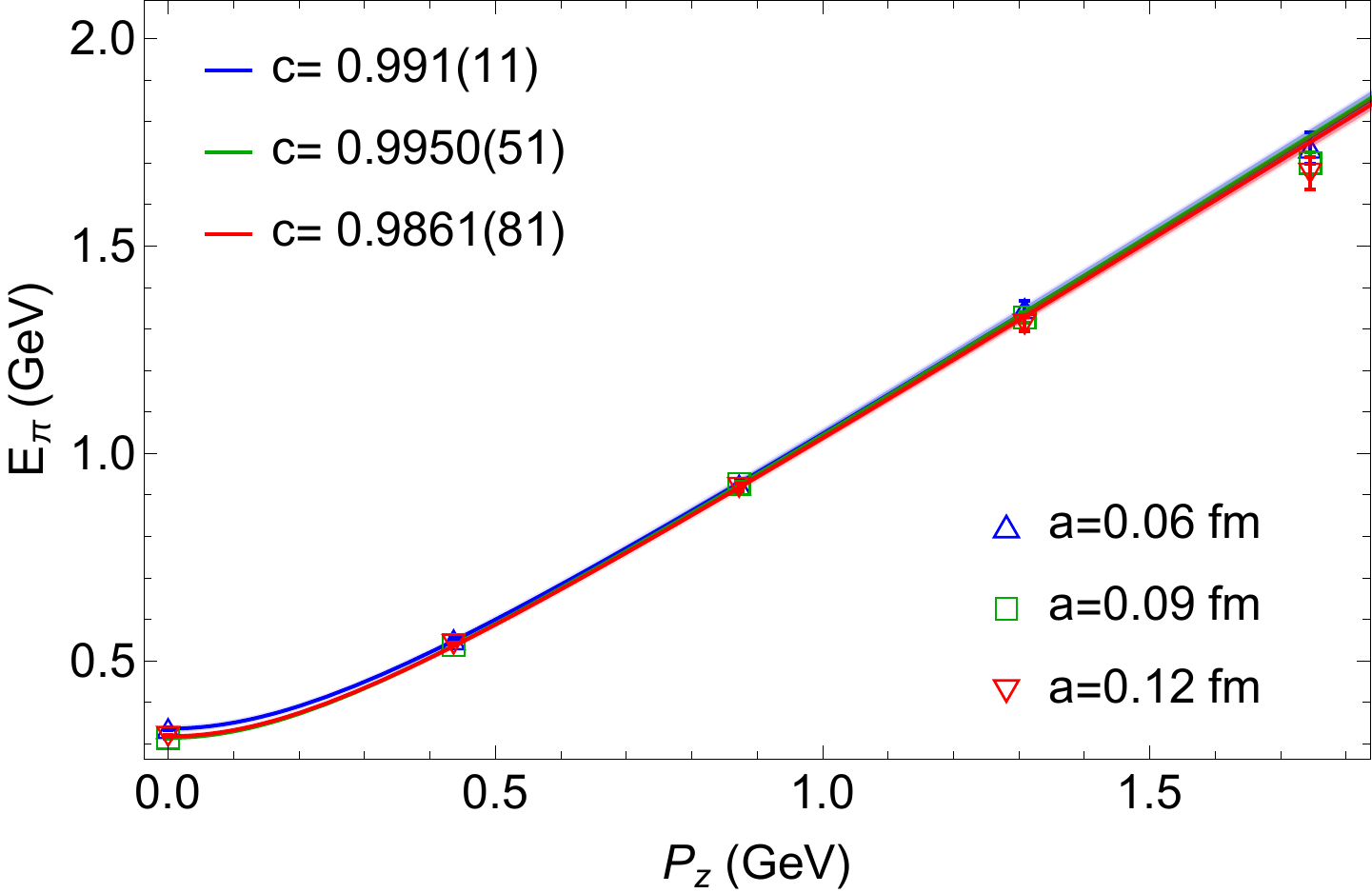}
    \includegraphics[width=0.32\linewidth]{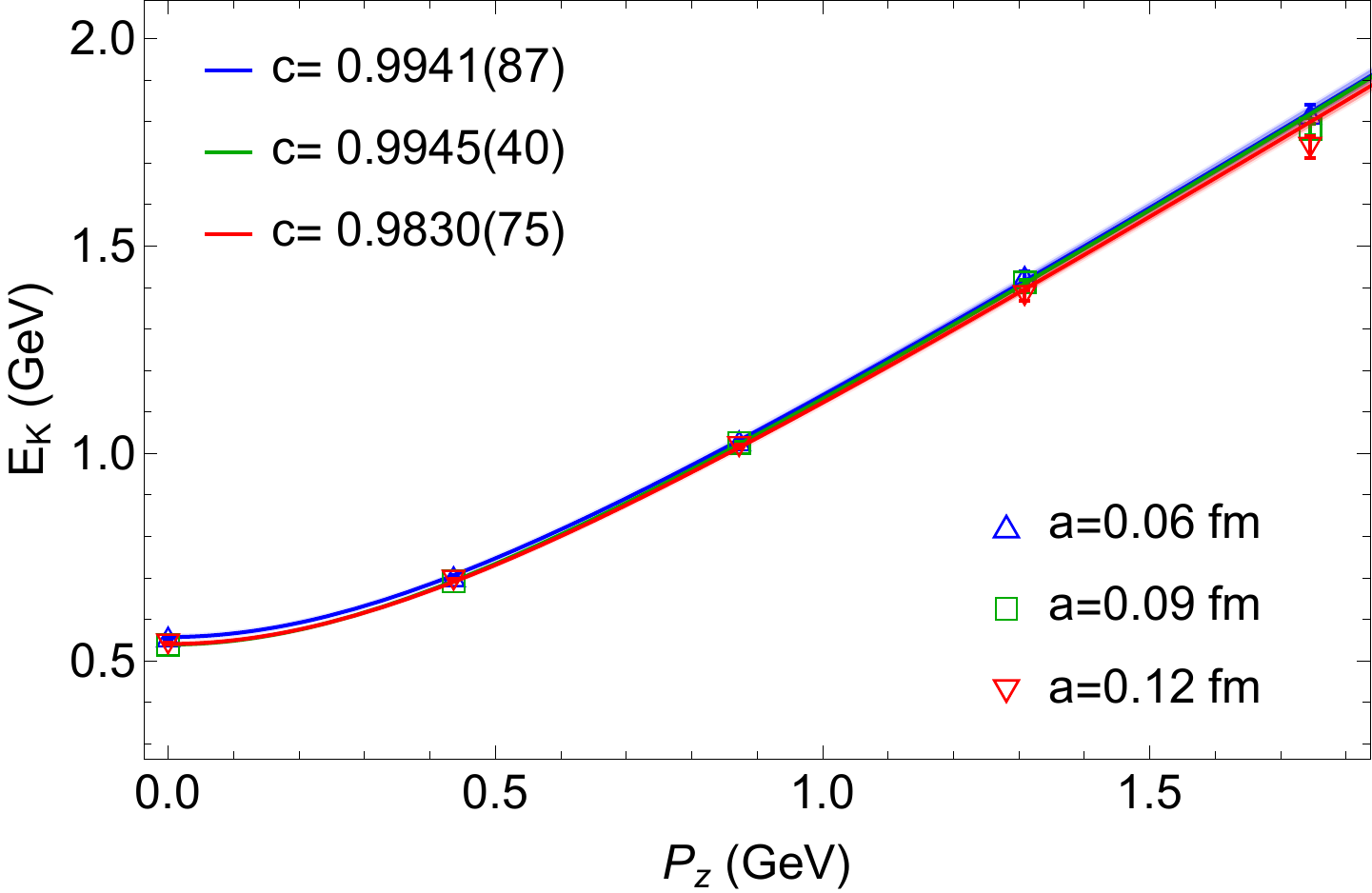}
    \includegraphics[width=0.32\linewidth]{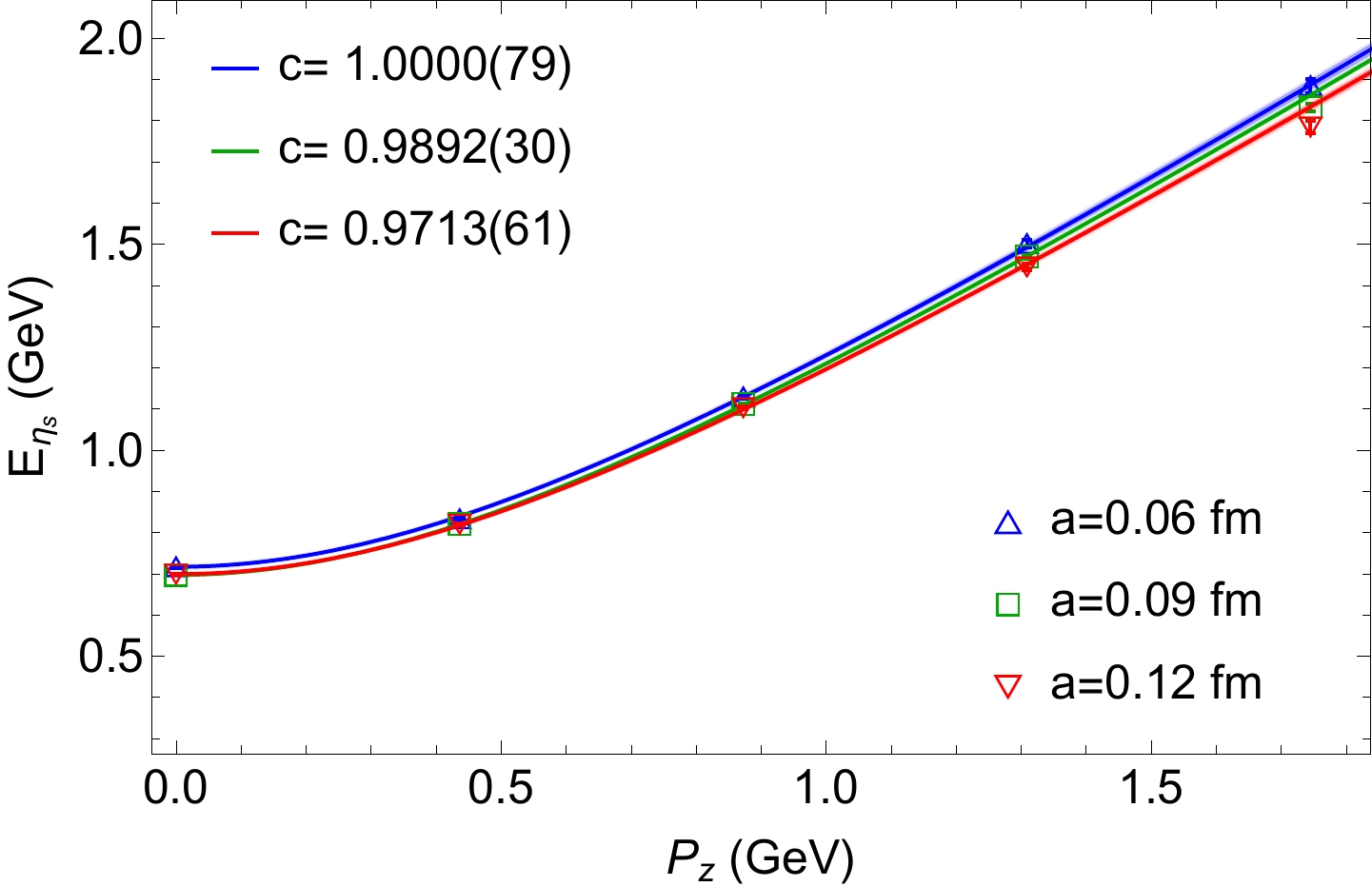}
	\caption{The $\pi$ (left), $K$ (middle) and $\eta_s$ (right) dispersion relations of the meson energy from the two-state fits for a12m310, a09m310, a06m310 ensembles, respectively. The speed of light gets closer to one at finer lattices.}
	\label{fig:DR}
\end{figure*}

The dispersion relation for three particles on three lattices are in Fig.~\ref{fig:DR}. We can see that the speed of light gets closer to one at finer lattice. On coarser lattices, heavier mesons show a larger deviation.

\begin{figure*}
	\centering
	\includegraphics[width=0.32\linewidth]{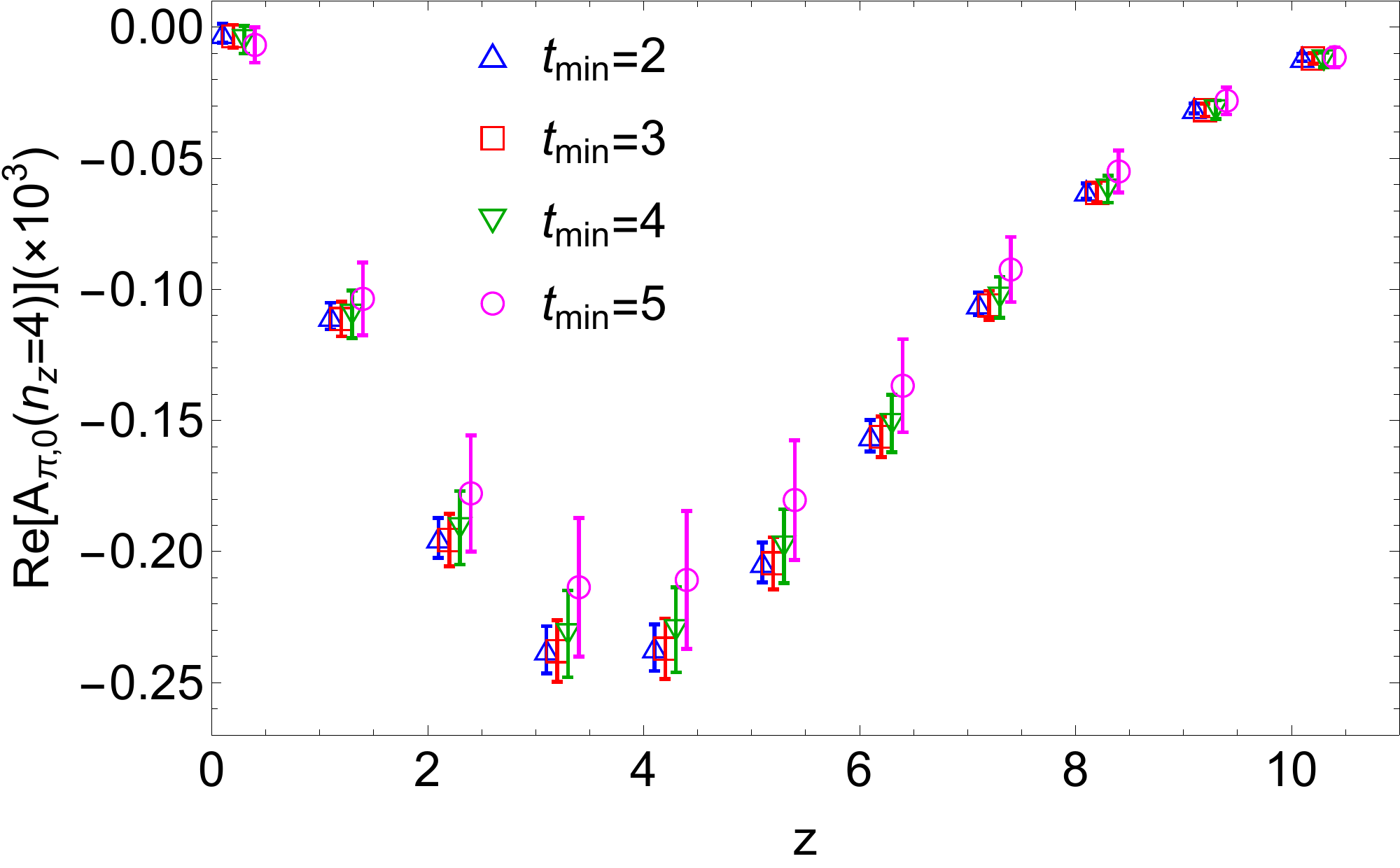}
	\includegraphics[width=0.32\linewidth]{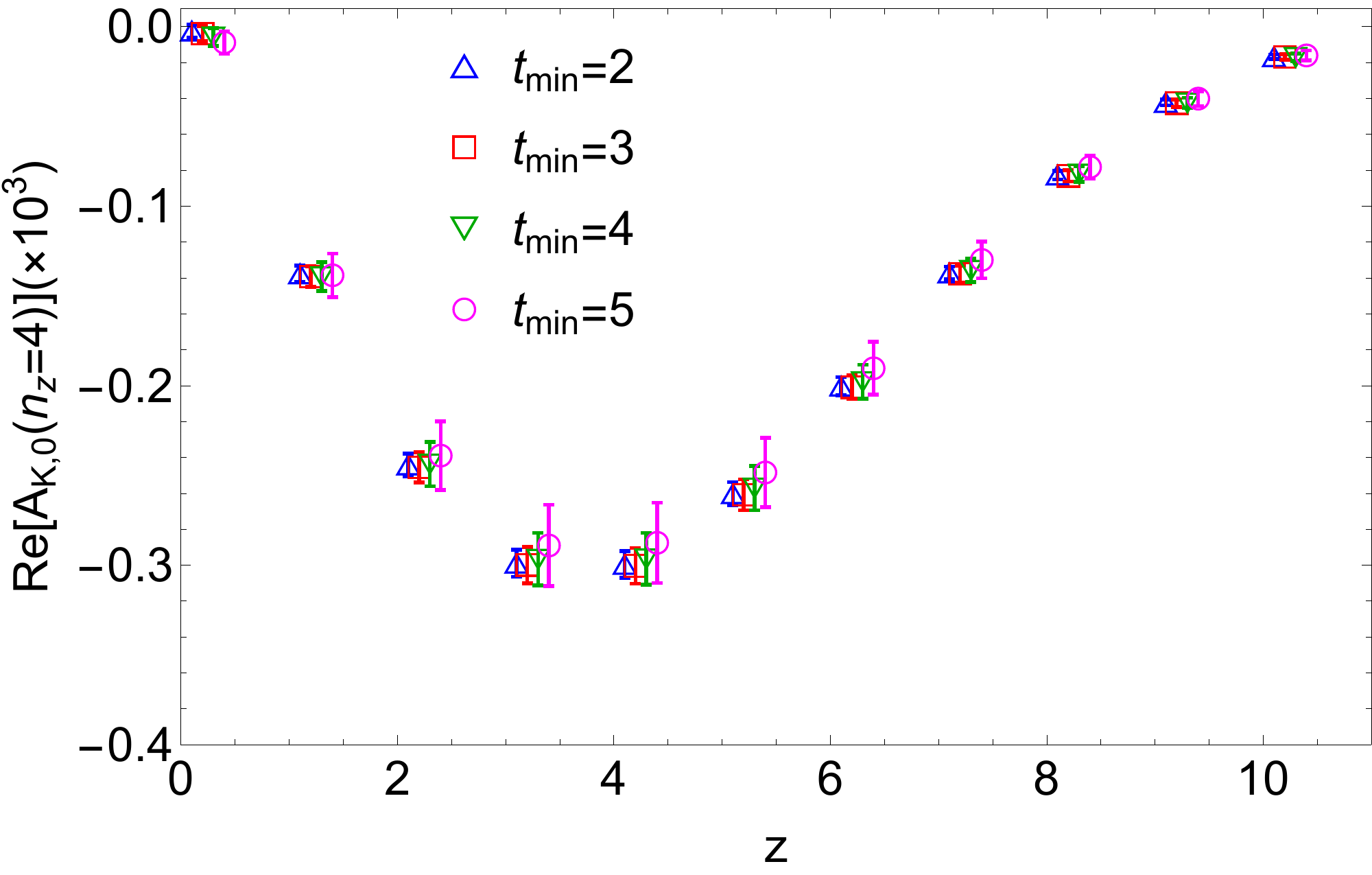}
	\includegraphics[width=0.32\linewidth]{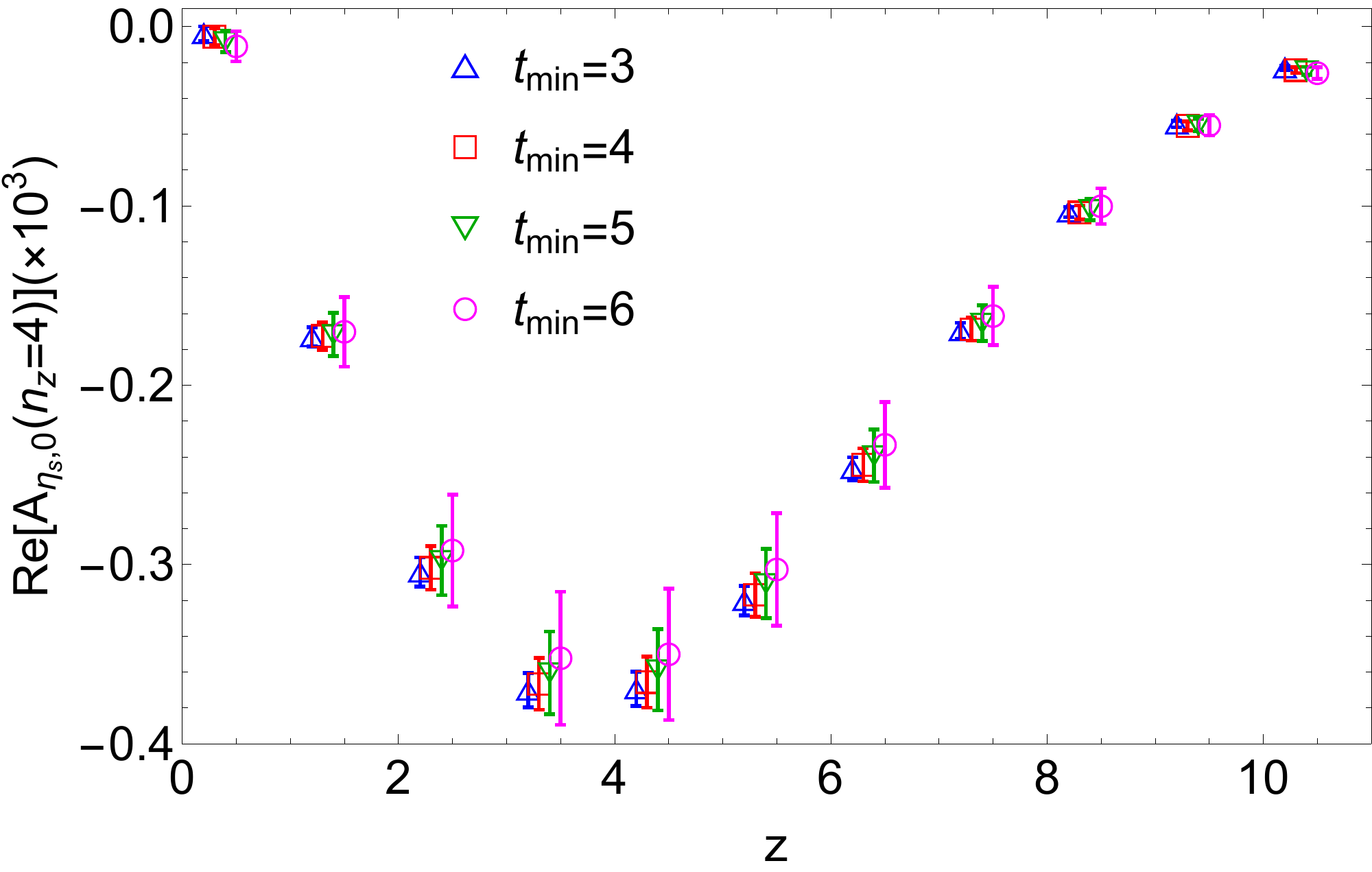}	
	\includegraphics[width=0.32\linewidth]{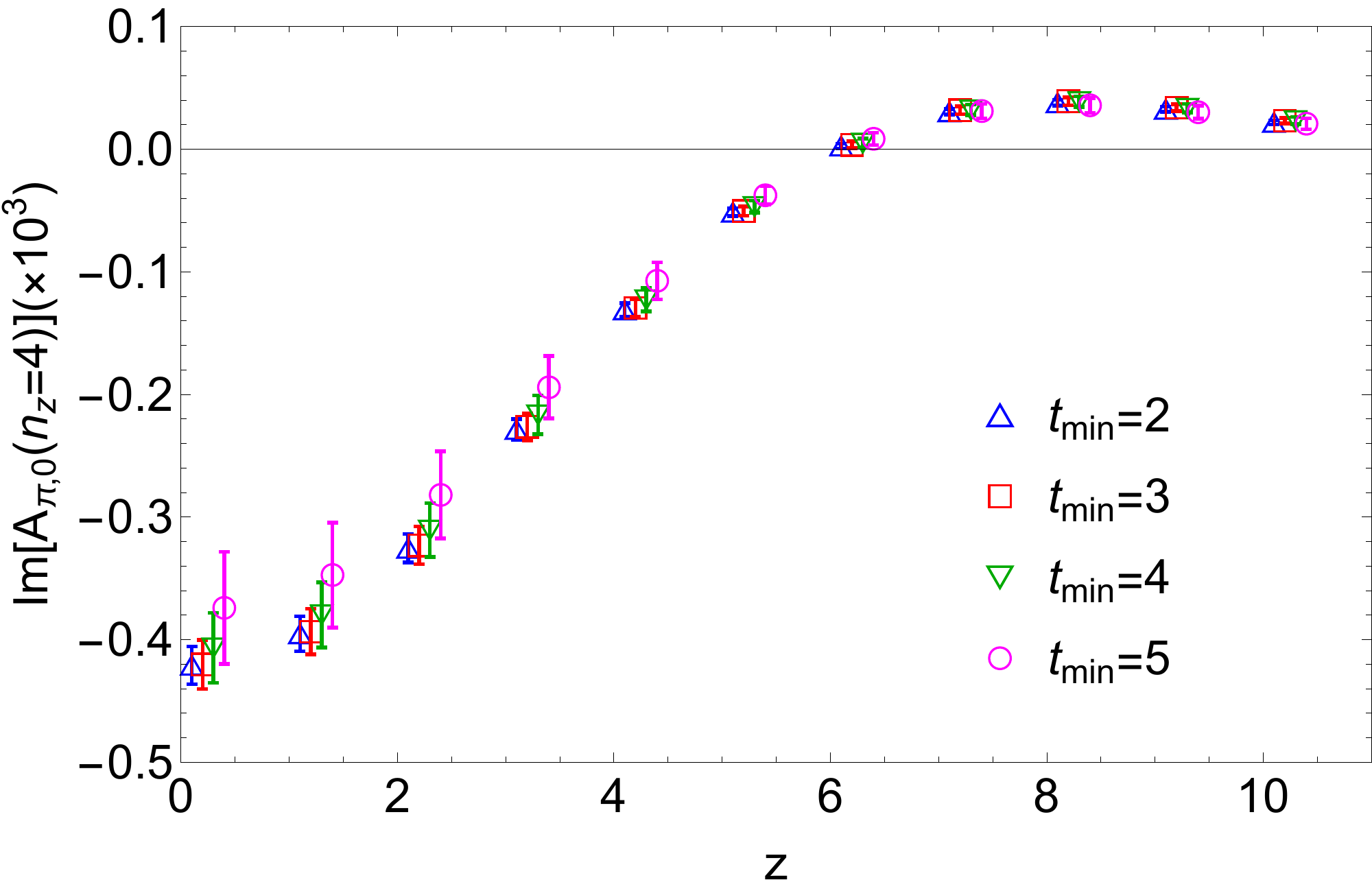}
	\includegraphics[width=0.32\linewidth]{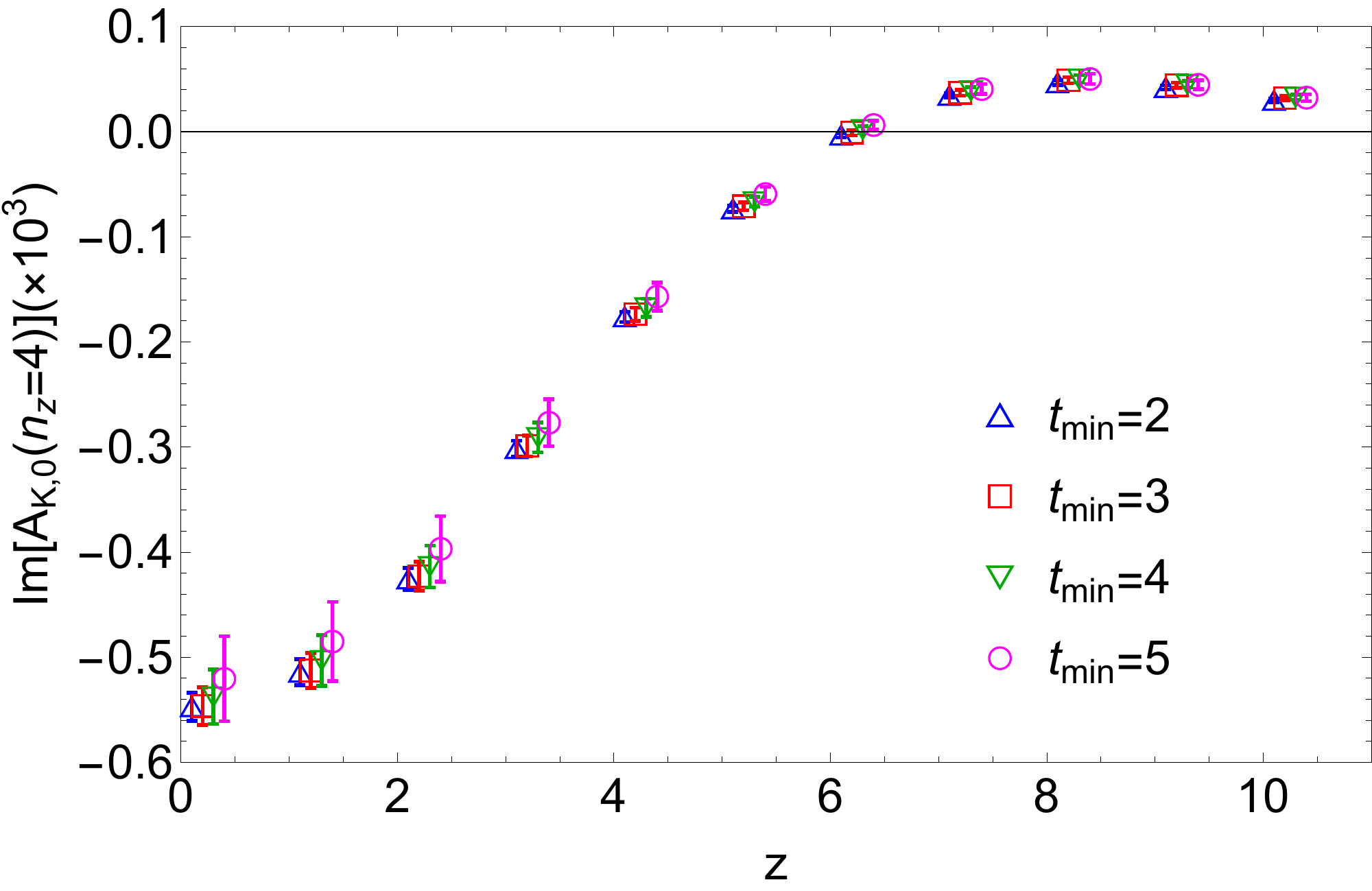}
	\includegraphics[width=0.32\linewidth]{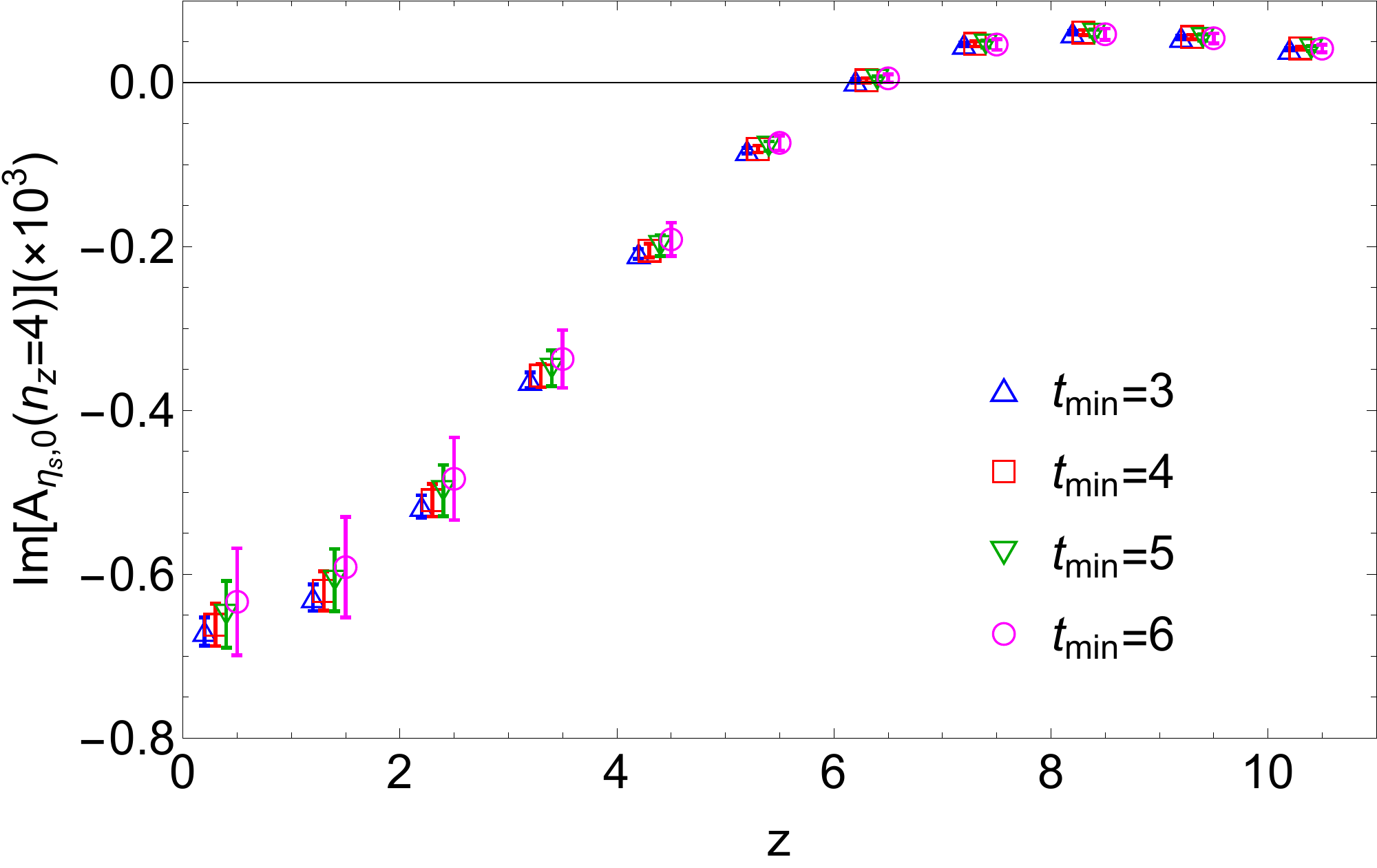}
	\caption{The real (top row) and imaginary (bottom row) ground-state amplitude $A_{M,0}$ as a function of $z$ at $P_z=4\frac{2\pi}{L}$ from two-state fits with different fit ranges $[t_{\text{min}}, 13]$ for $\pi$ (left column), $K$ (middle column) and $\eta_s$ (right column) on the a06m310 ensemble. The ground-state amplitude extracted from different $t_{\text{min}}$ are consistent with each other within error, while larger $t_{\text{min}}$ results in larger uncertainties. For $\pi$, $K$ and $\eta_s$, $t_{\text{min}}=\{4, 4, 5\}$ are used in the final analysis.}
	\label{fig:A0_z_a06}
\end{figure*}

\begin{figure*}
	\centering
	\includegraphics[width=0.32\linewidth]{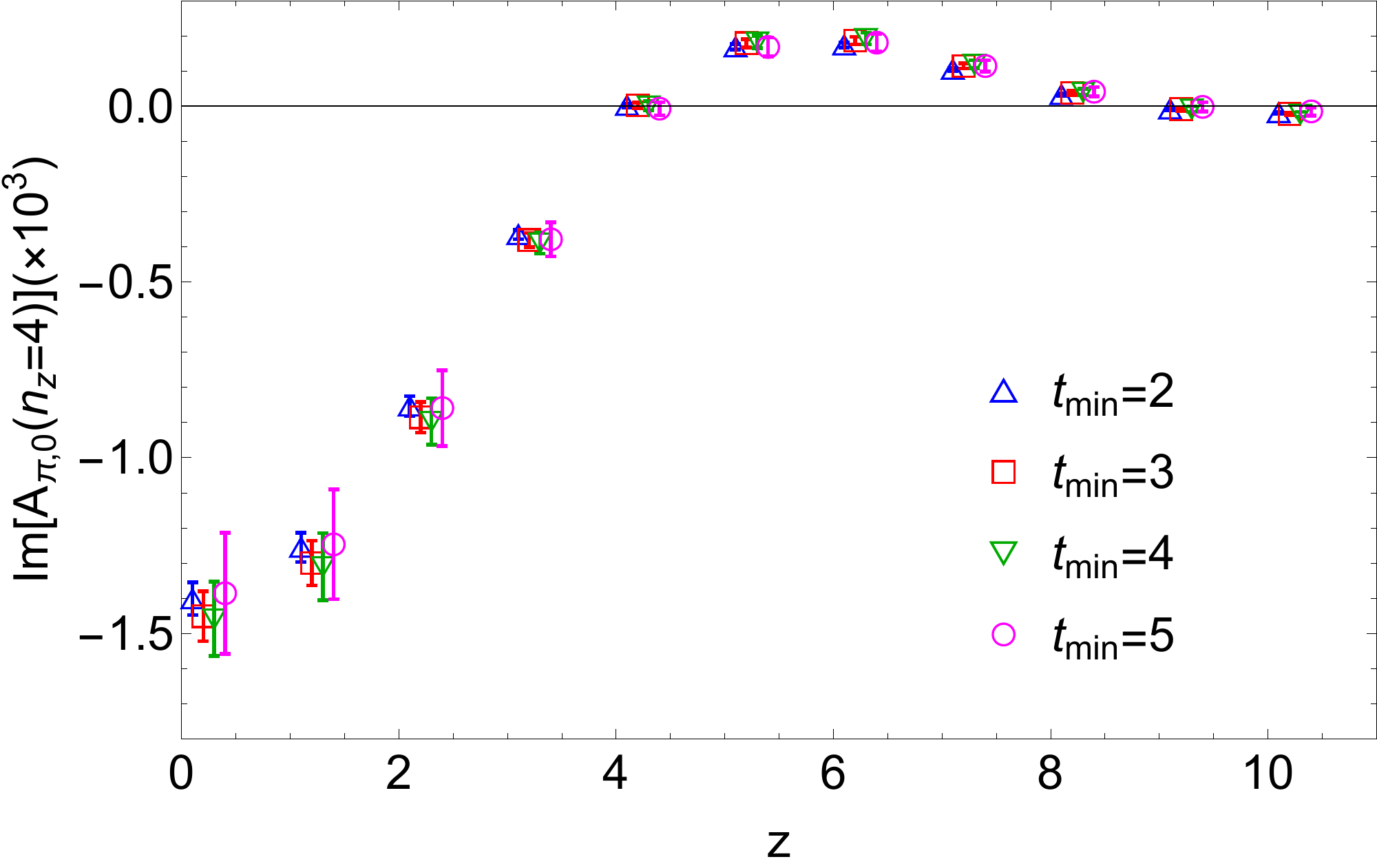}
	\includegraphics[width=0.32\linewidth]{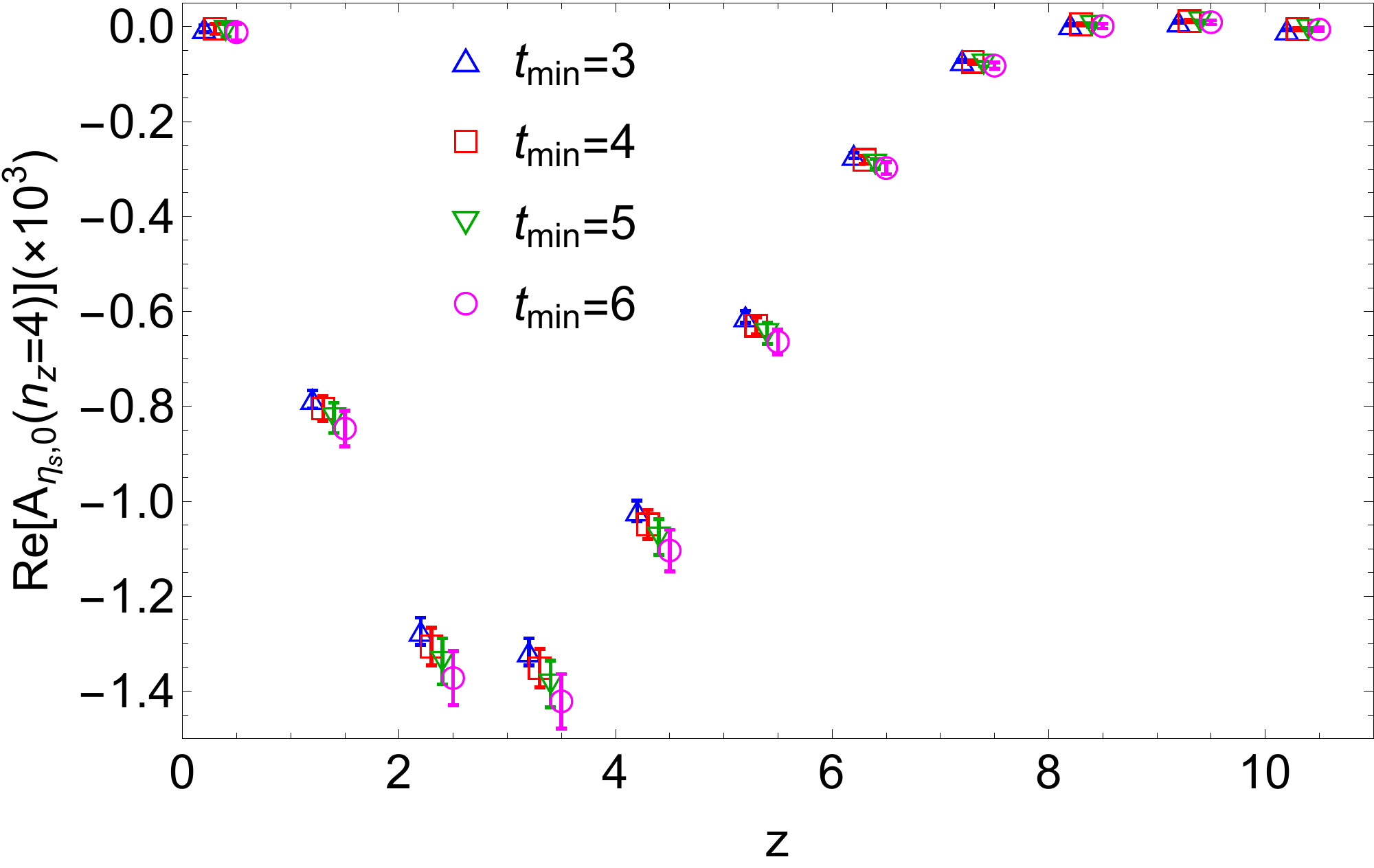}
	\includegraphics[width=0.32\linewidth]{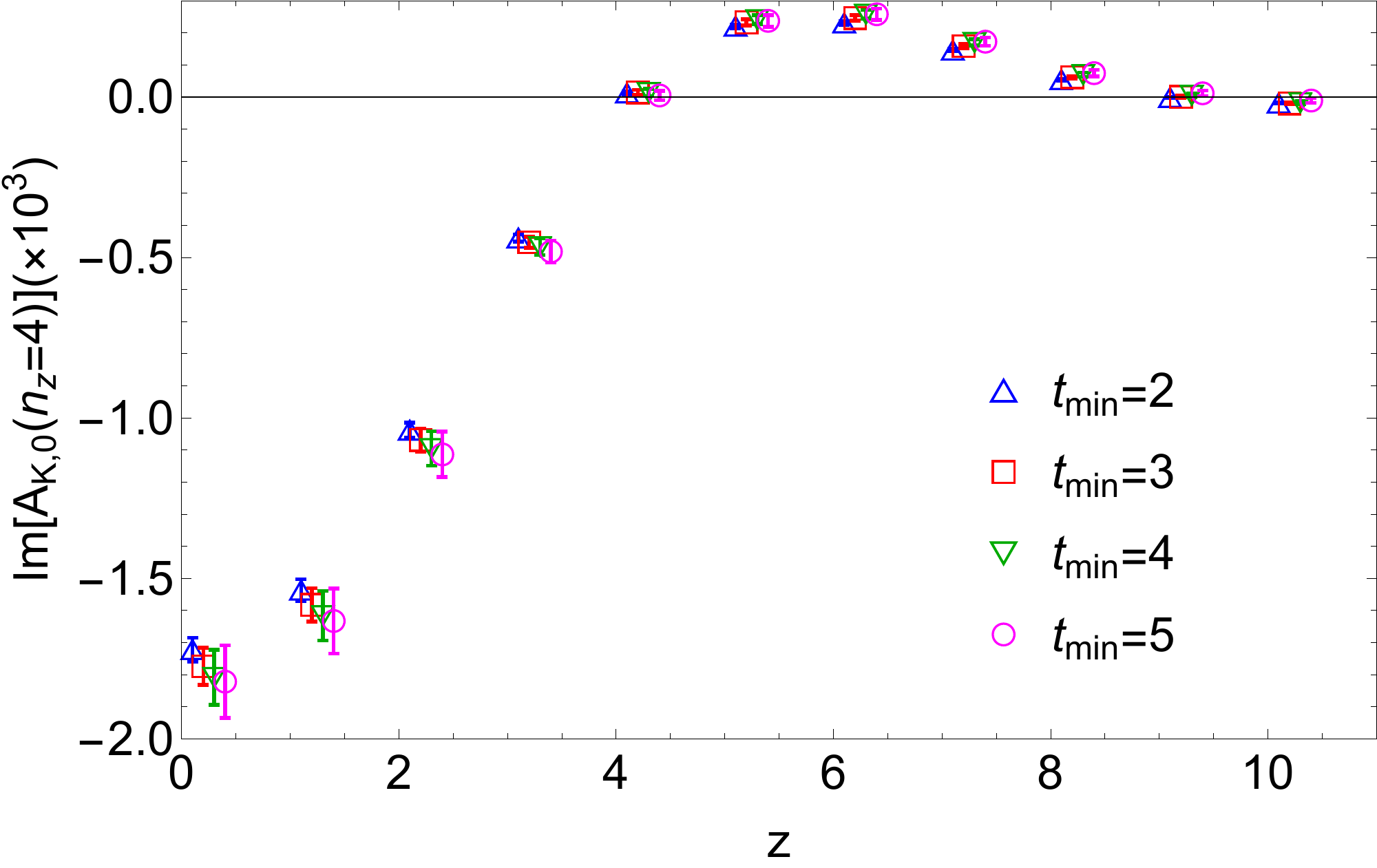}
	\includegraphics[width=0.32\linewidth]{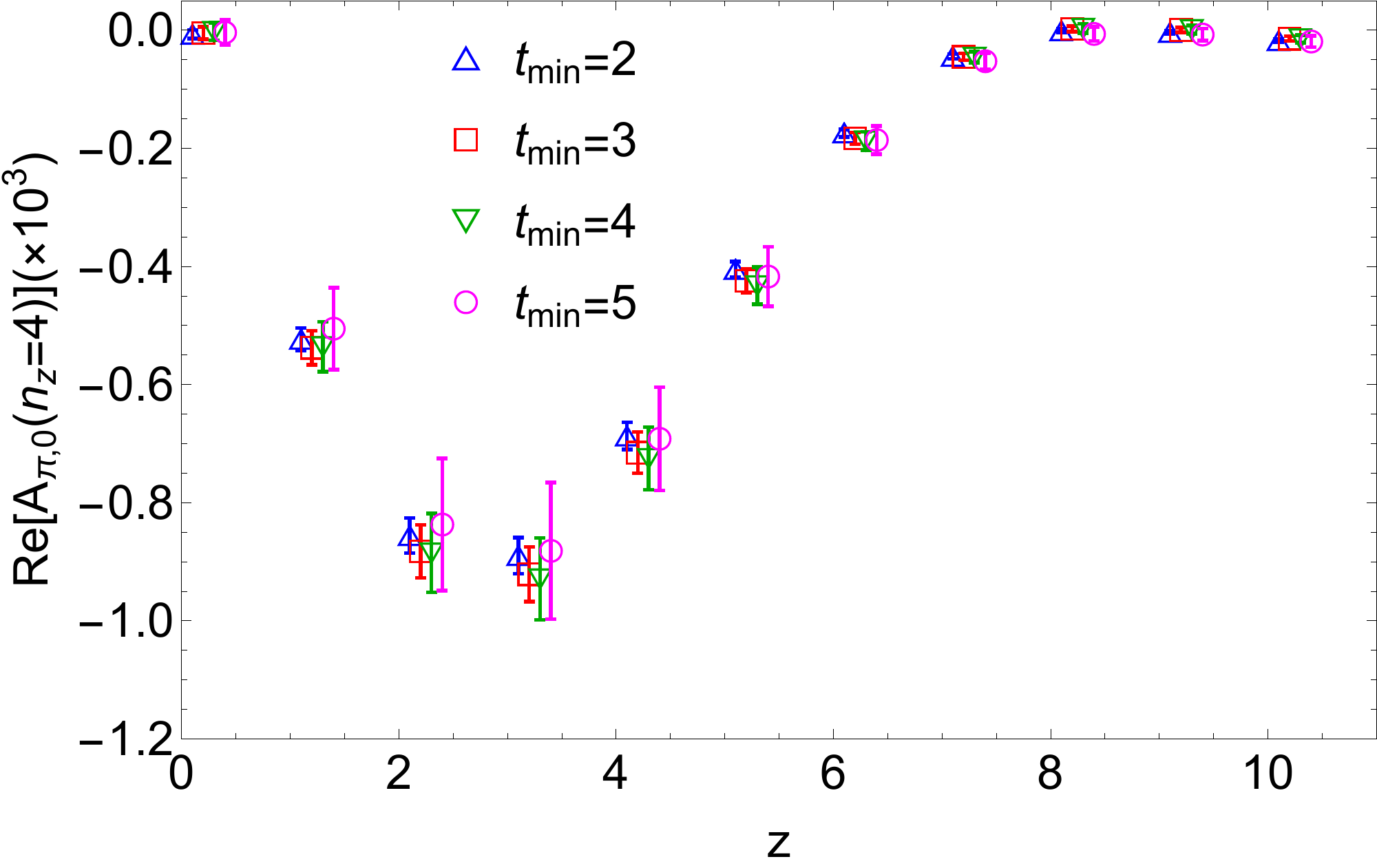}
	\includegraphics[width=0.32\linewidth]{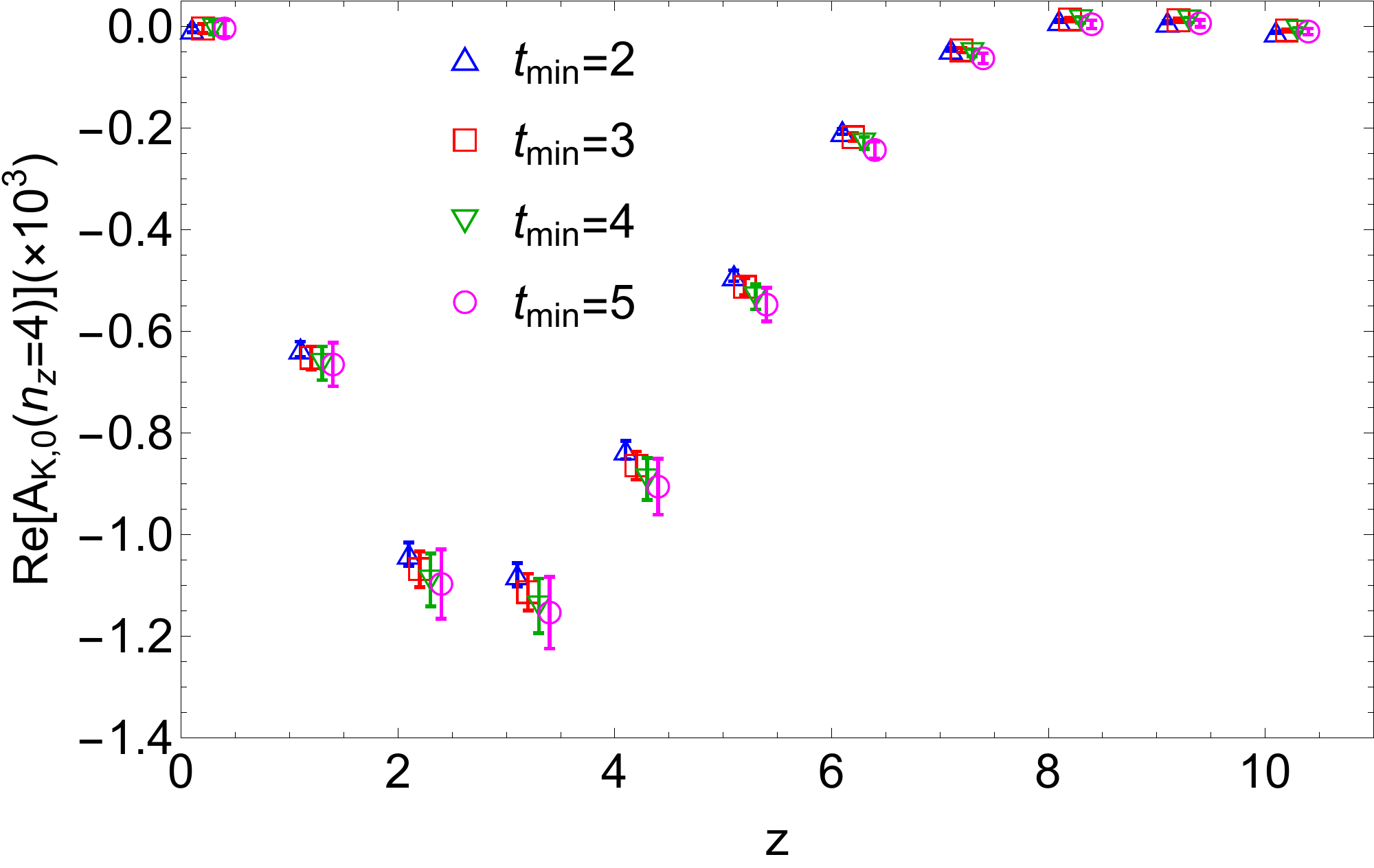}
	\includegraphics[width=0.32\linewidth]{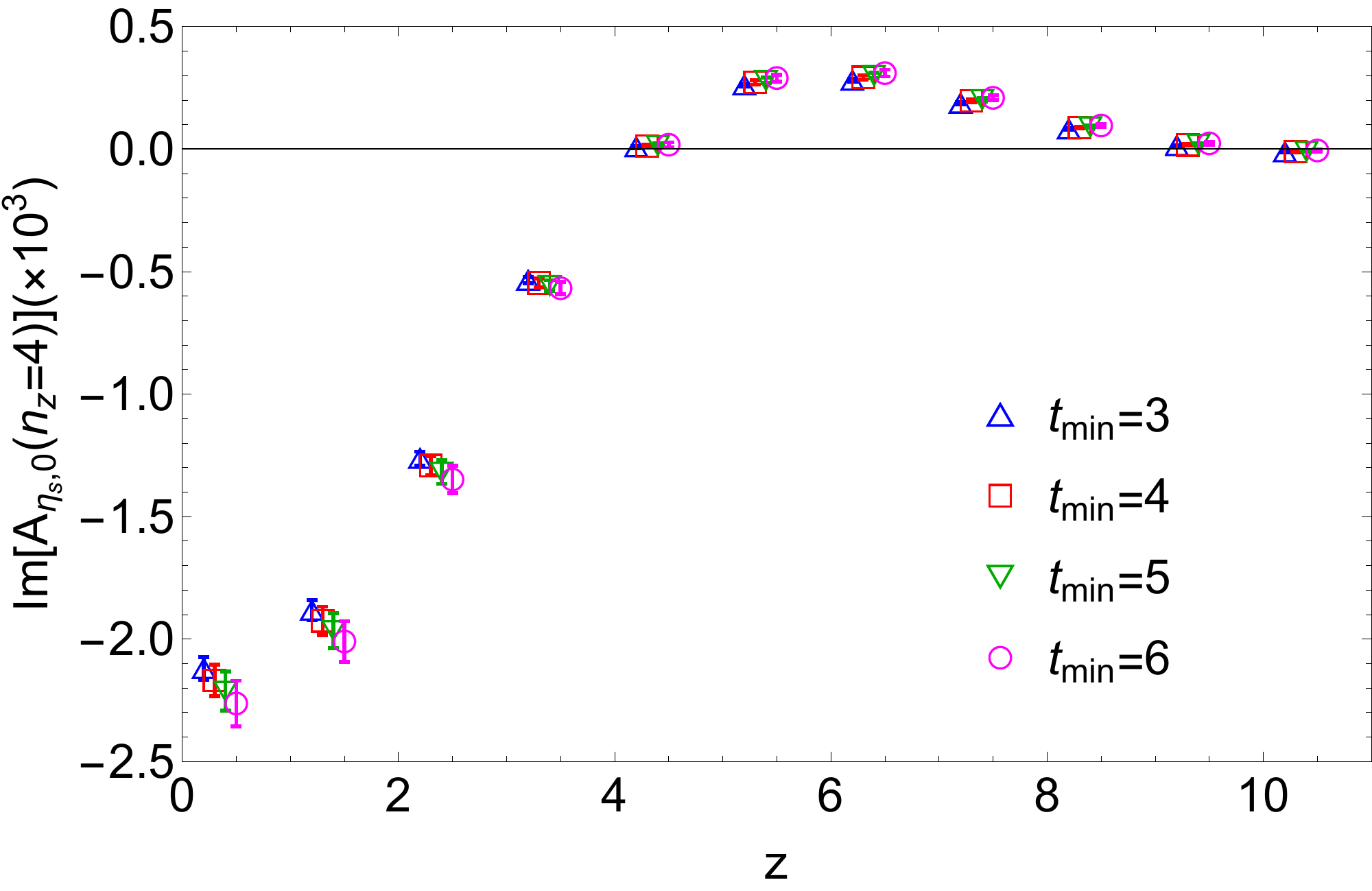}
	\caption{The real (top row) and imaginary (bottom row) ground-state amplitude $A_{M,0}$ as a function of $z$ at $P_z=4\frac{2\pi}{L}$ from two-state fits with different fit ranges $[t_{\text{min}}, 13]$ for $\pi$ (left column), $K$ (middle column) and $\eta_s$ (right column) on the a09m310 ensemble. The ground-state amplitude extracted from different $t_{\text{min}}$ are consistent with each other within error, while larger $t_{\text{min}}$ results in larger uncertainties. For $\pi$, $K$ and $\eta_s$, $t_{\text{min}}=\{5, 4, 5\}$ are used in the final analysis.}
	\label{fig:A0_z_a09}
\end{figure*}

\begin{figure*}
	\centering
	\includegraphics[width=0.32\linewidth]{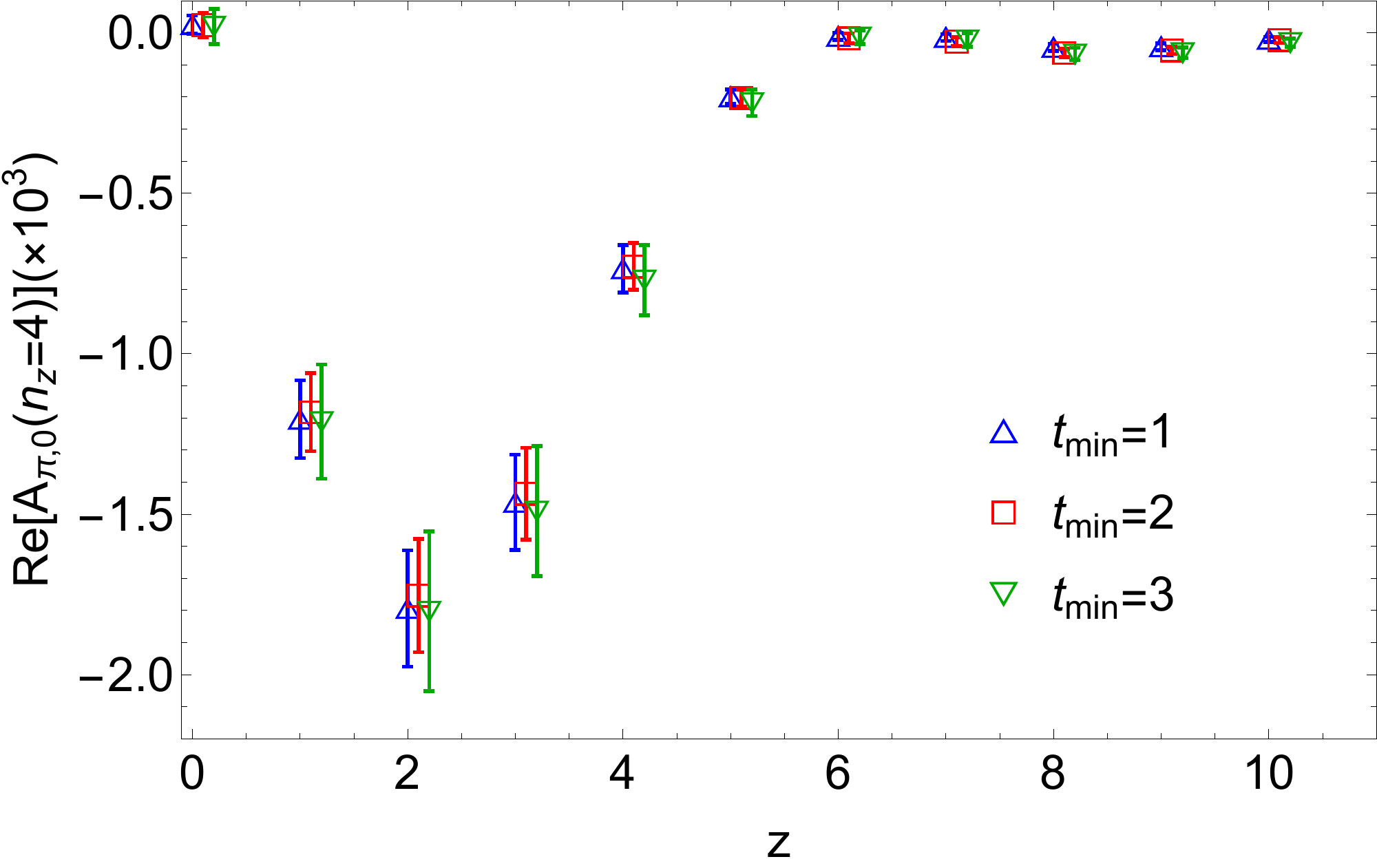}
	\includegraphics[width=0.32\linewidth]{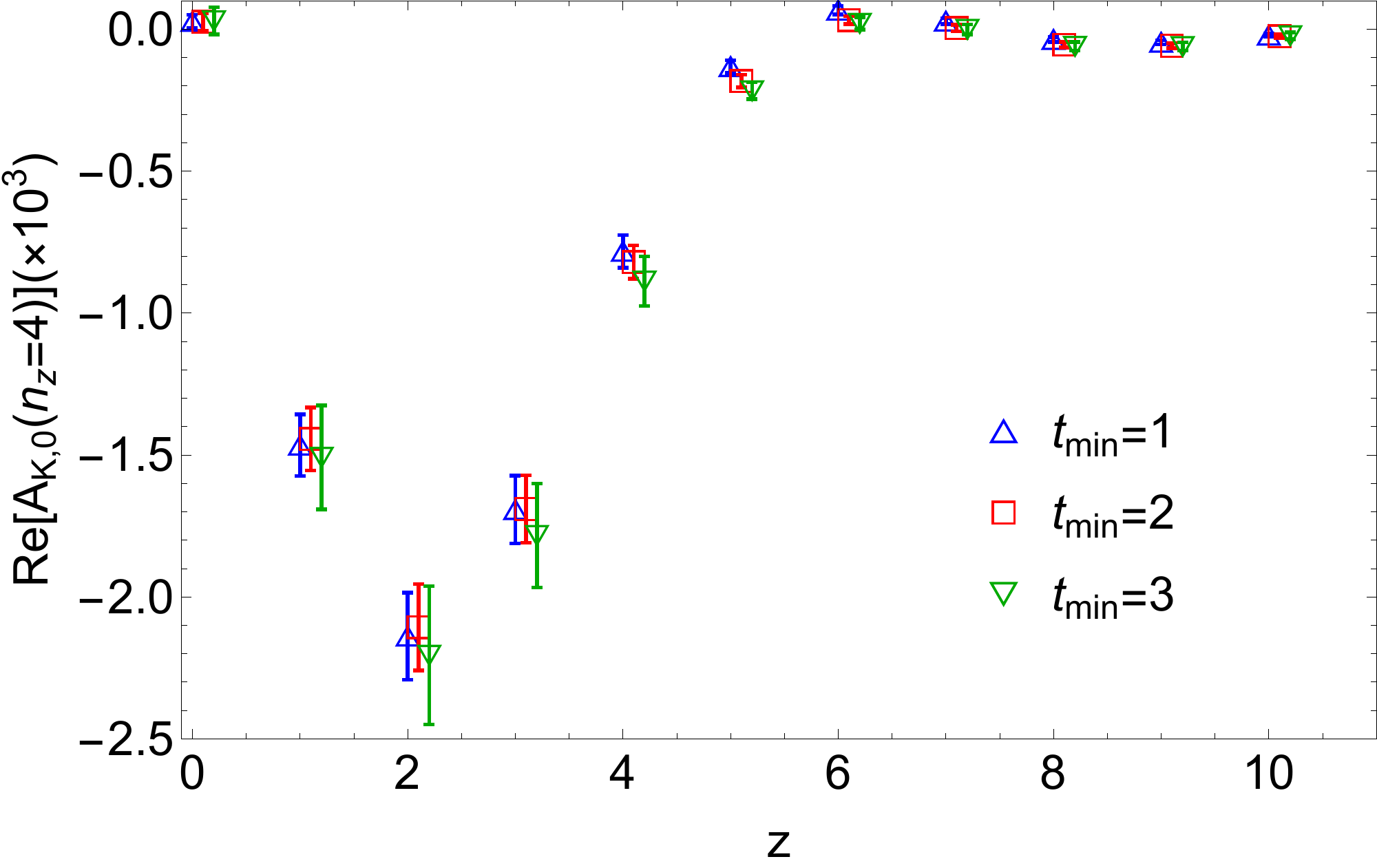}
	\includegraphics[width=0.32\linewidth]{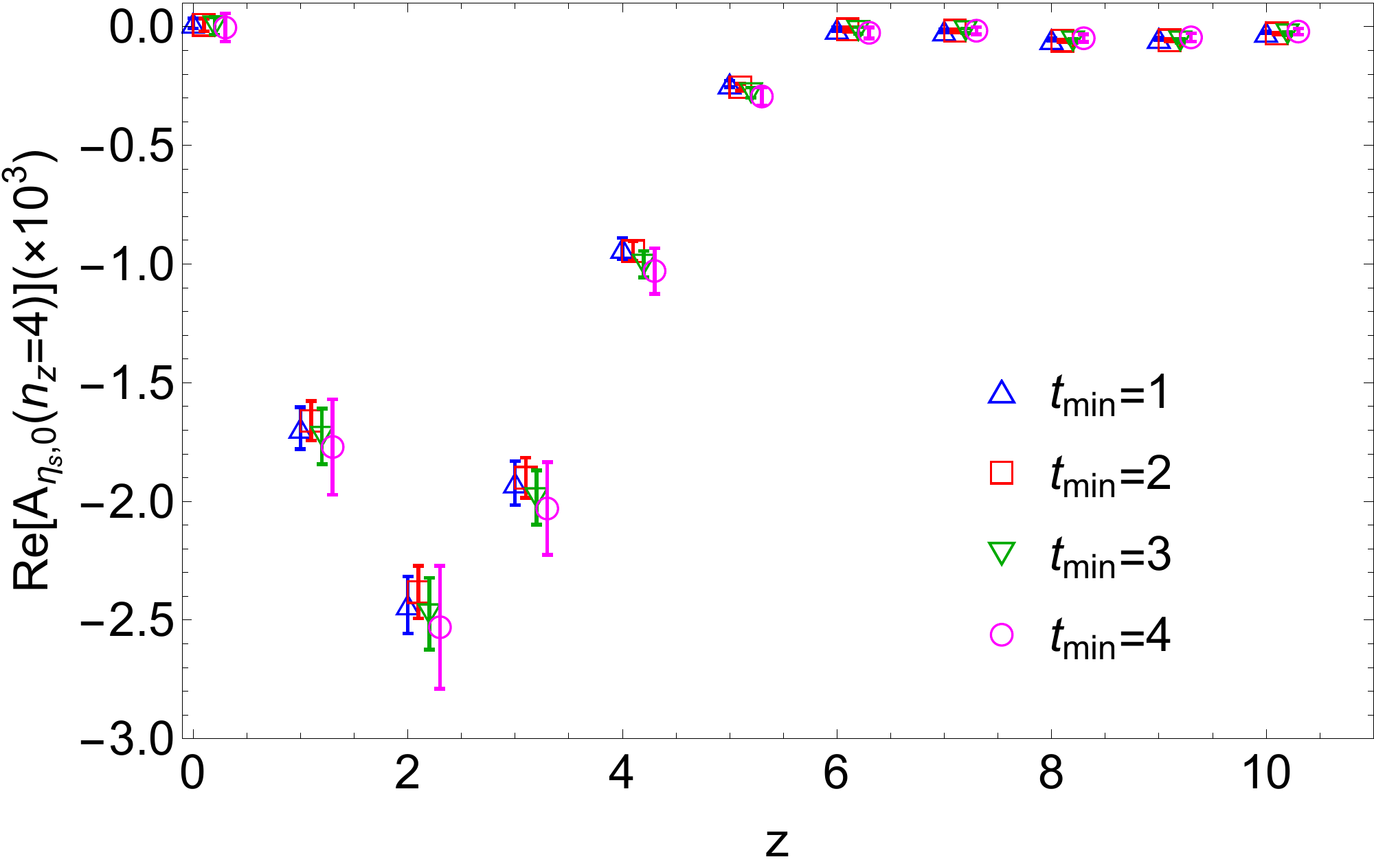}	
	\includegraphics[width=0.32\linewidth]{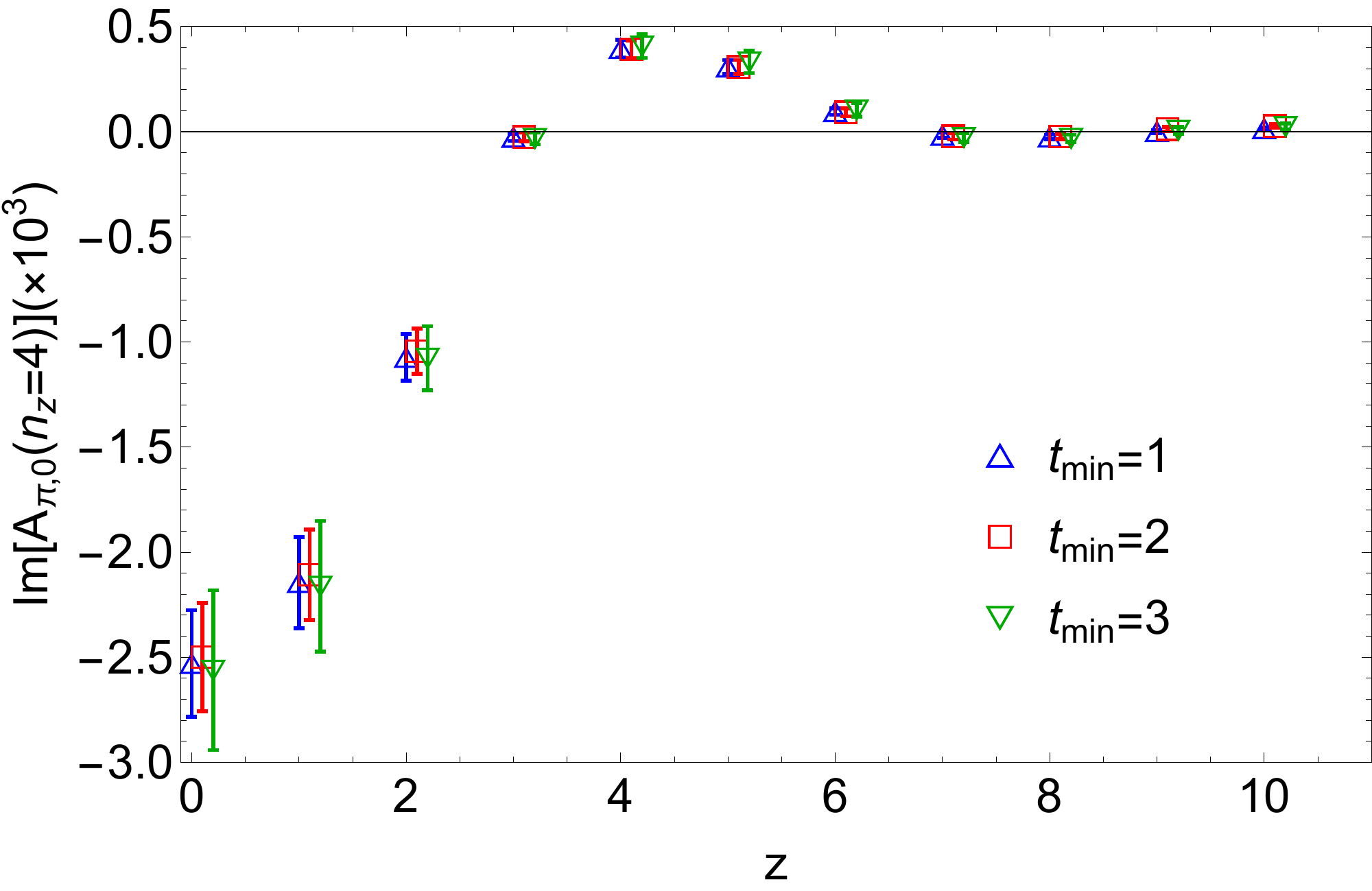}
	\includegraphics[width=0.32\linewidth]{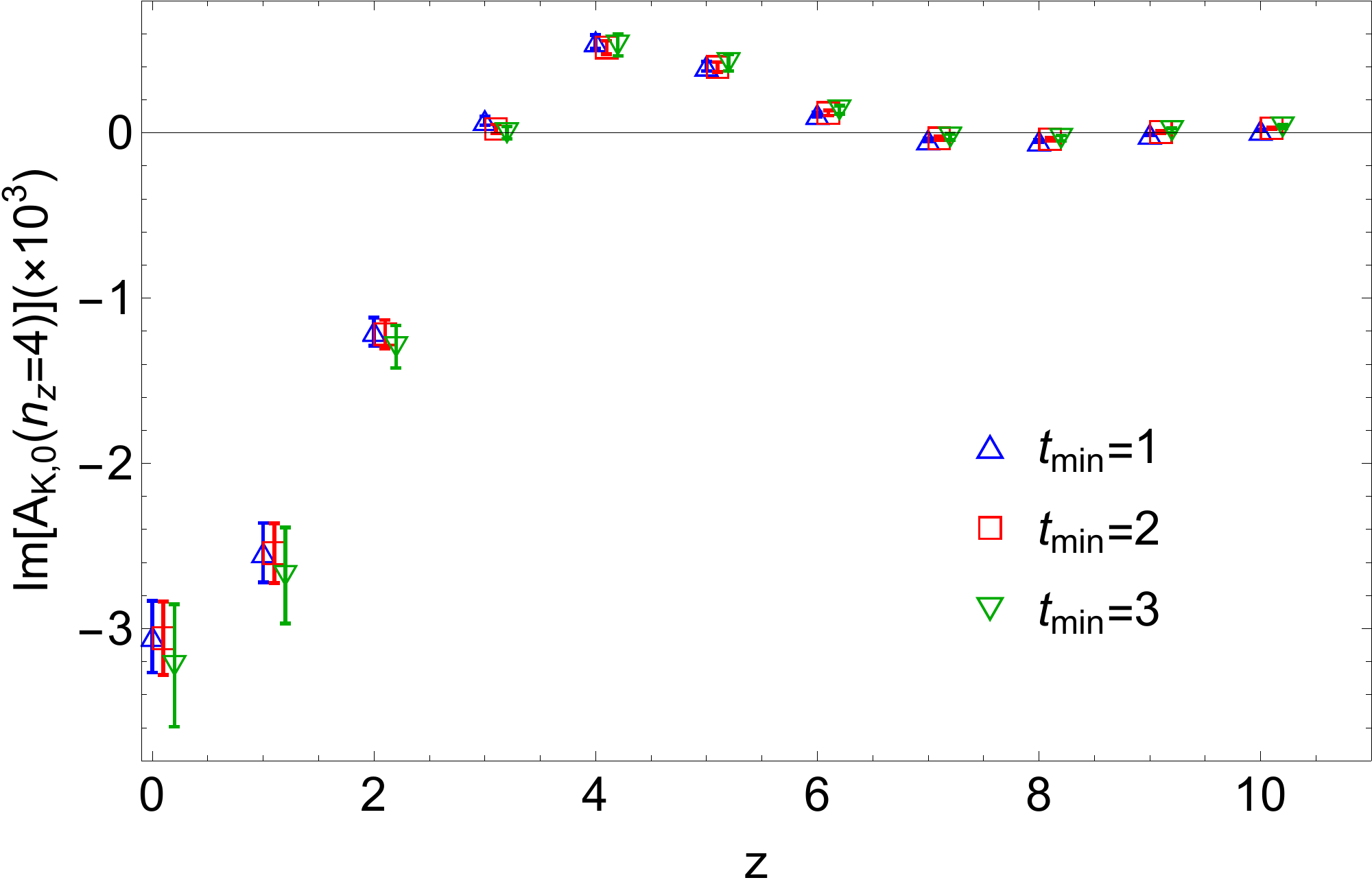}
	\includegraphics[width=0.32\linewidth]{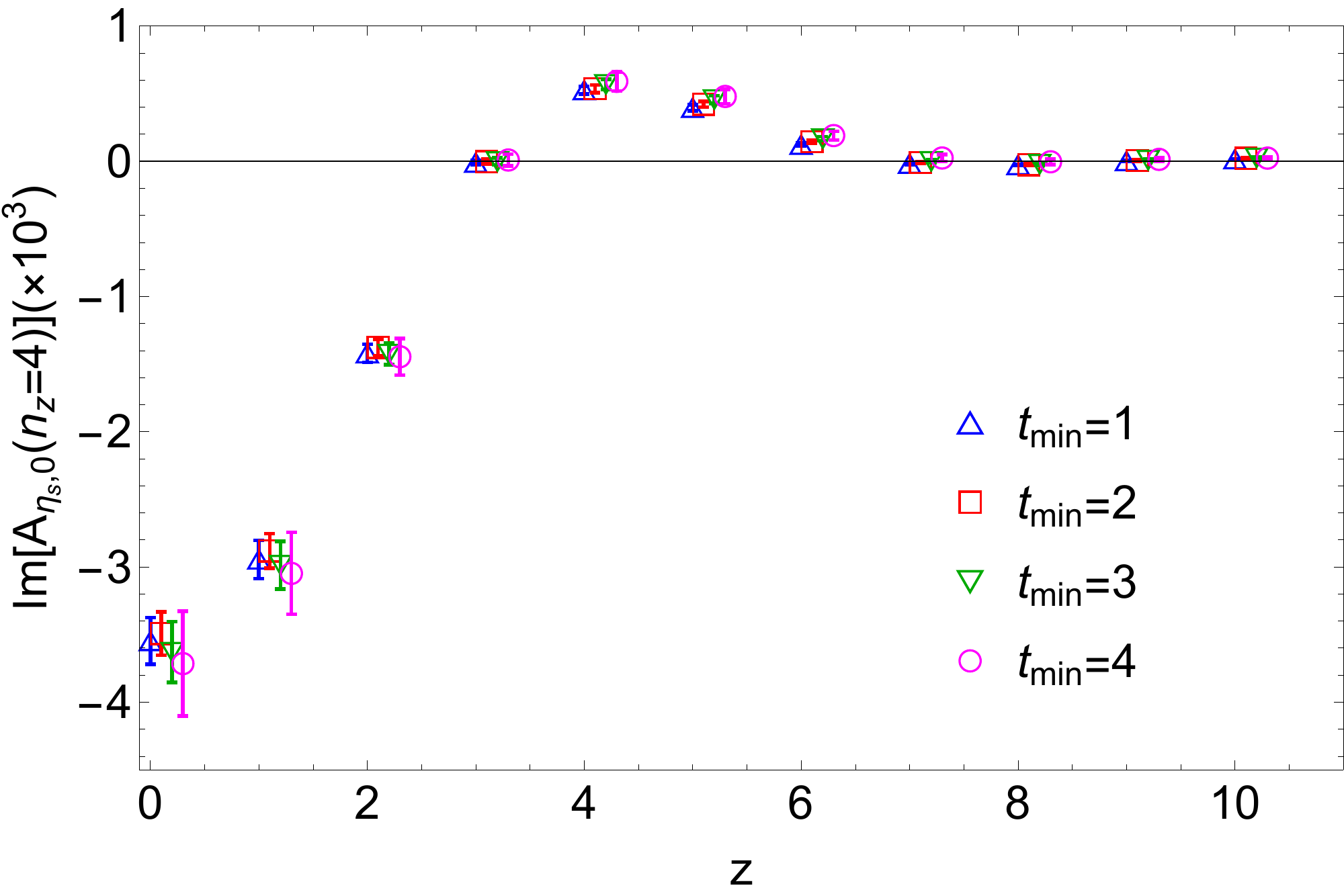}
	\caption{The real (top row) and imaginary (bottom row) ground-state amplitude $A_{M,0}$ as a function of $z$ at $P_z=4\frac{2\pi}{L}$ from two-state fits with different fit ranges $[t_{\text{min}}, 10]$ for $\pi$ (left column), $K$ (middle column) and $\eta_s$ (right column)) on the a12m310 ensemble. The ground-state amplitude extracted from different $t_{\text{min}}$ are consistent with each other within error, while larger $t_{\text{min}}$ results in larger uncertainties. For $\pi$, $K$ and $\eta_s$, $t_{\text{min}}=\{2, 2, 3\}$ are used in the final analysis for this ensemble.}
	\label{fig:A0_z_a12}
\end{figure*}

By varying the fit range for the two-point correlators, we obtain different sets of ground-state coefficients. These fit results on three lattices are shown in Fig.~\ref{fig:A0_z_a06}, Fig.~\ref{fig:A0_z_a09} and Fig.~\ref{fig:A0_z_a12}. Fit results from different ranges are generally consistent with each other. Taking both fit stability and fit qualities on all operators into account, we choose $t_{\text{min}}=\{4, 4, 5\}$ for $\pi$, $K$ and $\eta_s$ on a06m310 lattice, $t_{\text{min}}=\{5, 4, 5\}$ on a09m310 lattice, and $t_{\text{min}}=\{2, 2, 3\}$ on a06m310 lattice.

\begin{figure*}
	\centering
	\includegraphics[width=0.45\linewidth]{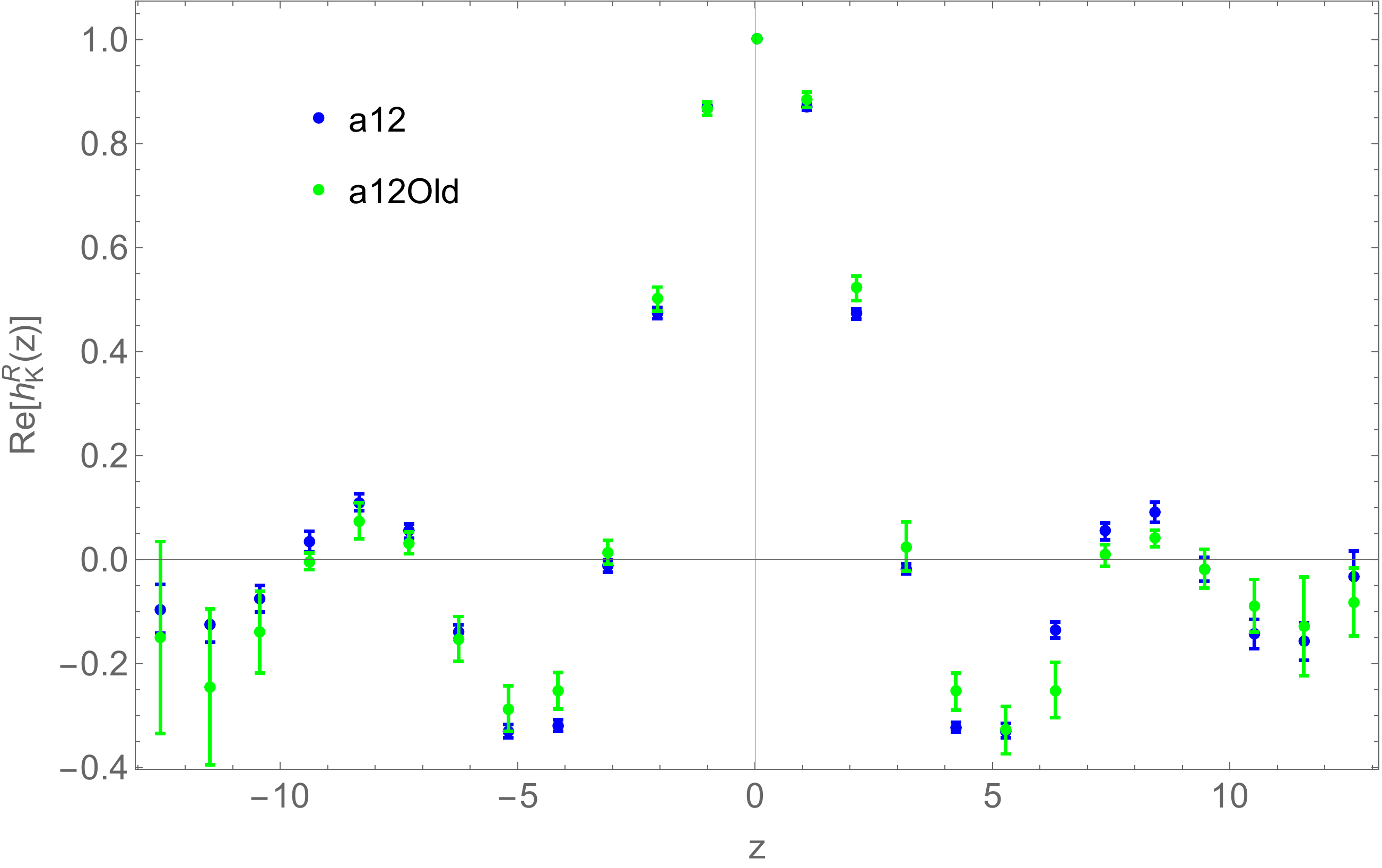}
	\includegraphics[width=0.45\linewidth]{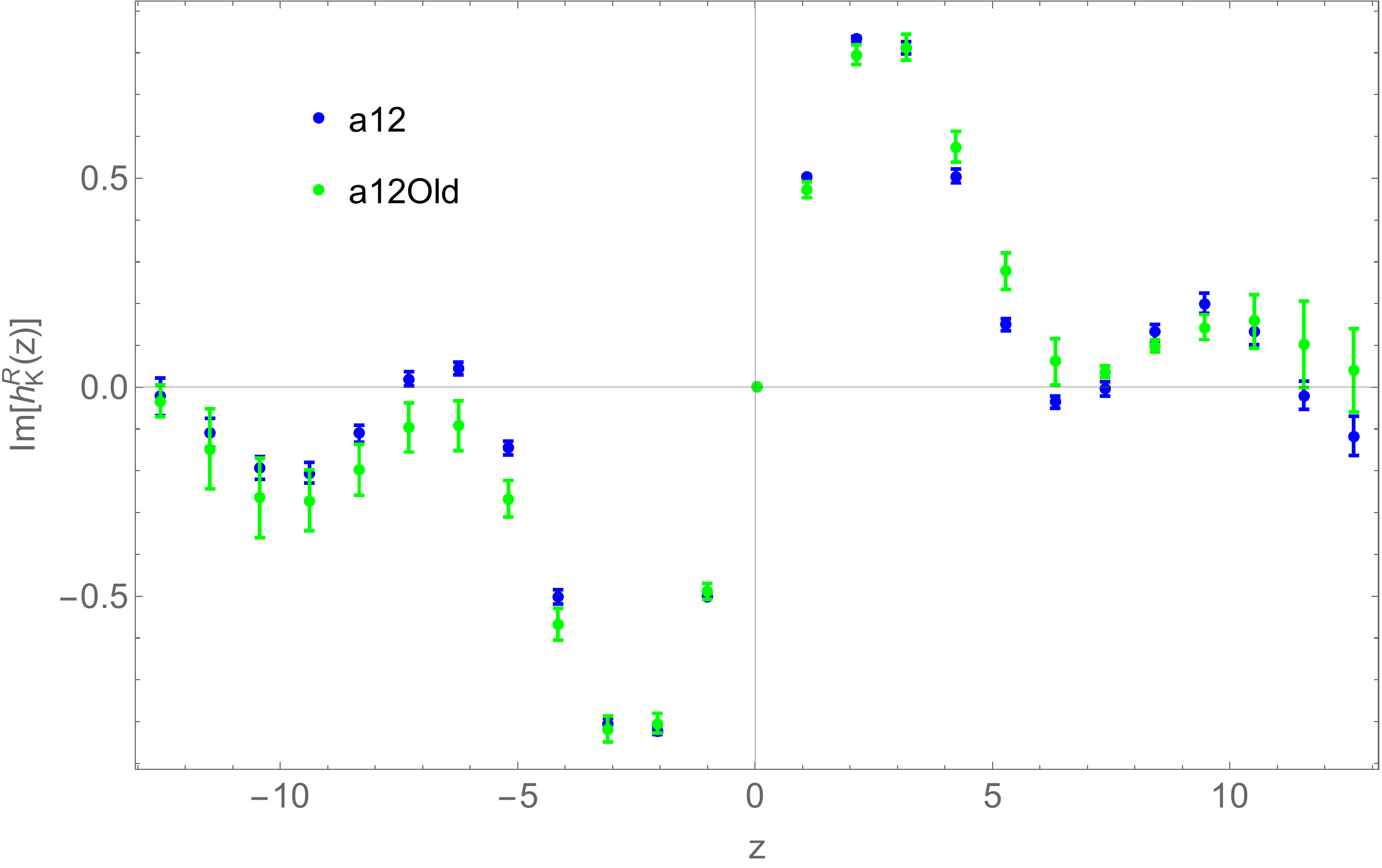}
	\caption{Comparison of the kaon ME (blue points) with previous results~\cite{Chen:2017gck}  (green points) on the $a\approx0.12$~fm ensemble with $n_z=4$.}
	\label{fig:old_compare}
\end{figure*}

We show a comparison of our new data and the data from the previous work~\cite{Chen:2017gck} in Fig.~\ref{fig:old_compare}. We see that they are consistent at most points; however, these slight deviations can result in very different asymmetry behavior, because the asymmetry is only a few percent of the overall magnitude.

\begin{figure*}
	\centering
	\includegraphics[width=0.48\linewidth]{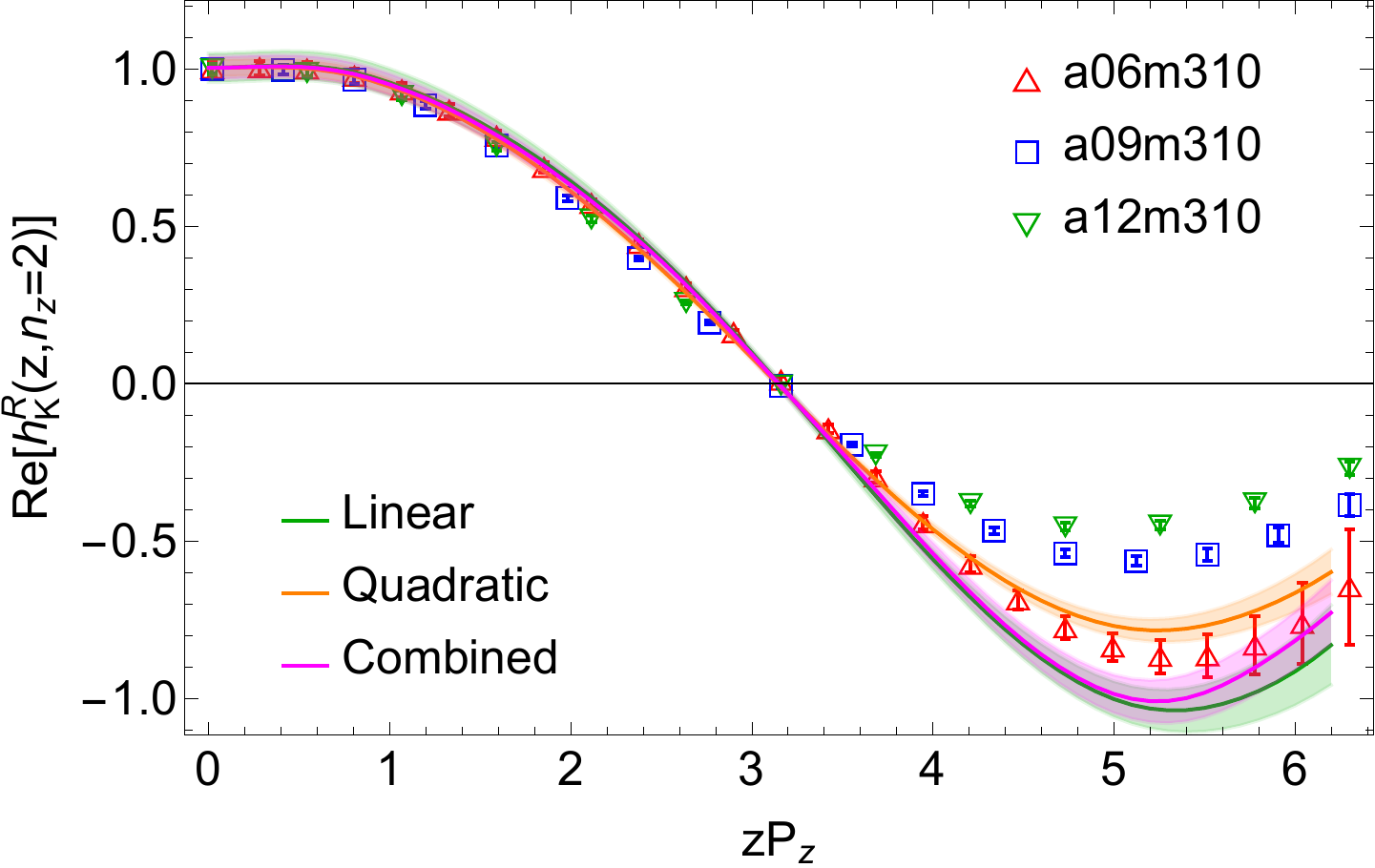}
	\includegraphics[width=0.48\linewidth]{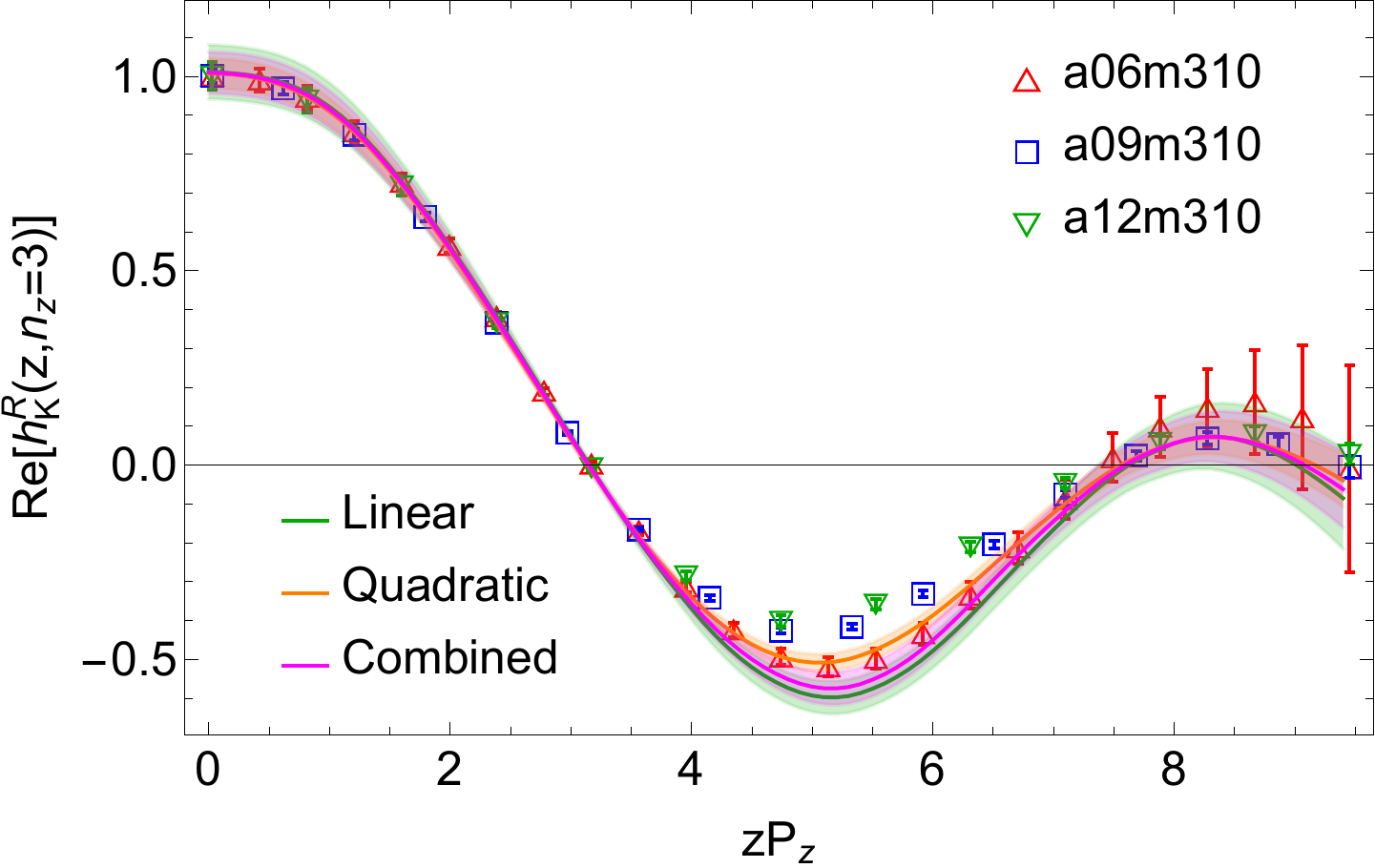}
	\caption{Extrapolation of the kaon renormalized matrix elements at $P_z=n_z\times \frac{2\pi}{L}$ with $n_z=2$ (left) and $n_z=3$ (right), $\mu^R=3.8$~GeV, $p_z^R=0$ to the continuum limit from two functional forms and their AIC combination. We observe larger discretization effect for small $P_z$ due to the non-negligible higher twist effects.
	}
	\label{fig:extrapolated_small_p}
\end{figure*}

The continuum extrapolation for smaller momenta $P_z=0.86\GeV$ and $P_z=1.29\GeV$ are shown in Fig.~\ref{fig:extrapolated_small_p}. There is a large discretization effect at $P_z=0.86\GeV$, which may come from higher-twist effects and the $\frac{1}{a^2}$ power divergent pole.

\bibliographystyle{apsrev4-1}

\end{document}